\documentclass{article}
\usepackage[utf8]{inputenc}
\usepackage{geometry}           
\usepackage[utf8]{inputenc}     
\usepackage{palatino}           
\usepackage[english]{babel}    
\usepackage{graphicx}           
\usepackage{indentfirst}        
\usepackage[unicode]{hyperref}  
\usepackage{amsmath,amsfonts,amssymb}
\usepackage{color}
\usepackage{ulem}
\usepackage{url}
\usepackage{bm}
\usepackage{bbm}
\usepackage{float}
\usepackage{mathtools}
\usepackage[toc,page]{appendix}
\usepackage{setspace}
\usepackage{lipsum}
\usepackage{needspace}
\usepackage{xpatch}
\xpretocmd{\section}{\needspace{6\baselineskip}}{}{}
\usepackage[section]{placeins}
\usepackage[T1]{fontenc}
\usepackage{caption}
\usepackage{subcaption}
\usepackage{dsfont}
\usepackage{enumerate}
\usepackage{hyperref}

\newcommand{\Sec}[2]{\section{#1\label{Sec:#2}}}
\newcommand{\sSec}[2]{\subsection{#1 \label{Sec:#2}}}
\newcommand{\ssSec}[2]{\subsubsection{#1 \label{Sec:#2}}}

\newcommand{\setR}{\mathbbm{R}}

\newcommand{\ang}[2]{\widehat{\pare{\vec{#1},\vec{#2}}}}
\newcommand{\conv}[1]{\stackrel{#1}{\ast}}
\newcommand{\pare}[1]{\left(\, #1 \, \right)}

\newcommand{\bra}[1]{\left[\, #1 \, \right]}
\newcommand{\abs}[1]{\left|\, #1 \, \right|}

\newcommand{\Set}[1]{\left\{\, #1 \, \right\}}

\newcommand{\vect}[1]{\pare{\begin{array}{ccc}#1\end{array}}}
\newcommand{\ve}{\vec{e}}
\newcommand{\vv}{\vec{v}}

\newcommand{\vz}{\vec{0}}

\newcommand{\vcF}{\vec{\cF}}
\newcommand{\vcP}{\vec{\cP}}
\newcommand{\vX}{\vec{X}}
\newcommand{\vcX}{\vec{\cX}}
\newcommand{\vgamma}{\vec{\gamma}}
\newcommand{\vnabla}{\vec{\nabla}}

\newcommand{\Cell}[2]{{#1}_{#2}} 
\newcommand{\V}[2]{V_{\Cell{#1}{#2}}} 
\newcommand{\K}[2]{\cK_{\Cell{#1}{#2}}}  
\newcommand{\KS}[2]{\cK_{\Cell{#1}{#2},S}}  

\newcommand{\A}[2]{A_{\Cell{#1}{#2}}}
\newcommand{\E}[2]{\cE^{#1}_{#2}}
\newcommand{\F}[2]{F_{\Cell{#1}{#2}}}
\newcommand{\N}[2]{\cN_{\Cell{#1}{#2}}}
\newcommand{\R}[2]{R_{\Cell{#1}{#2}}}
\newcommand{\W}[4]{W^{\Cell{#1}{#2}}_{\Cell{#3}{#4}}}
\newcommand{\Ps}[4]{P^{\Cell{#1}{#2}}_{\Cell{#3}{#4}}}
\newcommand{\Pst}[2]{P_{\Cell{#1}{#2}}}
\newcommand{\D}[4]{d\bra{\Cell{#1}{#2}, \, \Cell{#3}{#4}}}
\newcommand{\Dd}[4]{d^2\bra{\Cell{#1}{#2}, \, \Cell{#3}{#4}}}

\newcommand{\cD}{{\mathcal D}}
\newcommand{\cE}{{\mathcal E}}
\newcommand{\cG}{{\mathcal G}}
\newcommand{\cH}{{\mathcal H}}
\newcommand{\cI}{{\mathcal I}}
\newcommand{\cJ}{{\mathcal J}}
\newcommand{\cK}{{\mathcal K}}
\newcommand{\cL}{{\mathcal L}}
\newcommand{\cM}{{\mathcal M}}
\newcommand{\cN}{{\mathcal N}}
\newcommand{\cP}{{\mathcal P}}
\newcommand{\cS}{{\mathcal S}}
\newcommand{\cF}{{\mathcal F}}
\newcommand{\cX}{{\mathcal X}}
\newcommand{\cW}{{\mathcal W}}

\newcommand{\vk}{\vec{k}}

\newcommand{\vphi}{\vec{\phi}}

\newcommand{\vpsi}{\vec{\psi}}

\newcommand{\Prob}[1]{\mathds{P}\left[\, #1 \, \right]}
\newcommand{\Probc}[2]{\mathds{P}\bra{#1 \, \left| \, #2 \right.}}

\providecommand{\keywords}[1]
{
  \small	
  \textbf{\textit{Keywords---}} #1
}

\title{
On the potential role of lateral connectivity in retinal anticipation}
\author{Selma Souihel$^1$ and Bruno Cessac$^1$\\
$^1$ \small{Université Côte d'Azur, Inria, Biovision team and Neuromod Institute, France.}
\date{}
}

\begin{document}

\maketitle

\begin{abstract}
    We analyse the potential effects of lateral connectivity (amacrine cells and gap junctions) on motion anticipation in the retina.
    Our main result is that lateral connectivity can - under conditions analysed in the paper - trigger a wave of activity enhancing the anticipation mechanism provided by local gain control \cite{berry-brivanlou-etal:99,chen-marre-etal:13}. We illustrate these predictions by two examples studied in the experimental literature: differential motion sensitive cells \cite{baccus-meister:02} and direction sensitive cells where direction sensitivity is inherited from asymmetry in gap junctions connectivity \cite{trenholm-schwab-etal:13}.
    We finally present reconstructions of retinal responses to 2D visual inputs to assess the ability of our model to anticipate motion in the case of three different 2D stimuli. 
\end{abstract}


\Sec{Introduction}{Introduction}
Our visual system has to constantly handle moving objects. Static images do not exist for it, as the environment, our body, our head, our eyes are constantly moving. A "computational", contemporary view, assimilates the retina to an "encoder", converting the light photons coming from a visual scene into spike trains sent - via the axons of Ganglion cells (GCells) that constitute the optic nerve -  to the thalamus, and then to the visual cortex acting as a "decoder". In this view, comparing the size and the number of neurons in the retina - about $1$ million of GCells
 (humans) 
- to the size, structure, and number of neurons in the visual cortex (around $538$ million per hemisphere in the human visual cortex \cite{colonnier-okusky:81}) the "encoder" has to be quite smart to efficiently compress the visual information coming from a world made of moving objects. Although it has long been thought that the retina was not more than a simple camera, there are more and more evidences that the retina is "smarter than neuroscientists believed" \cite{gollisch-meister:10}. It is indeed  able to perform complex tasks and general motion features extractions such as approaching motion, differential motion, motion anticipation, allowing the visual cortex to process visual stimuli with more efficiency.

The process leading from the photons reception in the retina to the cortical response takes about $30-100$ milliseconds. Most of this delay is due to photo-transduction. Though this might look fast, it is actually too slow. A tennis ball moving at $30$ m/s - $108$ km/h (the maximum measured speed is about $250$ km/h) covers between $0.9$ and $3$ m during this time, so, without a mechanism compensating this delay it wouldn't be possible to play tennis (not to speak of survival, a necessary condition for a species to reach the level where playing tennis becomes possible). The visual system is indeed able to extrapolate the trajectory of a moving object to perceive it at its actual
location. This corresponds to anticipation mechanisms taking place in the visual cortex and in the retina, with different modalities \cite{valois-valois:91,nijhawan:94,baldo-klein:95,nijhawan:97}.

In the early visual cortex an object moving across the visual field triggers a wave of activity ahead of motion, thanks to the cortical lateral connectivity \cite{benvenuti-chemla-etal:20,subramaniyan-ecker-etal:18,jancke-erlaghen-etal:04}. Jancke et al. \cite{jancke-erlaghen-etal:04} first demonstrated the existence of anticipatory mechanisms in the cat primary visual cortex. They recorded cells in the central visual field of area 17 (corresponding to the primary visual cortex) of anaesthetized cats, responding to small squares of light, either flashed or moving in different directions, and with different speeds. When presented with the moving stimulus, cells show a reduction of neural latencies, as compared to the flashed stimulus. Subramaniyan et al. \cite{subramaniyan-ecker-etal:18} have reported the existence of similar anticipatory effects in the macaque primary visual cortex, showing that a moving bar is processed faster than a flashed bar. They give two possible explanations to this phenomenon : either a shift in the cells receptive fields induced by motion, or a faster propagation of motion signals as compared to the flash signal.

In the retina, anticipation takes a different form. One observes a peak in the firing rate response of GCells to a moving object, occurring \textit{before} the peak response to the same object when flashed. This effect can be explained by purely local mechanisms, at individual cells level  \cite{berry-brivanlou-etal:99,chen-marre-etal:13}. To our best knowledge, collective effects similar to the cortical ones - that is, a rise in the cell’s activity before the object enters in its receptive field due to a wave of activity ahead of the moving object - have not been reported yet. \\

\begin{figure}
\begin{center}
\includegraphics[width=12cm,height=9cm]{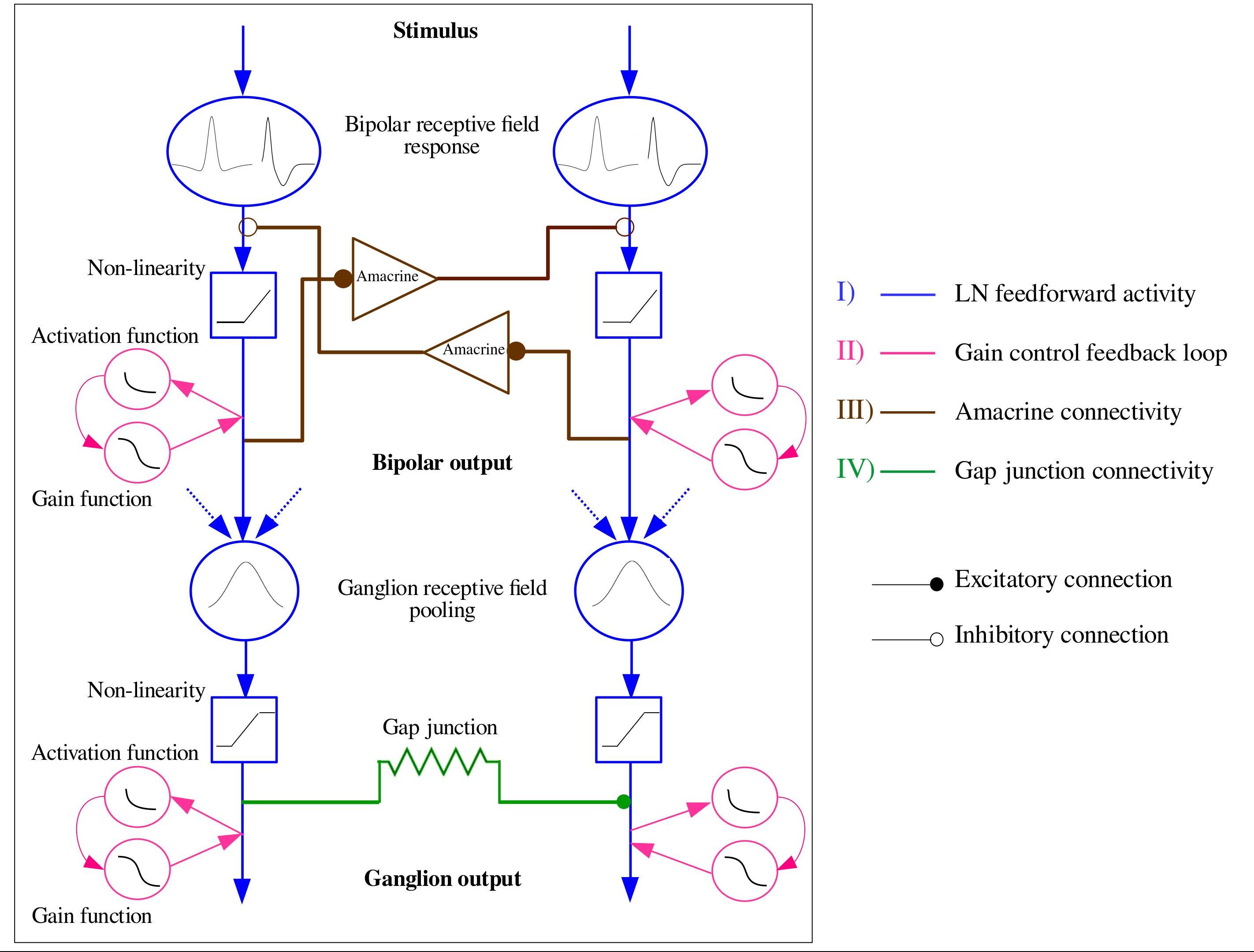} 
\caption{\textbf{Synthetic view of the retina model.} A stimulus is perceived by the retina, triggering different pathways. \textbf{Pathway I (blue)} corresponds to a feed-forward response where, from top to bottom:
The stimulus is first convolved with a spatio-temporal receptive field that mimics the Outer Plexiform Layer (OPL) ("Bipolar receptive field response"). This response is rectified by low voltage threshold (blue squares). Bipolar cells responses are then pooled (blue circles with blue arrows) and input Ganglion cells. The firing rate response of a Ganglion cell is a sigmoidal function of the voltage (blue square). 
Gain control can be applied at the Bipolar and Ganglion cells level (pink circles) triggering anticipation. 
This corresponds to the label \textbf{II (pink)} in the figure. 
Lateral connectivity is featured by \textbf{pathway III (brown)} through ACells, and \textbf{pathway IV (green) } through gap-junctions at the level of GCells. \label{Fig:architecture} 
}
\end{center}
\end{figure}

In a classical, Hubel-Wiezel-Barlow \cite{hubel-wiesel:60,barlow:61,olshausen-field:98} view of vision, each retinal ganglion cell carries a flow of information with an efficient coding strategy maximizing the available channel capacity by minimizing the redundancy between GCells. From this point of view, the most efficient coding is provided when GCells are independent encoders (parallel streaming identified by a "I" in Fig. \ref{Fig:architecture}). 
In this setting one can propose a simple and satisfactory mechanism explaining anticipation in the retina, based on gain control at the level of Bipolar cells (BCells) and GCells (label "II" in \ref{Fig:architecture}) \cite {berry-brivanlou-etal:99,chen-marre-etal:13}. 

Yet, some GCells are connected. Either directly, by electric synapses-gap junctions (pathway IV in Fig. \ref{Fig:architecture}), or indirectly, via specific Amacrine cells (ACells, pathway III in Fig. \ref{Fig:architecture}). It is known that these pathways are involved in motion processing by the retina.
AII ACells play a fundamental role in the interaction between the ON and OFF cone pathway \cite{nelson-kolb:04}. There are GCells able to detect the differential motion of an object onto a moving background \cite{baccus-meister:02}, thanks to ACells lateral connectivity. Some GCells are direction sensitive because they are connected via a specific, asymmetric, gap junctions connectivity \cite{trenholm-schwab-etal:13}. 
\textit{Could lateral connectivity play a role in motion anticipation, inducing a wave of activity ahead of the motion, similar to the cortical anticipation mechanism ?} While some studies hypothesize that local gain control mechanisms can be explained by the prevalence of inhibition in the retinal connectome \cite{johnston-lagnado:15}, the mechanistic aspects of the role of lateral connectivity on motion anticipation has not, to the best of our knowledge, been addressed yet on either experimental or computational grounds.\\

In this paper, we address this question from a modeller, computational neuroscientist, point of view. We propose here a simplified description of the pathways I, II, III, IV of Fig. \ref{Fig:architecture}, grounded on biology, but not sticking at it, to numerically study the potential effects of gain control combined  with lateral connectivity - gap junctions or ACells - on motion anticipation. The goal here is not to be biologically realistic but, instead, to propose from biological observations potential mechanisms enhancing the retina's capacity to anticipate motion and compensate the delay introduced by photo-transduction and feed-forward processing in the cortical response. We want the mechanisms to be as generic as possible, so that the detailed biological implementation is not essential. This has the advantage of making the model more prone to mathematical analysis.  

The first contribution of our work lies in the development of a model of retinal anticipation where GCells have gain control, orientation selectivity and are laterally connected. It is based on a model introduced by Chen et al. in \cite{chen-marre-etal:13} - itself based on \cite{berry-brivanlou-etal:99} - reproducing several motion processing features: anticipation, alert response to motion onset and motion reversal. The original model handles one dimensional motions and its cells are not laterally connected (only pathways I and II were considered). The extension proposed here features cells with oriented receptive field, although our numerical simulations do not consider this case (see discussion). Lateral connectivity is based on biophysical modelling and existing literature \cite{tauchi-masland:84,destexhe-mainen-etal:94b,baccus-meister:02,hosoya-baccus-etal:05,trenholm-schwab-etal:13}. In this framework, we study different types of motion. We start with a bar moving with constant speed and study the effect of contrast, bar size, and speed on anticipation, generalizing previous studies by Berry et al \cite{berry-brivanlou-etal:99} and Chen et al \cite{chen-marre-etal:13}.  We then extend the analysis to two dimensional motions, investigating e.g. angular motion and curved trajectories. Far from making an exhaustive study of anticipation in complex stimuli, the goal here is to calibrate anticipation, without lateral connectivity, so as to compare the effect when connectivity is switched on. 

The second contribution emphasizes a potential role of lateral connectivity (gap junctions and ACells) on anticipation. For this, we first make a general mathematical analysis concluding that lateral connectivity can induce a wave triggered by the stimulus which, under specific conditions can improve anticipation. The effect depends on the connectivity graph and is non linearly tuned by gain control. In the case of gap junctions, the wave propagation depends whether connectivity is symmetric (the standard case) or asymmetric, as proposed by Trenholm et al. in \cite{trenholm-schwab-etal:13} for a specific type of direction sensitive GCells. In the case of ACells, the connectivity graph is involved in the spectrum of a propagation operator controlling the time evolution of the network response to a moving stimulus. We instantiate this general analysis by studying differential motion sensitive cells \cite{baccus-meister:02} with two types of connectivity: nearest neighbours, and a random connectivity, inspired from biology \cite{tauchi-masland:84}, where only numerical results are shown.
In general, the anticipation effect depends on the connectivity graph structure and the intensity of coupling between cells as well as on the respective characteristic times of response of cells, in a way that we analyse mathematically and illustrate numerically.

We actually observe two forms of anticipation. The first one, discussed in the beginning of this introduction and already observed in \cite{berry-brivanlou-etal:99,chen-marre-etal:13}, is a shift in the peak of a retinal Gcell response, occurring before the object reaches the center of its receptive field. In our case, lateral connectivity can enhance the shift improving the mere effect of gain control. The second anticipation effect we observe is a raise in GCells activity before the bar reaches the receptive field of the cell, similarly to what is observed in the cortex \cite{benvenuti-chemla-etal:20}. To the best of our knowledge, this effect has not been studied in the retina and constitutes therefore a prediction of our model.\\

The paper is organized as follows. Section \ref{Sec:Model} introduces the model of retinal organization and cells types dynamics, ending up with a system of non linear differential equations driven by a time-dependent stimulus. Section \ref{Sec:Results} is divided in four parts. The first part analyses mathematically the potential anticipation effects in a general setting, before considering the role of ACells and lateral inhibition on anticipation (section \ref{Sec:Role_Amacrine}) and gap junctions (section \ref{Sec:Role_Gaps}). Both sections contain general mathematical results, as well as numerical simulations for one dimensional motion. 
The fourth part investigates examples of two dimensional motions. The last section is devoted to discussion and conclusion. In Appendix \ref{Sec:parameters}, we have added the values of parameters used in simulations, and, in Appendix \ref{Sec:Spatio-temp_filtering} the receptive fields mathematical form used in the paper, as well as the numerical method to compute efficiently the response of oriented two dimensional receptive fields to spatio-temporal stimuli. Appendix \ref{Sec:Random_Connectivity} presents a model of random connectivity from Amacrine to Bipolar cells inspired from biological data \cite{tauchi-masland:84}. Finally, Appendix (\ref{Sec:Linear_Analysis}) contains  mathematical results which constitute the skeleton of the work, but whose proof would be too long to integrate in the core of the paper. This work is based on Selma Souihel's PhD thesis where more extensive results can be found \cite{souihel:19}. In particular, there is an analysis of the conjugated effects of retinal and cortical anticipation, subject of a forthcoming paper, and briefly discussed in the conclusion. 

In all the following simulations, we use the CImg Library, an open-source C++ tool kit for image processing, in order to load the stimuli and reconstruct the retina activity. The source code is available on demand.

\Sec{Material and methods}{Model}
\sSec{Retinal organization}{Retinal_organization}

In the retinal processing light photons coming from a visual scene are converted into voltage variations by photoreceptors (cones and rods). The complex hierarchical and layered structure of the retina allows to convert these variations into spike trains, produced by Ganglion Cells (GCells) and conveyed to the thalamus via their axons. We considerably simplify this process here. Light response induces a voltage variations of Bipolar cells (BCells), laterally connected via Amacrine cells (ACells), and feeding GCells, as depicted in Fig. \ref{Fig:architecture}. We describe this structure in details here. Note that neither BCells nor ACells are spiking. They act synaptically on each other by graded variations of their potential.

\begin{figure}[H]
\centering
\includegraphics[width=10cm,height=8cm]{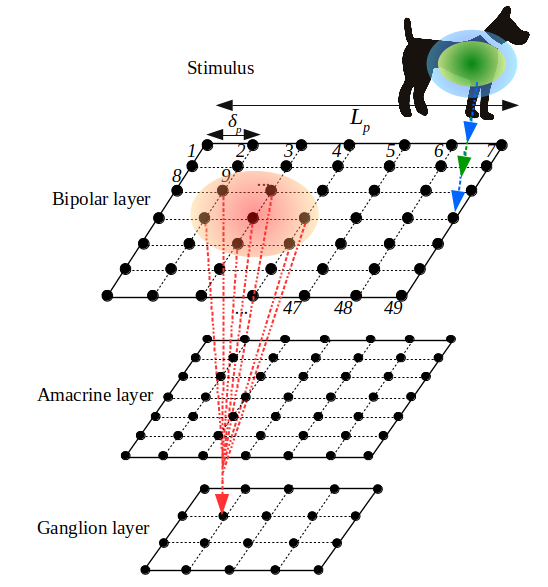}
\caption{\textbf{Example of a retina grid tiling and indexing.} 
 The green and blue ellipses denote respectively the positive center and the negative surround of the Bcell receptive field $\cK_S$. The center of RF coincides with the position of the cell (blue and green arrows). The red ellipse denotes the ganglion cell pooling over bipolar cells (eq. \eqref{eq:Pooling}). 
\label{Fig:tiling}}
\end{figure}

We assimilate the retina to a flat, two dimensional square of edge length $L$ mm. Therefore, we do not integrate the $3$ dimensional structure of the retina in the model, merely for mathematical convenience. Spatial coordinates are noted $x,y$ (see Fig. \ref{Fig:tiling} for the whole structure).
  
In the model, each cell population tiles the retina with a regular square lattice. The density of cells is therefore uniform for convenience but the extension to non uniform density can be afforded. 
For the population $p$ we note $\delta_p$ the lattice spacing in mm, and $N_p$ the total number of cells. Without loss of generality we assume that $L$, the retina's edge size, is a multiple of $\delta_p$. We note $L_p=\frac{L}{\delta_p}$, the number of cells $p$ per row or column so that $N_p=L_p^2$. Each cell in the population $p$ has thus Cartesian coordinates
$(x,y)=(i_x \delta_p,i_y \delta_p)$, $(i_x,i_y) \in \Set{1, \dots, L_p}^2$. To avoid multiples indices, 
we associate to each pair $(i_x,i_y)$ a unique index $i=i_x+(i_y-1) \, L_p$. The cell of population $p$, located at coordinates $(i_x \delta_p, i_y \delta_p)$ is then denoted by $\Cell{p}{i}$. 
We note $\D{p}{i}{p'}{j}$ the Euclidean distance between $\Cell{p}{i}$ and $\Cell{p'}{j}$.

We use the notation $\V{p}{i}$ for the membrane potential of cell $\Cell{p}{i}$.
Cells are coupled.
The synaptic weight from cell $\Cell{p}{j}$  to cell $\Cell{q}{i}$ reads $\W{p}{j}{q}{i}$. Thus, the pre-synaptic neuron is expressed in the upper index; the post-synaptic, in the lower index. 
Dynamics of cells is voltage-based. This is because our model is constructed from Chen et al model \cite{chen-marre-etal:13} itself derived from Berry et al \cite{berry-brivanlou-etal:99} where a voltage-based description is used. Implicitly, voltage is measured with respect to the rest state of the cell ($\V{p}{i}=0$ when the cell receives no input).

\sSec{Bipolar cells layer}{Bipolar_cells_layer}

The model consists first of a set of $N_B$ BCells, regularly spaced by a distance $\delta_B$, with spatial coordinates $x_i,y_i$, $i=1 \dots N$. Their voltage, a function of the stimulus, is computed as follows.

\ssSec{Stimulus response and receptive field}{StimRF}

The projection of the visual scene on the retina ("stimulus") is a function $\cS(x,y,t)$ where $t$ is the time coordinate. As we don't consider color sensitivity here $\cS$ characterizes a black and white scene, with a control on the level of contrast $\in [0,1]$. 
A Receptive Field (RF) is a region of the visual field (the physical space) in which stimulation alters the voltage of a cell. Thus, BCell $i$ has a spatio-temporal receptive field $\K{B}{i}$, featuring the biophysical processes occurring at the level of the Outer Plexiform Layer (OPL), that is photo-receptors (rod-cones) response modulated by Horizontal Cells (HCells). As a consequence, in our model, the voltage of BCell $i$ is stimulus-driven by the term:
\begin{equation}\label{eq:Vdrive}
V_{i_{drive}}(t)= \bra{\K{B}{i} \conv{x,y,t} \cS}(t) =
\int_{x=-\infty}^{+\infty} \,\int_{y=-\infty}^{+\infty} \,
\int_{s=-\infty}^{t} \,  \cK(x-x_i,y-y_i,t-s) \, \cS(x,y,s) dx \, dy \, ds,
\end{equation}
where $\conv{x,y,t}$ means space-time convolution.
We consider only one family of BCells so that the kernel $\cK$ is the same for all BCells. What changes is the center of the RF, located at $x_i,y_i$, which also corresponds to the coordinates of the BCell $i$ .   
We consider in the paper separable kernel
$\cK(x,y,t)=\cK_S(x,y) \, \cK_T(t)$ where $\cK_S$ is the spatial part and $\cK_T$ the temporal part. The detailed form of $\cK$ is given in Appendix \ref{Sec:Spatio-temp_filtering}.

We have :
\begin{equation}\label{eq:dVdrive}
\frac{d V_{i_{drive}}}{d t}=\bra{\K{B}{i} \conv{x,y,t}  \frac{d \cS}{dt}}(t),
\end{equation}
resulting from the condition $\K{B}{i}(x,y,0)=0$ (see Appendix \ref{Sec:Spatio-temp_filtering}).
Note that the exponential decay of the spatial and temporal part at infinity ensures the existence of the space-time integral. 
The spatial integral $\int_{\setR^2} \cK_S(x,y) S(x,y,u) \, dx \,dy$ is  numerically computed using error function in the case of circular RF, and a computer vision method from Geusenroek et al. \ \cite{geusebroek-smeulders-etal:03} in the case of anisotropic RF, allowing to integrate generalized Gaussians with an efficient computational time. This method is described in the Appendix, section \ref{Sec:Spatio-temp_filtering}.

%



For explanations purposes, we will often use the approximation
of $V_{i_{drive}}$ by a Gaussian pulse, with width $\sigma$, propagating at constant speed $v$ along the direction $\ve_x$:
\begin{equation}\label{eq:VdriveAppGauss}
V_{i_{drive}}(t)=\frac{A_0}{\sqrt{2 \, \pi} \, \sigma} \, e^{-\frac{1}{2} \, \frac{\pare{x - v t}^2}{\sigma^2}} \equiv \frac{V_0}{\sqrt{2 \, \pi}} \, e^{-\frac{1}{2} \, \frac{\pare{x - v t}^2}{\sigma^2}},
\end{equation}
where $x=k \, \delta_B$ is the horizontal coordinate of BCell $i$ and where $\sigma$ is in $mm$, $A_0$ is in $mV.mm$ (and is proportional to stimulus contrast), $V_0$ is in $mV$.

\ssSec{BCells voltage and Gain control}{Bipolar_cells_voltage}

In our model, the BCell voltage is the sum of the external drive \eqref{eq:Vdrive} received by the BCell and of a post-synaptic potential $\Pst{B}{i}$ induced by connected ACells:
\begin{equation}\label{eq:VBipTot}
\V{B}{i}(t)=V_{i_{drive}}(t)  + \Pst{B}{i}(t).
\end{equation}
The form of $\Pst{B}{i}$ is given by eq. \eqref{eq:PSP_A_Bip} in the section \ref{Sec:Connections_amacrines_to_bipolars}. $\Pst{B}{i}(t)=0$ when no ACells are considered.

BCells have voltage threshold \cite{berry-brivanlou-etal:99}:  
\begin{equation} \label{eq:NLinRectBip}
\cN_B(\V{B}{i}) =\left\{
	\begin{array}{ll}
		0, \quad &\mbox{if} \quad \V{B}{i} \le \theta_B; \\
		\V{B}{i}-\theta_B, \quad &\mbox{else}. 
	\end{array}
	\right.
\end{equation}
Values of parameters are given in appendix \ref{Sec:parameters}.

BCells have gain control, a desensitization when activated by a steady illumination \cite{yu-sing-lee:05}. This desensitization is mediated by a rise in intracellular calcium $Ca^{2+}$, at the origin of a feedback inhibition preventing thus prolonged signalling of the ON BCell \cite{snellman-kaur-etal:08,chen-marre-etal:13}. Following Chen et al., we introduce 
the dimensionless activity variable $\A{B}{i}$ obeying the differential equation:
\begin{equation}\label{eq:dA}
\frac{d\A{B}{i}}{dt} =  -\frac{\A{B}{i}}{\tau_a} + h_B \, \cN(\V{B}{i}(t)).
\end{equation}

Assuming an initial condition $\A{B}{i}(t_0)=0$ at initial time $t_0$ the solution is:
\begin{equation}\label{eq:ActivityInteg}
\A{B}{i}(t)=h_B \, \int_{t_0}^t e^{-\frac{t-s}{\tau_a}} \, \cN(\V{B}{i}(s)) \, ds.
\end{equation}

\begin{figure}
\centering
\includegraphics[width=9cm,height=6cm]{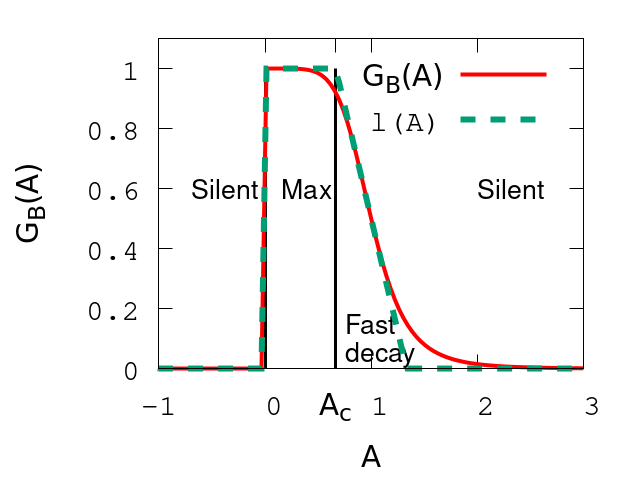}
\caption{\textbf{Gain control \eqref{eq:gain_control_bip} as a function of activity $A$.} The function $l(A)$, in dashed line, is a piecewise linear approximation of $\cG_{B}(A)$ from which $3$ regions are roughly defined. In the region "Silent" the gain vanishes so the cell does not respond to stimuli; in the region "Max", the gain is maximal so that cell behaviour does not show any difference with a not gain-controlled  cell; the region "Fast decay" is the one which contributes to anticipation by shifting the peak in the cell's activity (see section \ref{Sec:MechanismAnticipation}). The value $A_c=\frac{2}{3}$ corresponds to the value of activity where gain control, in the piecewise linear approximation, becomes effective.}
\label{Fig:Gain}
\end{figure}

The bipolar output to ACells and GCells is then characterized by a non linear response to its voltage variation,
 given by : 
\begin{equation} \label{eq:ResponseBip}
\R{B}{i}\pare{\V{B}{i}, \A{B}{i}}=
\cN_B\pare{\V{B}{i}} \, \cG_B\pare{\A{B}{i}},
\end{equation}
where :
\begin{equation} \label{eq:gain_control_bip}
\cG_B(\A{B}{i})=\left\{
	\begin{array}{ll}
		0, & \mbox{if} \hspace{0.2cm} \A{B}{i} \le 0; \\
		\frac{1}{1+\A{B}{i}^6}, & \mbox{else}. 
	\end{array}
	\right.
\end{equation}
Note that $\R{B}{i}$ has the physical dimension of a voltage, whereas, from eq. \eqref{eq:gain_control_bip}, the activity $\A{B}{i}$ is dimensionless. As a consequence, the parameter $h_B$ in eq. \eqref{eq:dA} must be expressed in $ms^{-1}mV^{-1}$. The form \eqref{eq:gain_control_bip} and its $6$-th power are based on experimental fits made by Chen et al. Its form is shown in Fig. \ref{Fig:Gain}. 

In the course of the paper we will use the following piecewise linear approximation also represented in Fig. \ref{Fig:Gain}:
\begin{equation} \label{eq:GainPL}
\cG_{B}(A) = \left\{
	\begin{array}{lll}
		0, & \mbox{if } \hspace{0.2cm} A \in  ]-\infty,0[ \, \cup \, [\frac{4}{3}, +\infty[, \quad &\mbox{Silent \, region}; \\
		1, & \mbox{if } \hspace{0.2cm} A \in  [0,\frac{2}{3}], \quad &\mbox{Maximal \, gain}; \\
		-\frac{3}{2}A+ 2, \quad & \mbox{if } \hspace{0.2cm} A \in  [\frac{2}{3},\frac{4}{3}], \quad &\mbox{Fast \, decay}. 
	\end{array}
	\right.
	\end{equation}
Thanks to this approximation we roughly distinguish $3$ regions for the gain function $\cG_{B}(A)$. This shape is useful to understand the mechanism of anticipation (section \ref{Sec:MechanismAnticipation}).

\sSec{Amacrine cells layer}{Amacrine_cells_layer}

There is a wide variety of ACells (about 30-40 different types for humans) \cite{remington:12}. Some specific types are well studied such as Starburst Amacrine Cells, which are involved in direction sensitivity \cite{euler-detwiler-etal:02, tukker-taylor-etal:04, enciso-rempe-etal:10}, as well as contrast impression and suppression of GCells response \cite{niu-yuan:07}, or AII, a central element of the vertebrate rod-cone pathway \cite{nelson-kolb:04}. 

Here, we don't want to consider specific types of ACells with a detailed biophysical description. Instead, we want to point out the potential role they can play in motion anticipation, thanks to the inhibitory lateral connectivity they induce. We focus on a specific circuitry 
involved in differential motion: an object with a different motion from its background induces more salient activity. The mechanism, observed in mice and rabbit retinas \cite{olveczky-baccus-etal:03,gollisch-meister:10} is featured in Fig. \ref{Fig:architecture}, pathway III. When the left pathway receives a different illumination from the right pathway (corresponding e.g. to a moving object), this asymmetry is amplified by  the ACells' mutual inhibition, enhancing the response of the left pathway in a "push-pull" effect. We want to propose that such a mutual inhibition circuit,  deployed in a lattice through the whole retina, can generate - under specific conditions mathematically analysed - a wave of activity propagation triggered by the moving object. \\

In the model, ACells tile the retina with a lattice spacing $\delta_A$. We index them with $j=1 \dots N_A$.

\ssSec{Synaptic connections between ACells and BCells}{Connections_amacrines_to_bipolars}

 We consider here a simple model of ACells. We assimilate them to passive cells (no active ionic channels) acting as a simple relay between BCells. This aspect is further discussed later in the paper.
The ACell $\Cell{A}{j}$, connected to the BCell $\Cell{B}{i}$, induces on the latter the post-synaptic potential :
%
%
$$
\Ps{A}{j}{B}{i}(t) = \W{A}{j}{B}{i}(t) \int_{-\infty}^t \gamma_{B}(t-s) \, \V{A}{j}(s) ds; 
 \qquad \gamma_{B}(t)=e^{-\frac{t}{\tau_{B}}} H(t),
$$
%
where the Heaviside function $H$ ensures causality. Thus, the post synaptic potential is the mere convolution of the pre synaptic ACell voltage, with an exponential $\alpha$-profile \cite{destexhe-mainen-etal:94b}. In addition, we assume the propagation to be instantaneous.

Here, the synaptic weight $\W{A}{j}{B}{i} < 0$ mimics the inhibitory connection from ACell to BCell (glycine or GABA) with the convention that $\W{A}{j}{B}{i}=0$ if there is no connection from $\Cell{A}{j}$ to $\Cell{B}{i}$. 

In general, several ACells input the  BCell $\Cell{B}{i}$ giving a total PSP:

\begin{equation}\label{eq:PSP_A_Bip}
\Pst{B}{i}(t) = \sum_{j=1}^{N_B} \W{A}{j}{B}{i} \int_{-\infty}^t \gamma_{B}(t-s) \, \V{A}{j}(s) ds.
\end{equation}

Conversely, the BCell $\Cell{B}{i}$ connected to $\Cell{A}{j}$  induces, on this cell, a synaptic response characterized by a post-synaptic potential (PSP) $\Pst{A}{j}(t)$. As ACells are passive elements their voltage $\V{A}{j}(t)$ is equal to this PSP. We have thus:
\begin{equation}\label{eq:PSP_Bip_A}
\V{A}{j}(t) = \sum_{i=1}^{N_A} \W{B}{i}{A}{j} \int_{-\infty}^t \gamma_{A}(t-s) \, \R{B}{i}(s) ds,
\end{equation}
with $\gamma_{A}(t)=e^{-\frac{t}{\tau_{A}}} H(t)$. Here, $\W{B}{i}{A}{j} > 0$ corresponding to the excitatory effect of BCells on ACells, through a glutamate release. Note that the voltage of the BCell is rectified and gain-controlled.

\ssSec{Dynamics}{Dynamics}

The coupled dynamics of Bipolar and Amacrine cells can be described by a dynamical system that we derive now. 

\paragraph{Bipolar voltage.} By differentiating \eqref{eq:PSP_A_Bip},
 \eqref{eq:VBipTot},
%
%
and introducing:
\begin{equation}\label{eq:FBip}
\F{B}{i}(t)= \bra{\K{B}{i} \conv{x,y,t} \pare{\frac{\cS}{\tau_B} + \frac{d \cS}{dt}}}(t) = \frac{V_{i_{drive}}}{\tau_B} \, + \, \frac{d V_{i_{drive}}}{d t},
\end{equation}
we end up with the following equation for the bipolar voltage:
\begin{equation}\label{eq:dVBip}
\frac{d\V{B}{i}}{d t} = - \frac{1}{\tau_{B}} \V{B}{i} + \sum_{j=1}^{N_A} \W{A}{j}{B}{i}\, \V{A}{j} + \F{B}{i}(t),
\end{equation}
where we have used \eqref{eq:dVdrive}. This is a differential equation driven by the time dependent term $\F{B}{i}$ containing the stimulus and its time derivative. 

To illustrate the role of $\F{B}{i}$, let us consider an object moving with a speed $\vv$ depending on time, thus with a non zero acceleration $\vgamma=\frac{d \vv}{dt}$. This stimulus has the form $\cS(t)=g\pare{\vX-\vv(t) \,t}$, with $\vX=\vect{x\\y}$,  so that $\frac{d \cS}{dt}=- \vnabla g\pare{\vX-\vv(t) \,t}.\pare{\vv + \vgamma t }$
where $\vnabla$ denotes the gradient. Therefore, thanks to the eq. \eqref{eq:dVBip}, BCells are \textit{sensitive} to changes in directions, thereby justifying a study of $2$ dimensional stimuli (Section \ref{Sec:2D_Stimuli}).
Note that this property is inherited from the simple, differential structure of the dynamics, the term $\frac{d V_{i_{drive}}}{d t}$ resulting from the differentiation of $\V{B}{i}$. This term does not appear in the classical formulation \eqref{eq:Vdrive} of the bipolar response, without amacrine connectivity. It appears here because synaptic response involves an implicit time derivative via the convolution \eqref{eq:PSP_Bip_A}.

\paragraph{Coupled dynamics.} 

Likewise, differentiating \eqref{eq:PSP_Bip_A} gives:

\begin{equation}\label{eq:dVa}
\frac{d \V{A}{j}}{d t} = - \frac{1}{\tau_{A}} \V{A}{j}+ \sum_{i=1}^{N_B} \W{B}{i}{A}{j}  \R{B}{i}.
\end{equation}

 Eq. \eqref{eq:dA} (activity), \eqref{eq:dVBip} and \eqref{eq:dVa} define a set of $2 N_B +N_A$ differential equations, ruling the behaviour of coupled BCells and ACells, under the drive of the stimulus, appearing in the term $\F{B}{i}(t)$. We summarize the differential system here:
\begin{equation}\label{eq:Diff_Syst}
\left\{
	\begin{array}{lll}
	\frac{d\V{B}{i}}{d t} &=& - \frac{1}{\tau_{B}} \V{B}{i} + \sum_{j=1}^{N_A} \W{A}{j}{B}{i}\, \V{A}{j} + \F{B}{i}(t),\\
	&&\\
\frac{d \V{A}{j}}{d t} &=& - \frac{1}{\tau_{A}} \V{A}{j}+ \sum_{i=1}^{N_B} \W{B}{i}{A}{j}  \R{B}{i},\\
	&&\\
	\frac{d\A{B}{i}}{dt} &=&  -\frac{\A{B}{i}}{\tau_a} + h_B \, \cN(\V{B}{i}).
	\end{array}
	\right.
\end{equation}
We have used the classical dynamical systems convention where time appears explicitly only in the driving term $\F{B}{i}(t)$ to emphasize that \eqref{eq:Diff_Syst} is non-autonomous.
Note that BCells act on ACells via a rectified voltage (gain control and piecewise linear rectification), in agreement with fig. \ref{Fig:architecture}, pathway III. We analyse this dynamics in section \ref{Sec:Mathematical_study}.

\ssSec{Connectivity graph}{Connectivity_graph}

The way ACells connect to BCells, and reciprocally, have a deep impact on the dynamics \eqref{eq:Diff_Syst}. In this paper, we want to point out the role of relative excitation (from BCells to ACells) and inhibition (from ACells to BCells) as well as the role of the network topology. For mathematical convenience  - dealing with square matrices - we assume from now on that there are as many BCell as ACells and we set $N \equiv N_A=N_B$. At the core of our mathematical studies is  a matrix, $\cL$, defined in section \ref{Sec:Mathematical_study}, whose spectrum conditions the evolution of the BCells-ACells network under the influence of a stimulus. It is interesting and relevant to relate the spectrum of $\cL$ to the spectrum of the connectivity matrices ACells to BCells and BCells to ACells. There is not such general relation for arbitrary matrices of connectivity. A simple case holds when the two connectivity matrices commute. Here, we choose an even simpler situation, based on the fact that we compare the role of the direct feed-forward pathway on anticipation in the presence of ACells lateral connectivity. We feature the direct pathway by assuming that a BCell connects only one ACell 
with a weight $w^+$ uniform for all BCell, so that $\W{B}{}{A}{} = w^+ \, I_{N,N}$, $ w^+>0$, where $I_{N,N}$ is the $N$-dimensional identity matrix.  
In contrast, we assume that ACells connect to BCells with a connectivity matrix $\cW$, not necessarily symmetric, with a uniform weight $- w^-$, $ w^->0$, so that $\W{A}{}{B}{} = -w^- \, \cW$.\\

We consider then two types of network topology for $\cW$:
\begin{enumerate}
    \item \textbf{Nearest neighbours.} An ACell connects its $2d$ nearest BCell neighbours  where $d=1,2$ is the lattice dimension. 
    \item \textbf{Random ACell connectivity.} This model is inspired from the paper  \cite{tauchi-masland:84} on the shape and arrangement of starburst ACells in the rabbit retina. Each cell (ACell and BCell) has a random number of branches (dendritic tree), each of which has a random length and a random angle with respect to the horizontal axis. The length of branches $L$ follow an exponential distribution with  spatial scale $\xi$.  
The number of branches $n$ is also a random variable, Gaussian with mean $\bar{n}$ and variance $\sigma_n$. 
The angle distribution is taken to be isotropic in the plane, i.e. uniform on $[0,2 \pi[$. 
When a branch of an ACell A intersects a branch of a BCell B there is a chemical synapse from A to B. The probability that two branches intersect follows a nearly exponential probability distribution that can be analytically computed (see Appendix, section \ref{Sec:Random_Connectivity}). 

\end{enumerate}

\sSec{Ganglion cells}{Ganglion_cells}

There are many different types of GCells in the retina, with different physiologies and functions \cite{baden-berens-etal:16,sernagor-hennig:13}. In the present computational study we focus on specific subtypes associated to the pathways I-II (Fast OFF cells with gain control), III (Differential Motion Sensitive cells), IV (Direction selective cells),   in Fig. \ref{Fig:architecture}. All these have common features: BCells pooling and gain control.

\ssSec{BCells pooling}{Bipolar_cells_pooling}

In the retina, GCells of the same type cover the surface of the retina, forming a mosaic. The degree of overlap between GCells indicates the extent to which their dendritic arbours are entangled in one another. This overlap remains however very limited between cells of the same type \cite{segev-puchalla-etal:06}. We note $k$ the index of the GCells, $k=1 \dots N_G$ and $\delta_G$ the spacing between two consecutive GCells lying on the grid (Fig. \ref{Fig:tiling}).

In the model, GCell $k$ pools over the output of BCells in its neighbourhood \cite{chen-marre-etal:13}. Its voltage reads:
\begin{equation}\label{eq:Pooling}
\V{G}{k}^{(P)} = \sum_i \W{B}{i}{G}{k} \R{B}{i}, 
\end{equation}
where the superscript "P" stands for "Pool". We use this notation to differentiate this voltage from the total GCell voltage,  $\V{G}{k}$, when they are different. This happens in the case when GCells are directly coupled by gap junctions (sections \ref{Sec:Gap_junctions_connectivity}, \ref{Sec:Role_Gaps}). When there is no ambiguity we will drop the superscript "P". In eq. \eqref{eq:Pooling}, the weights $\W{B}{i}{G}{k}$ are Gaussian:
\begin{equation}\label{eq:GaussianPooling}
\W{B}{i}{G}{k}=a_p \, e^{-\frac{\Dd{B}{i}{G}{k}}{2 \,\sigma_{p}^2}}.
\end{equation} 
where $\sigma_{p}$ has the dimension of a distance and $a_p$ is dimensionless.  

\ssSec{Ganglion cells response}{Ganglion_cells_response}

The voltage $\V{G}{k}$  is processed through a gain control loop similar to the BCell layer \cite{chen-marre-etal:13}. As GCells are spiking cells, a non-linearity is fixed so as to impose an upper limit over the firing rate. Here, it is modelled by a sigmoid function, e.g. :
\begin{equation} \label{eq:non_linearity_gang}
\cN_{G}\pare{V}=\left\{
    \begin{array}{ll}
		0, \quad &\mbox{if} \quad V \le \theta_G; \\ 
		\alpha_{G}(V-\theta_{G}), \quad &\mbox{if} \quad \theta_{G} \le V \le N_{G}^{max}/\alpha_{G} + \theta_{G}; \\ 
		N_{G}^{max}, \quad &\mbox{else}.
	\end{array}
	\right.
\end{equation}
This function corresponds to a probability of firing in a time interval. Thus, it is expressed in $Hz$. Consequently, $\alpha_G$ is expressed in $Hz \, mV^{-1}$ and $N_{G}^{max}$ in $Hz$.
Parameters values can be found in the appendix \ref{Sec:parameters}. 

Gain control is implemented with an activation function $\A{G}{k}$, solving the following differential equation:
\begin{equation}\label{eq:ActivityGang}
\frac{d\A{G}{k}}{dt} =  -\frac{\A{G}{k}}{\tau_{G}} + h_{G} \, \cN_{G}\pare{\V{G}{k}},
\end{equation}
and a gain function : 
\begin{equation} \label{eq:gain_control_gang}
\cG_{G}(A)=\left\{
	\begin{array}{ll}
		0, &\quad \mbox{if} \quad A < 0; \\
		\frac{1}{1+A}, &\quad \quad \mbox{else}. 
	\end{array}
	\right.
\end{equation}

Note that the origin of this gain control is different from the BCell gain control \eqref{eq:gain_control_bip}. Indeed, Chen et al. hypothesize that the biophysical mechanisms that could lie behind ganglion gain control are spike-dependent inactivation of $Na^+$ and $K^+$ channels, while the study by Jacoby et al. \cite{jacoby-zhu-etal:15} hypothesize that GCells gain control is mediated by feed-forward inhibition that they receive from ACells. The specific forms of the non-linearity and the gain control function used in this paper match however the first hypothesis, namely the suppression of the $Na^+$ current \cite{chen-marre-etal:13}.

Finally, the response function of this GCell type is:
\begin{equation} \label{eq:ResponseGang}
\R{G}{}\pare{\V{G}{k},\A{G}{k}}=\cN_{G}(\V{G}{k}) \, \cG_{G}(\A{G}{k}).
\end{equation}
In contrast to BCell response $\R{B}{}$, \eqref{eq:ResponseBip}, which is a voltage, here $\R{G}{}$ is a firing rate.

Gain control has been reported for OFF GCells only \cite{berry-brivanlou-etal:99} \cite{chen-marre-etal:13}. Therefore, we restrict our study to OFF cells, i.e with a negative center of the spatial RF kernel. However, on mathematical grounds, it is easier to carry our explanation when the RF center is positive. Thus, for convenience, we have adopted a change in convention in terms of contrast measurement. We take the reference value 0 of the stimulus to be white rather than black, black corresponding then to 1. The spatial RF kernel is also inverted, with a positive center and a negative surround. The problem is therefore mathematically equivalent to an ON Cell submitted to positive stimulus.

\ssSec{Differential Motion Sensitive Cells }{DMSC}

We consider here a class of GCells, connected to ACells according to pathways III in fig. \ref{Fig:architecture}, acting as differential motion detectors. They are able to respond saliently to an object moving over a stationary surround, while being strongly inhibited by global motion. Here, stationary is meant in a general, probabilistic sense. This can be a uniform background, or a noisy background where the probability distribution of the noise is time-translation invariant. These cells are hence able to filter head and eye movements. Baccus et al. \cite{baccus-meister:02} emphasized a pathway accountable for this type of response, involving polyaxonal ACells which selectively suppress GCells response to global motion and enhance their response to differential motion as shown in Fig. \ref{Fig:architecture}, pathway III. 
The GCell receives an excitatory input from the BCells lying in its receptive field which respond to the central object motion, and an indirect inhibitory input from ACells that are connected to BCells which respond to the background motion. When motion is global, the excitatory signal is equivalent to the inhibitory one, resulting in an overall suppression. However, when the object in the center moves distinctively from the surrounding background, the cell in the center responds strongly. 

There are here two concomitant effects. When a moving object (say, from left to right) enters the BCell pool connected to a central GCell $k_D$, the BCells in the periphery of the pool respond first, with no significant change on the GCell response, because of the Gaussian shape \eqref{eq:GaussianPooling} of the pooling: weights are small in the periphery. Those BCells excite however the ACells they are connected to, with the effect of inhibiting the BCells of neighbouring GCells pools. This has the effect of decreasing the voltage of these BCells, which in turn excite less ACells which, in turn, inhibit less the BCells of the pool $k_D$. Thus, the response of the GCell $k_D$ is enhanced, while the cells on the background are inhibited. We call this effect  "push-pull" effect.
Note that propagation delays ought to play an important role here, although we are not going to consider them in this paper.


\ssSec{Direction selective GCells and gap junctions connectivity}{Gap_junctions_connectivity}

These cells correspond to the pathway IV in Fig. \ref{Fig:architecture}. They are only coupled via electric synapses (gap junctions). In several animals, like the mouse, this enables the corresponding GCells to be direction sensitive. Note that other mechanisms, involving lateral inhibition via Starburst Amacrine Cells have also been widely reported \cite{euler-detwiler-etal:02,tukker-taylor-etal:04,enciso-rempe-etal:10,wei-hamby-etal:10,vaney-sivyer-etal:12,sethuramanujam-mclaughlin-etal:16,sethuramanujam-awatramani-etal:18}. Here we focus on gap junctions direction sensitive cells (DSGCs).
There exist four major types of these DSGCs, each responding to edges moving in one of the four cardinal directions. Trenhlom et al. \cite{trenholm-schwab-etal:13} have emphasized the role of these cells coupling in lag normalization: uncoupled cells begin responding when a bar enters their receptive field, i.e, their dendritic field extension, whereas coupled cells start responding \textit{before} the bar reaches their dendritic field. This anticipated response is due to the effective propagation of activity from neighbouring cells through gap junctions, and is particularly interesting when comparing the responses for different velocities of the bar. Trenhlom et al. have shown that the uncoupled DSGCs detect the bar at a position which is further shifted as the velocity grows, while coupled cells respond at an almost constant position, regardless of the velocity. In our work, we analyse this effect in terms of a propagating wave driven by the stimulus and show that, temporally, this spatial lag normalization induces a motion extrapolation that confers to the retina more than just the ability to compensate for processing delays, but to anticipate motion. 

Classical, symmetric bidirectional gap junctions coupling between neighbouring cells would involve a current of the form $-g (\V{G}{k}-\V{G}{k-1})-g(\V{G}{k}-\V{G}{k+1})$ where $g$ is the gap junction conductance. In contrast, here, the current takes the form
$-g (\V{G}{k}-\V{G}{k-1})$. This is due to the specific asymmetric structure of the direction selective GCell dendritic tree \cite{trenholm-schwab-etal:13}. The experimental results of these authors suggest that the effect of the possible gap junction input from downstream cells, in the direction of motion, can be neglected due to offset inhibition and gain control suppression. This, along with the asymmetry of the dendritic arbour, justify the approximation whereby the cell k+1 doesn't influence the current in the cell k. This induces a strong difference in the propagation of a perturbation. Indeed, consider the case $\V{G}{k}-\V{G}{k-1}=\V{G}{k}-\V{G}{k+1}=\delta$. In the symmetric form the total current vanishes whereas in the asymmetric form the current is $-g \delta$. Still, the current can have both directions depending on the sign of $\delta$.
This has a strong consequence on the way GCells connected by gap junctions respond to a propagating stimulus, as shown in section \ref{Sec:Role_Gaps}.

The total GCell voltage is the sum of the pooled BCell voltage $\V{G}{k}^{(P)}$ and of the effect of neighbours GCells connected to $k$ by gap junctions:
$$
\V{G}{k} (t) = \V{G}{k}^{(P)} - \frac{g}{C} \int_{-\infty}^t (\V{G}{k} (s) - \V{G}{k-1} (s))ds
$$
where $C$ is the membrane capacitance. 
Deriving the previous equation with respect to time, we obtain the following differential equation governing the GCell voltage: 
\begin{equation}\label{eq:dVdtgapbefore}
\frac{d\V{G}{k}}{dt} = \frac{d\V{G}{k}^{(P)}}{dt} - w_{gap} \, \bra{\V{G}{k} (t) - \V{G}{k-1} (t)},
\end{equation}
where:
\begin{equation}\label{eq:wgap}
w_{gap}=\frac{g}{C}. 
\end{equation}
Gain control is then applied on $\V{G}{k}$ as in \eqref{eq:ResponseGang}. An alternative is to consider that gain control occurs before gap junctions effect. We investigated this effect as well (not shown, see \cite{souihel:19}). Mainly, the anticipatory effect is enhanced when the gain control is applied after the gap junction coupling, thus, from now, we focus on the formulation \eqref{eq:dVdtgapbefore} in the paper. 


Note that our voltage-based model of gap junctions takes a different from as Trenholm et. al (expressed in terms of currents), because we had to adapt it so as to deal with the pooling voltage form \eqref{eq:Pooling}. Still, our model reproduces the lag normalization as in the original model as we checked (not shown, see \cite{souihel:19}).




\Sec{Results}{Results}

\sSec{The mechanism of motion anticipation and the role of gain control}{MechanismAnticipation}

The (smooth) trajectory of a moving object can be 
extrapolated from its past position and velocity to obtain an estimate of its current location \cite{nijhawan:94,baldo-klein:95,nijhawan:97}. When human subjects are shown a moving bar travelling at constant
velocity, while a second bar is briefly flashed in alignment with the moving bar, the subjects report seeing the flashed bar trailing behind the
moving bar. This led Berry et al \cite{berry-brivanlou-etal:99} to investigate the potential role of the retina in anticipation mechanisms. Under constraints on the bar' speed and contrast they were able to exhibit a positive anticipation time, defined as the time lag between 
the peak in the retinal GCell response to a flashed bar and the corresponding peak when the stimulus is a moving bar.

In this paper we adopt a slightly different definition although inspired from it. Indeed, the goal of this modelling paper is to dissect the various potential stages of retinal anticipation as developed in the next subsections. 

Several layers and mechanisms are involved in the model, each one defining a response time and potentially contributing to anticipation, under conditions that we now analyse.
%
%
%

\ssSec{Anticipation at the level of a single, isolated, BCell; the local effect of gain control}{Anticipation_BCell}

\begin{figure}
\centering
 \includegraphics[width=11cm,height=10cm]{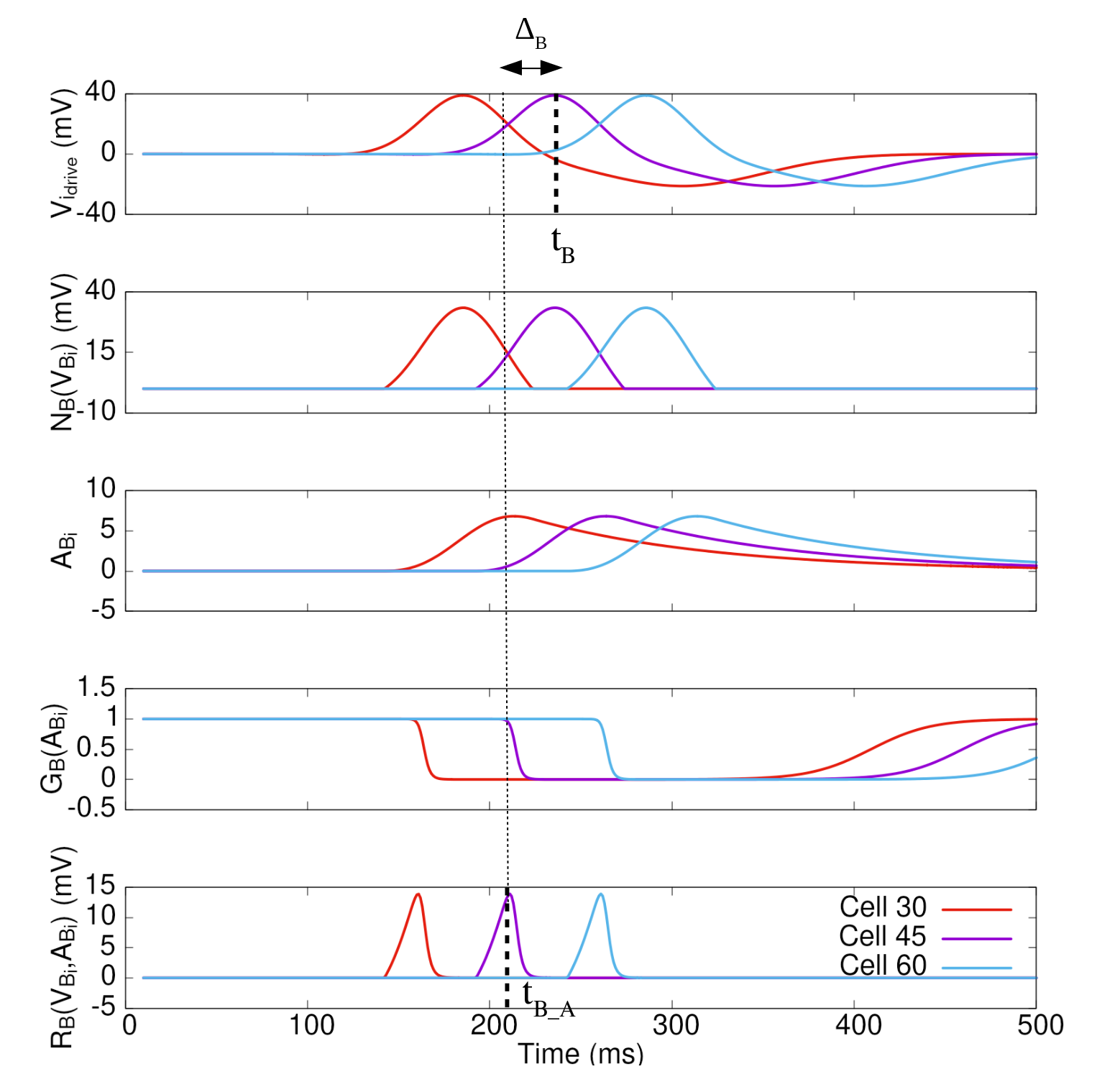}\\ 
\vspace{0.1cm}
\caption{\textbf{The mechanism of motion anticipation and the role of gain control.} The figure illustrates the bipolar anticipation time $\Delta_B$, without lateral connectivity. We see the response of OFF BCells with gain control to a dark moving bar. The curves correspond to three cells spaced by $450 \, \mu m$. The first line (a) shows the linear filtering of the stimulus, corresponding to $V_{drive}(t)$ (eq. \eqref{eq:Vdrive}). The  line (b) corresponds to the threshold non-linearity $\Cell{\cN}{B}$ applied to the linear response; (c) represents the adaptation variable \eqref{eq:Diff_Syst}, and (d) shows the gain control time curse. Finally, the last line (e) corresponds to the response $\R{B}{i}$ of the BCell. The two dashed lines correspond respectively to $t_B$ and $t_{B_A}$, the peak in the response of the (purple) Bcell without pooling.  
\label{Fig:AnticipationBConly}
}
\end{figure}

We consider first a single BCell, without lateral connectivity so that $\V{B}{i}=V_{i_{drive}}$. The very mechanism of anticipation at this stage is illustrated in Fig.  \ref{Fig:AnticipationBConly}.  The  peak response time of the convolution of the stimulus with the RF of one BCell occurs at a time $t_B$ (dashed line in Fig. \ref{Fig:AnticipationBConly} a).
The increase in $V_{i_{drive}}$
leads to an increase in activity (Fig. \ref{Fig:AnticipationBConly}, c) and an increase of $\R{B}{}$ (Fig. \ref{Fig:AnticipationBConly}, e). When activity becomes large enough, gain control switches on (Fig. \ref{Fig:AnticipationBConly}  d) leading to a sharp decrease of the response $\R{B}{}$ (Fig. \ref{Fig:AnticipationBConly} e) and a peak in $\R{B}{}$ occurring at time $t_{B_A}$ (dashed line in Fig. \ref{Fig:AnticipationBConly} e) \textit{before} $t_B$. The \textit{bipolar anticipation time}, $\Delta_B= t_B - t_{B_{A}}$, is therefore  positive. %

Mathematically, $\Delta_B > 0$ results from the intermediate value theorem using that $\frac{d V_{i_{drive}}}{d t} \geq 0$ on $\bra{0,t_B}$ and that $t_{B_A}$ is defined by:
%
%
$$
\left. \frac{d V_{i_{drive}}}{dt}\right|_{t=t_{B_A}} = -  V_{i_{drive}}(t_{B_A}) \, \frac{\cG'_B\pare{\A{B}{i}}}{\cG_B\pare{\A{B}{i}}} \,  \left. \frac{d\A{B}{i}}{dt}\right|_{t=t_{B_A}},
$$
%
where the right hand side is positive provided that the parameters $h_B, \tau_a$ are tuned\footnote{From \eqref{eq:dA} $\frac{d \A{B}{i}}{dt}>0$
if  $A_i(t) < h_B \, \tau_a \, V_{i_{drive}}(t)$. This essentially requires $\tau_a$ to be slow enough.} such that $\frac{d \A{B}{i}}{dt} \geq 0$ on $[0,t_B]$.
An important consequence is that the amplitude of the response at the peak is \textit{smaller} in the presence of gain control (compare the amplitude of the voltage in Fig. \ref{Fig:AnticipationBConly}, a to \ref{Fig:AnticipationBConly}, e). 

The anticipation time at the BCells level depends on parameters such as $h_B, \tau_a$. It depends as well on characteristics of the stimulus such as contrast, size and speed. An easy way to figure this out is to consider that the peak in BCell response (Fig. \ref{Fig:AnticipationBConly} d, e) arises when the gain control function $\cG_B\pare{\A{B}{i}}$ starts to drop off (Fig. \ref{Fig:AnticipationBConly}  e), which, from the piecewise linear approximation \eqref{eq:GainPL} of BCell arises when $A=\frac{2}{3}$. When $V_{i_{drive}}$ has the form  \eqref{eq:VdriveAppGauss} this gives, using $\cN(V_{i_{drive}})=V_{i_{drive}}$, \eqref{eq:ActivityInteg}, and letting the initial time $t_0 \to -\infty$ (which corresponds to assuming that the initial state was taken in a distant past, quite longer than the time scales in the model):
\begin{equation}\label{eq:SeuilA}
\A{B}{i}(t_{B_A}) = A_0 \, \frac{h_B}{v} \,  e^{\frac{1}{2} \, \frac{\sigma^2}{\tau_a^2 \, v^2}} \, 
e^{\frac{1}{\tau_a v} \pare{x-v \,t_{B_A}}} \, \bra{1 - \Pi\pare{\frac{x-v \,t_{B_A}}{\sigma}+ \frac{\sigma}{\tau_a \, v}}} = \frac{2}{3},
\end{equation}
where $\Pi(x)$ is the cumulative distribution function of the standard Gaussian probability (see definition, eq. \eqref{eq:Pi} in the appendix).
This establishes an explicit equation for the time $t_{B_A}$
as a function of contrast ($A_0$), size ($\sigma$), and speed ($v$) as well as the parameters $h_B$ and $\tau_a$. We do not show the corresponding curves here (see \cite{souihel:19} for a detailed study)  preferring to illustrate the global anticipation at the level of GCells, illustrated in Fig. \ref{Fig:ParametersAnticipation} below.

\ssSec{Anticipation time of the BCells pooled voltage}{tG}

The main effects we want to illustrate in the paper (impact of lateral connectivity on GCells anticipation) are 
evidenced by the shift of the peak in activity of the  BCells pooled voltage, occurring at time $t_G$. We focus on this time here, postponing to section \ref{Sec:Anticipation_time_GCs} the subsequent effect of GCells gain control. We assume therefore here that $h_G=0$ so that  $\A{G}{k}=0$ and $\cG_{G}(\A{G}{k})=1$ in \eqref{eq:non_linearity_gang}. Thus, the firing rate of Gcell $k$ is $\cN_{G}(\V{G}{k})$.
For mathematical simplicity we will consider that the firing rate function \eqref{eq:NLinRectBip} of $G$ is a smooth, monotonously increasing sigmoid function so that $\cN'_G(\V{G}{k}) >0$. 
We define $t_G$ as the time when $ \V{G}{k}$ is maximum,  after the stimulus is switched on. This corresponds to $\frac{d \V{G}{k}}{dt}=0$ and $\frac{d^2 \V{G}{k}}{dt^2} < 0$.  Equivalently, from equations \eqref{eq:Pooling}, \eqref{eq:dVdtgapbefore}:
\begin{equation}\label{eq:Anticipation_General}
\sum_i \W{B}{i}{G}{k}  \frac{d \R{B}{i}}{dt}
= \sum_i \W{B}{i}{G}{k}  \bra{\cG_B\pare{\A{B}{i}} \, \cN'_B(\V{B}{i}) \, \frac{d \V{B}{i}}{dt} + \cN_B(\V{B}{i}) \, \cG'_B\pare{\A{B}{i}} \,  \frac{d\A{B}{i}}{dt}}
=  w_{gap}\, \bra{\V{G}{k} - \V{G}{k-1}},
\end{equation}
where this equation holds at time $t=t_G$ (we have not written explicitly $t_G$ to alleviate notation).
This is the most general equation for the anticipation time at the level of BCells pooling. 

%
%
In the sum $\sum_i$, there are two types of BCells. The inactive ones where $\V{B}{i} \leq \Theta_B$, $\cN_B(\V{B}{i})=0$ and $\frac{d \R{B}{i}}{dt}=0$ so they do not contribute to the activity. The active BCells, $\V{B}{i} > \Theta_B$, obey $\cN_B\pare{\V{B}{i}} = \V{B}{i}$. For the moment we assume that, at time $t_G$, \textit{there is no Bcell switching from one state (active/inactive) to the other}, postponing this case to the end of the section. Then, eq. \eqref{eq:Anticipation_General} reduces to:
%
%
%
\begin{equation}\label{eq:ExtremaGen}
%
\begin{array}{lll}
&\sum_i \underbrace{\W{B}{i}{G}{k}}_{(V)} \, \underbrace{\cG_B\pare{\A{B}{i}}}_{(II)} \,  \, \pare{- \frac{1}{\tau_{B}} \V{B}{i} + \underbrace{\sum_{j=1}^{N_A} \W{A}{j}{B}{i}\, \V{A}{j}}_{(III)} + \underbrace{\F{B}{i}(t)}_{(I)}}\\
&\\
 = -&\sum_i \underbrace{\W{B}{i}{G}{k}}_{(V)} \,\underbrace{\cG'_B\pare{\A{B}{i}}}_{(II)} \,  \V{B}{i}(t) \, \frac{d \A{B}{i}}{dt} \, + \, \underbrace{w_{gap}\, \bra{\V{G}{k} - \V{G}{k-1}}}_{(IV)}.
\end{array}
\end{equation}
This general equation emphasizes the respective role of (I), stimulus (term $\F{B}{i}(t)$); (II), gain control (terms $\cG_B\pare{\A{B}{i}},\cG'_B\pare{\A{B}{i}}$); (III),  ACell lateral connectivity (term $\W{A}{j}{B}{i}$); (IV), gap junctions (term $w_{gap}\, \bra{\V{G}{k} (t'_{G_{A}}) - \V{G}{k-1} (t'_{G_{A}})}$); (V), pooling (terms $\W{B}{i}{G}{k}$). Note that we could as well consider a symmetric gap junctions connectivity where we would have a term $w_{gap}\, \bra{-\V{G}{k+1} + 2 \, \V{G}{k} - \V{G}{k-1}}$ in IV.
The equation terms has been arranged this way for reasons that become clear in the next lines. It is not possible to solve this equation in full generality but it can be used to understand the respective role of each component. \\

In the absence of gain control and lateral connectivity ($\W{A}{j}{B}{i}=0$, $w_{gap}=0$) the peak in GCell $\Cell{G}{k}$ voltage, at time $t'_G$  is given by:
\begin{equation}\label{eq:tG}
\sum_i \W{B}{i}{G}{k}   \frac{d V_{i_{drive}}}{d t} 
= 0, 
\end{equation}
This generalizes the definition of $t_B$, time of peak of a single BCell, to a set of pooled BCells and we will proceed along the same lines as section \ref{Sec:Anticipation_BCell}.
We fix as reference time $0$ the time when the pooled voltage becomes positive. It increases then until the time $t'_G$ when $\sum_i \W{B}{i}{G}{k}   \frac{d V_{i_{drive}}}{d t}=0$. 
Thus, $\sum_i \W{B}{i}{G}{k}   \frac{d V_{i_{drive}}}{d t}$ is positive on $[0,t'_G[ $ and vanishes at $t'_G$.


We now show that, in the presence of gain control, the peak occurs at time $t_G < t'_G$ leading to anticipation induced by gain control and generalizing the effect observed for one Bcell in section \ref{Sec:Anticipation_BCell}. Indeed,
equation \eqref{eq:ExtremaGen} reads now:
\begin{equation}\label{eq:ExtremaGainNoConn}
\sum_{i} \W{B}{i}{G}{k} \cG_B\pare{\A{B}{i}} \, \frac{d V_{i_{drive}}}{d t} = -\sum_{i } \W{B}{i}{G}{k} \cG'_B\pare{\A{B}{i}} \,  V_{i_{drive}}(t) \, \frac{d \A{B}{i}}{dt}.
\end{equation}

Because $0 \leq \cG_B\pare{\A{B}{i}} \leq 1$, $\sum_{i} \W{B}{i}{G}{k} \cG_B\pare{\A{B}{i}} \, \frac{d V_{i_{drive}}}{d t} \leq \sum_{i} \W{B}{i}{G}{k}\, \frac{d V_{i_{drive}}}{d t}$ so that the left hand side in \eqref{eq:ExtremaGainNoConn} reaches $0$ at a time $ t_G \leq t'_G$. The right hand side is positive for the same reasons as in section \ref{Sec:Anticipation_BCell}. The same mathematical argument holds as well, using the intermediate value theorem,  to show that $ t_G < t'_G$.

We now investigate eq. \eqref{eq:ExtremaGen} with the two terms of lateral connectivity: (III), ACells and, (IV) gap junctions. 
The effect of gap junctions is straightforward. A positive term $w_{gap}\, \bra{\V{G}{k} - \V{G}{k-1}}$ increases the right hand side of eq. \eqref{eq:ExtremaGen}. As developed in section \ref{Sec:Role_Gaps} this arises when the stimulus propagates in the preferred direction of the cell inducing a wave of activity propagating ahead of the stimulus. In view of the qualitative argument developed above using the intermediate value theorem, this can enhance the anticipation time. This deserves however a deeper study developed in section \ref{Sec:Role_Gaps}.
 
The effect of ACells cells is less evident, as the term $\pare{- \frac{1}{\tau_{B}} \V{B}{i} + \sum_{j=1}^{N_A} \W{A}{j}{B}{i}\, \V{A}{j} + \F{B}{i}(t)}$ can have any sign, so that network effect can either anticipate or delay the ganglion response, as illustrated in several examples in the next section. As we show, this term is in general related to a wave of activity, enhancing or weakening the anticipation effect as shown in section \ref{Sec:Role_Amacrine}.\\

Let us finally discuss what happens when some BCell switches from one state (active/inactive) to the other (i.e. $\V{B}{i} =  \Theta_B$). In this case, taking into account the definition \eqref{eq:NLinRectBip}, the derivative $\cN'_B(\V{B}{i})=\frac{1}{2}$. Thus, when a BCell reaches the lower threshold, there is a big variation in $\cN'_B(\V{B}{i})$ thereby leading to a positive contribution in \eqref{eq:Anticipation_General} and an additional term 

$\frac{1}{2} \, \sum_i \W{B}{i}{G}{k} \, \cG_B\pare{\A{B}{i}}  \, \pare{- \frac{1}{\tau_{B}} \V{B}{i} + \sum_{j=1}^{N_A} \W{A}{j}{B}{i}\, \V{A}{j} + \F{B}{i}(t)}$
 in the left hand side of \eqref{eq:ExtremaGen}, where the sum holds on switching state cells. As we see in section \eqref{Sec:Role_Amacrine} this can have an important impact on the anticipation time. 

\ssSec{Anticipation time at the GCells level}{Anticipation_time_GCs}

We now show that the firing rate of the GCell $k$, given by \eqref{eq:ResponseGang},
reaches its maximum at a time $t_{G_A} < t_{G}$. From \eqref{eq:ResponseGang}, at time $t_{G_A}$:
\begin{equation}\label{eq:ExtremaGainGCells}
\frac{d \V{G}{k}}{dt}= \frac{\V{G}{k}}{1+\A{G}{k}} \, \frac{d \A{G}{k}}{dt}.
\end{equation}
$\V{G}{k}$ starts from $0$ and increases on the time interval $\bra{0,t_{G}}$  thus $\frac{d \V{G}{k}}{dt}$ is positive on $\bra{0,t_{G}}$ and vanishes at $t_{G}$. Thus, there is a time $t_d <t_{G}$ such that $\frac{d \V{G}{k}}{dt}$ increases on $[0,t_d]$ and decreases on $\bra{t_d,t_{G}}$.  The right hand side of \eqref{eq:ExtremaGainGCells} starts from $0$ at $t=0$ and stays strictly positive until, either $\V{G}{k}$ vanishes which occurs for $t > t_{G}$, or until $\frac{d \A{G}{k}}{dt}$ vanishes. We choose the characteristic time $\tau_{G}$ and the intensity $h_{G}$ in \eqref{eq:ActivityGang} so that $\frac{d \A{G}{k}}{dt}>0$ on $\bra{0,t_{G}}$ . Thus, $\frac{\V{G}{k}}{1+\A{G}{k}} \, \frac{d \A{G}{k}}{dt} >0$ on $\bra{0,t_{G}}$. Therefore, in the time interval  $\bra{t_d,t_{G}}$,
$\frac{d \V{G}{k}}{dt}$ decreases to $0$ while $\frac{\V{G}{k}}{1+\A{G}{k}} \, \frac{d \A{G}{k}}{dt}$ increases from $0$. From the intermediate value theorem these two curves have to intersect at a time $t_{G_A} < t_{G}$.

We finally define the total anticipation time of a Gcell as:
\begin{equation}\label{eq:Delta}
 \Delta= t_{B_c} - t_{G_{A}},
\end{equation} 
where $t_{B_c}$ is the peak of the BCell at the center of the BCells pooling to that GCell. 

 \ssSec{Anticipation variability : stimulus characteristics}{Anticipation_Variability}

In general, $\Delta$ depends on gain control, lateral connectivity, as well as characteristics of the stimulus such as speed and contrast. This has been shown mathematically in eq.  \eqref{eq:SeuilA} for a single BCell. Here, we investigate numerically the dependence of the total anticipation time of a Gcell when the stimulus is a bar of infinite height, width $\sigma$ mm, travelling in one dimension at speed $v$ mm/s with contrast $C \in \bra{0,1}$. Results are shown in fig. \ref{Fig:ParametersAnticipation}.
This figure is a calibration later used to compare to the effects induced by lateral connectivity. 

We first observe that anticipation increases with contrast, as it has experimentally been observed \cite{berry-brivanlou-etal:99}. Indeed, increasing the contrast increases $V_{i_{drive}}(t)$ thereby accelerating the growth of $A_i$ so that gain control takes place earlier (Fig \ref{Fig:ParametersAnticipation} a).
We also notice that anticipation increases with the width of the object until a maximum (Fig \ref{Fig:ParametersAnticipation} b).
Finally, the model shows a decrease in anticipation as a function of velocity, as it was evidenced experimentally  \cite{berry-brivanlou-etal:99,johnston-lagnado:15} (Fig \ref{Fig:AnticipationBConly} c). Indeed, when the velocity increases, $V_{drive}$ varies faster than the characteristic  activation time $\tau_a$ and the adaptation peak value is lower. Consequently, gain control has a weaker effect and the peak activity is less shifted than when the bar is slow.

A large part of these effects can be understood from eq. \eqref{eq:SeuilA}. Note however here that simulation of Fig. \ref{Fig:ParametersAnticipation} takes into account the convolution of a moving bar with the receptive field, the pooling effect, and gain control at the stage of GCells.

In Fig. \ref{Fig:ParametersAnticipation} we also show the evolution of GCells maximum firing rate as a function of the moving bar velocity, contrast and size. We observe that it increases with these  parameters, an expected result.

\begin{figure}[h]
%
\begin{tabular}{ccc}
   \includegraphics[width=0.32\textwidth,height=4.cm]{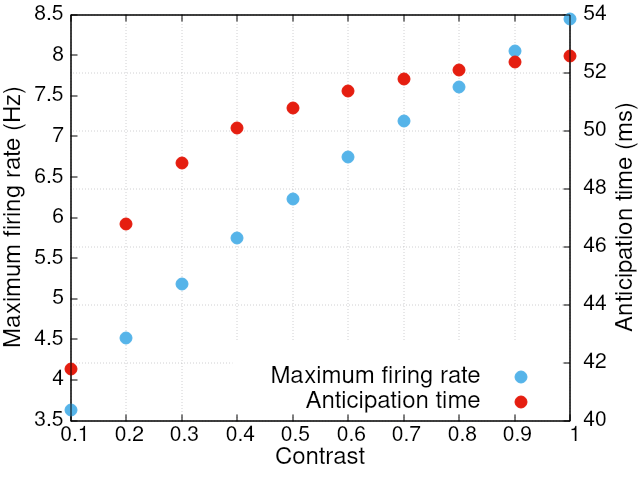} &
   \includegraphics[width=0.32\textwidth,height=4.cm]{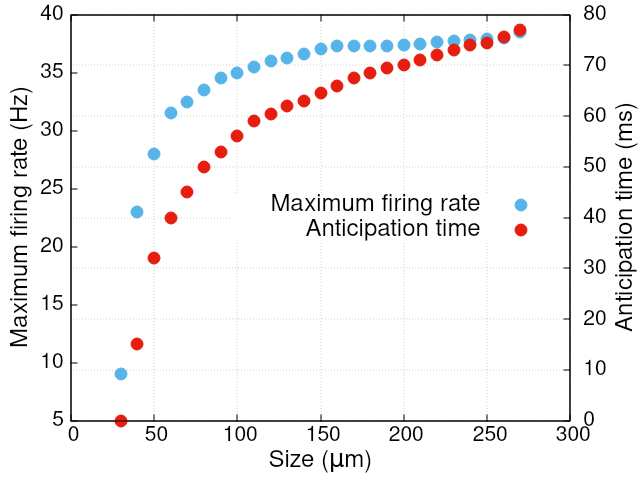} &
   \includegraphics[width=0.32\textwidth,height=4.cm]{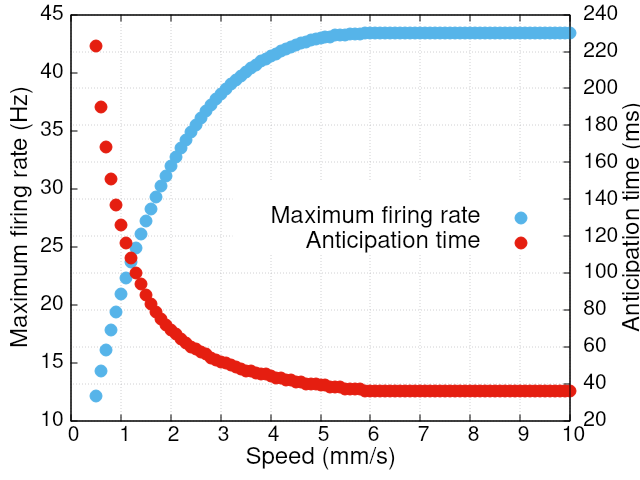} \\
\end{tabular}
\caption{\textbf{Maximum firing rate and anticipation time variability} with stimulus parameters in the gain control layer of the model. \textbf{Left:} contrast (with $v = 1$ mm/s et size = $90 \, \mu \, m$); \textbf{Middle:} size (with $v = 2$ mm/s et contrast = $1$); \textbf{Right:} speed (with contrast $1$ and size = $162 \, \mu \, m$)
  \label{Fig:ParametersAnticipation}.
}
\end{figure}

\sSec{The potential role of ACells lateral inhibition on anticipation}{Role_Amacrine}

In this section we study the potential effect of ACells (pathway III of Fig. \ref{Fig:architecture}) on motion anticipation. We restrict to the case where there are as many BCells as ACells ($N_B=N_A \equiv N$) so that the matrices $\W{A}{}{B}{}$ and $\W{B}{}{A}{}$  are square matrices. We first derive general mathematical results (for the full derivation, see appendix section \ref{Sec:Linear_Analysis}) before considering the two types of connectivity described in section \ref{Sec:Connectivity_graph}.

\ssSec{Mathematical study}{Mathematical_study}

\paragraph{Dynamical system.} We study mathematically the dynamical system \eqref{eq:Diff_Syst} that we write in a more convenient form. We use Greek indices $\alpha,\beta,\gamma = 1 \dots 3N$ and define the state vector $\cX$ as  
$$\vcX_\alpha =
\left\{
\begin{array}{llll}
&\V{B}{i}, \quad \alpha=i, &i=1 \dots N; \\
&\V{A}{i}, \quad \alpha=N+i, &i=1 \dots N; \\
&A_i, \quad \alpha=2N+i, &i=1 \dots N.
\end{array}
\right.
$$
Likewise, we define the stimulus vector $\vcF_\alpha=\F{B}{i}$, if $\alpha=1 \dots N$ and  $\vcF_\alpha=0$ otherwise.
Then, the dynamical system \eqref{eq:Diff_Syst} has the general form:
\begin{equation}\label{eq:Diff_Syst_Vect_Gen}
\frac{d \vcX}{dt} = \cH(\vcX)  + \vcF(t).
\end{equation}
where $\cH(\vcX)$ is a non linear function, via the function $\R{B}{i}\pare{\V{B}{i}, \A{B}{i}}$ 
of eq.  \eqref{eq:ResponseBip}, featuring gain control and low voltage threshold. The non linear problem can be simplified using the piecewise linear approximation \eqref{eq:GainPL}.
 Indeed, there is a domain  of $\setR^{3N}$:
\begin{equation}\label{eq:Omega}
\Omega=\Set{\V{B}{i} \geq \theta_B, \A{B}{i} \in \bra{0,\frac{2}{3}}, i=1 \dots N},
\end{equation}
where $\R{B}{i}\pare{\V{B}{i}, \A{B}{i}}=\V{B}{i}$ so that \eqref{eq:Diff_Syst} is linear and can be written in the form:
\begin{equation}\label{eq:Diff_Syst_Vect_Lin}
\frac{d \vcX}{dt} = \cL.\vcX  + \vcF(t).
\end{equation}
with:
\begin{equation}\label{eq:LX}
\cL=
\pare{\begin{array}{cccccc}
&-\frac{I_{N,N}}{\tau_B} & &\W{A}{}{B}{}&& 0_{N,N}\\
&\W{B}{}{A}{} & & -\frac{I_{N,N}}{\tau_A} && 0_{N,N}\\
& h_B \, I_{N,N}  && 0_{N,N} &&-\frac{I_{N,N}}{\tau_a} 
\end{array}
}
\end{equation}
where $I_{N,N}$ is the $N \times N$ identity matrix and $0_{N,N}$ is the $N \times N$ zero matrix.
 This corresponds to intermediate activity, where neither BCells gain control \eqref{eq:gain_control_bip} nor low threshold \eqref{eq:NLinRectBip} are active. We first study this case and describe then what happens when trajectories of \eqref{eq:Diff_Syst_Vect_Gen} get out of this domain, activating low voltage threshold or gain control.

The idea of using such a phase space decomposition with piecewise linear approximations has been used, in a different context by S. Coombes et al \cite{coombes-lai-etal:18} and in \cite{cessac:08,cessac-vieville:08,cessac:11}.

%
%
We consider the evolution of the state vector $\vcX(t)$ from an initial time $t_0$. Typically, $t_0$ is a reference time where the network is at rest, before the stimulus is applied. So, the initial condition $\vcX(t_0)$ will be set to $0$ without loss of generality. 


\paragraph{Linear analysis.}
The general solution of \eqref{eq:Diff_Syst_Vect_Lin} is:
\begin{equation}\label{eq:GenSolSDLin}
\vcX(t)=\int_{t_0}^t e^{\cL(t-s)}.\vcF(s) \, ds,
\end{equation}
The behaviour of the solution \eqref{eq:GenSolSDLin} depends on the eigenvalues $\lambda_\beta, \beta=1 \dots 3N$ of $\cL$ and its eigenvectors, $\vcP_\beta$, with entries $\cP_{\alpha\beta}$. The matrix $\cP$ transforms $\cL$ in Jordan form ($\cL$ is not diagonalizable when $h_B \neq 0$, see appendix \ref{Sec:General_Solutions}). Whatever the form of the connectivity matrices $\W{B}{}{A}{},\W{A}{}{B}{}$ the $N$ last eigenvalues are always $\lambda_\beta=-\frac{1}{\tau_a}, \beta=2N+1 \dots 3N$.\\

In appendix \ref{Sec:General_Solutions} we show the following general result (not depending on the specific form of $\W{B}{}{A}{},\W{A}{}{B}{}$, they just need to be square matrices and to be diagonalizable):
\begin{equation}\label{eq:XalphaDriven}
\cX_\alpha(t) = V_{\alpha_{drive}}(t) \, + \E{B}{B,\alpha}(t)+\E{B}{A,\alpha}(t)+\E{B}{a,\alpha}(t), \quad \alpha =1 \dots 3N,
\end{equation}
where the drive term \eqref{eq:Vdrive} is extended here to $3N$-dimensions with $V_{\alpha_{drive}}(t)=0$ if $\alpha > N$.
The other terms have the following definition and meaning:
\begin{equation}\label{eq:EBB}
\E{B}{B,\alpha}(t)=\sum_{\beta=1}^{N} \pare{\frac{1}{\tau_B} \, + \, \lambda_\beta } \, \sum_{\gamma=1}^{N} \cP_{\alpha\beta} \cP^{-1}_{\beta \gamma} \int_{t_0}^t e^{\lambda_\beta(t-s)} \, V_{\gamma_{drive}}(s)\, ds, \quad \alpha=1 \dots N,
\end{equation}
corresponds to the \textit{indirect} effect, via the ACells connectivity, of the BCells drive on BCells voltages (i.e. the drive excites BCell $i$ which acts on BCell $j$ via the ACells network);
\begin{equation}\label{eq:EBA}
\E{B}{A,\alpha}(t)=\sum_{\beta=N+1}^{2N} \pare{\frac{1}{\tau_B} \, + \, \lambda_\beta } \, \sum_{\gamma=1}^{N} \cP_{\alpha\beta} \cP^{-1}_{\beta \gamma} \int_{t_0}^t e^{\lambda_\beta(t-s)} \, V_{\gamma_{drive}}(s)\, ds, \quad \alpha=N+1 \dots 2N,
\end{equation}
corresponds to the effect of BCell drive on ACell voltages, and:
\begin{equation}\label{eq:EBa}
\E{B}{a,\alpha}(t)= h_B \, \pare{\sum_{\beta=1}^{2N} \sum_{\gamma=1}^{N} 
\cP_{\alpha-2N \beta} \cP^{-1}_{\beta \gamma}
\, \frac{\lambda_\beta+\frac{1}{\tau_B}}{\lambda_\beta+\frac{1}{\tau_a}}  \, \int_{t_0}^t e^{\lambda_\beta(t-s)} \, V_{\gamma_{drive}}(s)\, ds
 + \frac{-\frac{1}{\tau_B} + \frac{1}{\tau_a}}{\lambda_\beta+\frac{1}{\tau_a}}\, A^0_{\alpha-2N}(t)}, \quad \alpha=2N+1 \dots 3N,
\end{equation}
corresponds to the effect of the BCells drive on the dynamics of BCell activity variables. The first term of \eqref{eq:EBa}
corresponds to the action of BCells and ACells on the activity of BCells, via lateral connectivity. In the second term:
\begin{equation}\label{eq:A0}
 A^0_{\alpha-2N}(t)=  \int_{t_0}^t  e^{-\frac{t-s}{\tau_a}} \,  V_{\alpha-2N_{drive}}(s)\, ds
\end{equation}
corresponds to the direct effect of the BCell voltage with index $\alpha-2N$ on its activity (see eq. \eqref{eq:ActivityInteg}). 

To sum up, equation \eqref{eq:XalphaDriven} describes the direct effect of a time dependent stimulus (first term) and the indirect lateral network effects it induces.
%
The term $\E{B}{a,\alpha}(t)$ is what activates the gain control. In the piecewise linear approximation \eqref{eq:GainPL}, the BCell $i$ triggers its gain control when its activity:
\begin{equation}\label{eq:TriggerGain}
\E{B}{a,\alpha}(t) > \frac{2}{3}, \quad \alpha=2N+i.
\end{equation}
This relation extends the computation,  made in section \ref{Sec:Anticipation_BCell} for isolated BCells, to the case of a BCell under the influence of ACells. On this basis, let us now discuss how the network effect influences the activation of gain control and, thereby, anticipation.\\

\begin{figure}
\centering
\includegraphics[width=0.6\columnwidth,height=6cm]{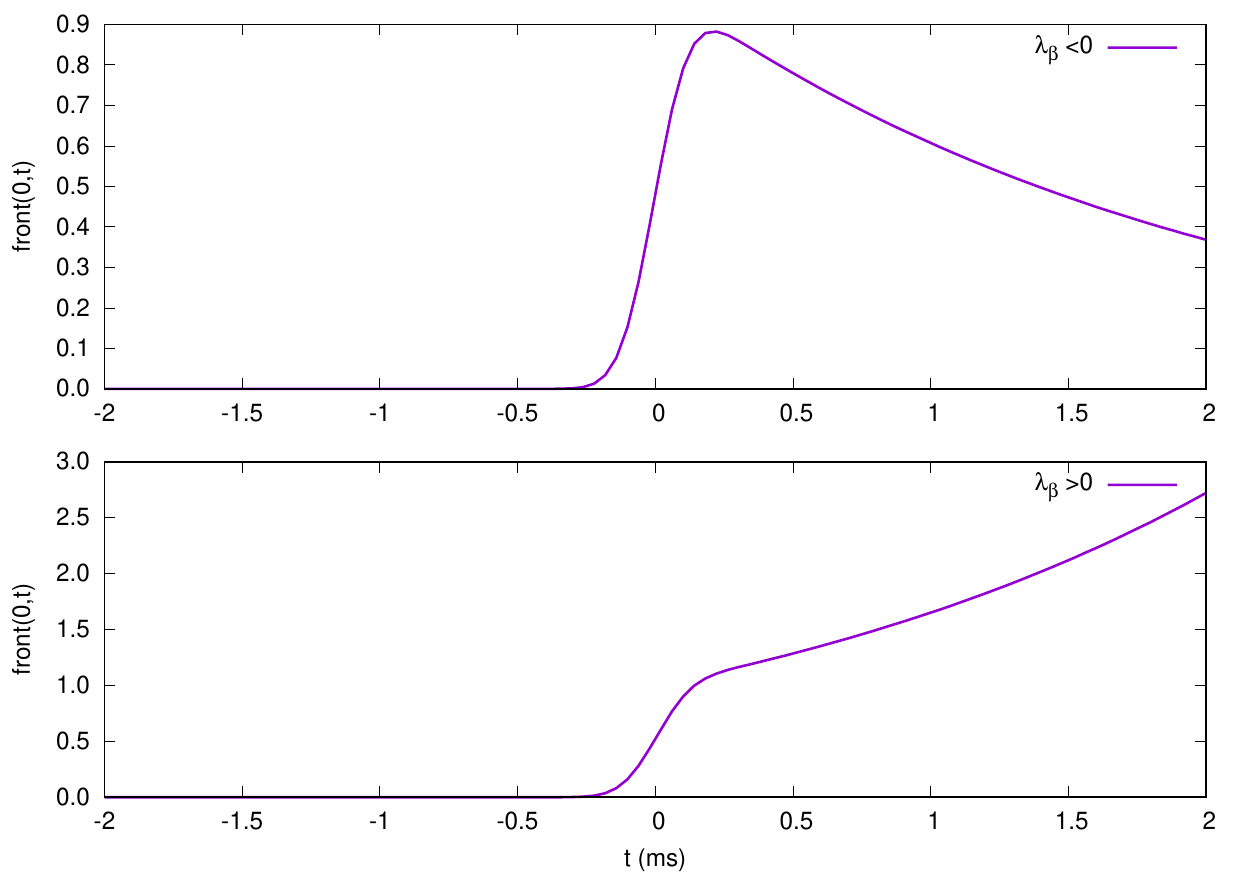}
\caption{\textbf{Front \eqref{eq:ActivationMode}} for different values of $\lambda_\beta$ (Purple) as a function of time, for the cell $\gamma=0$. All figures are drawn with  $v=2$ mm/s; $\sigma=0.2$ mm. \textbf{Top.} $\lambda_\beta=-0.5$ $ms^{-1}$; \textbf{Bottom.} $\lambda_\beta=0.5$ $ms^{-1}$.}
\label{Fig:ActivationMode}
\end{figure}

The structure of the terms \eqref{eq:EBB}, \eqref{eq:EBA} \eqref{eq:EBa} is interpreted as follows. The drive (index $\gamma=1 \dots N$) excites the eigenmodes $\beta=1 \dots 3N$ of $\cL$, with a weight proportional to $\cP^{-1}_{\beta \gamma}$. The mode $\beta$, in turn excites the variable $\alpha=1 \dots 3N$ with a weight proportional to $\cP_{\alpha\beta}$. The time dependence and the effect of the drive are controlled by the integral $\int_{t_0}^t e^{\lambda_\beta(t-s)} \, V_{\gamma_{drive}}(s)\, ds$. For example, when the stimulus has the Gaussian form \eqref{eq:VdriveAppGauss} and cells are spaced with a distance $\delta$ so that cell $\gamma$ is located at $x=\gamma \, \delta$, we have, 
%
taking $t_0 \to -\infty$:
\begin{equation}\label{eq:ActivationMode}
\int_{-\infty}^t e^{\lambda_\beta(t-s)} V_{\gamma_{drive}}(s)\, ds= \frac{A_0}{v}  \, e^{\frac{1}{2} \, \frac{\sigma^2 \, \lambda_\beta^2}{v^2}} \, e^{-\frac{\lambda_\beta}{v} \, \pare{\gamma\,\delta \, - \, v \, t}} \, \Pi\bra{\frac{\lambda_\beta \sigma}{v} \, - \, \frac{1}{\sigma} \pare{\gamma\,\delta \, - \, v \, t}},
\end{equation}
where $\Pi(x)$ is the cumulative distribution function of the standard Gaussian probability (see definition, eq. \eqref{eq:Pi} in the appendix). 
This is actually the same computation as \eqref{eq:SeuilA} with $\lambda_\beta=-\frac{1}{\tau_a}$.
%
%
Eq. \eqref{eq:ActivationMode} corresponds to a front,  separating a region where $\Pi\bra{\dots}=0$ from a region where $\Pi\bra{\dots}=1$, propagating at speed $v$
with an interface of width $\frac{1}{\sigma}$ multiplied by an exponential factor $e^{-\frac{\lambda_\beta}{v} \, \pare{\gamma\,\delta \, - \, v \, t}}$. Here, the sign of the real part of $\lambda_\beta$, $\lambda_{\beta,r}$ is important. If $\lambda_{\beta,r} < 0$ the front has the shape depicted in Fig. 
\ref{Fig:ActivationMode} top. It decays exponentially fast as $t \to +\infty$, with a time scale $\frac{1}{\abs{\lambda_{\beta,r}}}$. 
On the opposite, it increases exponentially fast, with a time scale $\frac{1}{\lambda_{\beta,r}}$ as $t \to +\infty$ when $\lambda_{\beta,r} > 0$, thereby enhancing the network effect and accelerating the activation of non linear effect (low threshold or gain control) leading the trajectory out of $\Omega$. Remark that the peak of the drive occurs at $\gamma\,\delta \, - \, v \, t=0$. The inflexion point of the function $\Pi(x)$ is at $x=0$. Thus, when $\lambda_\beta < 0$ the front is a bit behind the drive, whereas it is a bit ahead when $\lambda_\beta > 0$.

Having unstable eigenvalues is not the only way to get out of $\Omega$. Indeed, even if all eigenvalues are stable the drive itself can lead some cells to get out of this set.
When the trajectory of the dynamical system \eqref{eq:Diff_Syst_Vect_Lin} gets out of $\Omega$ two cases are then possible:
\begin{enumerate}[(i)]
\item Either a BCell $i$ is such that $\V{B}{i} < \theta_B$. In this case, $\R{B}{i}\pare{\V{B}{i}, \A{B}{i}}=0$. Then, in the matrix $\cL$ there is a line of zeros replacing the line $i$ in the matrix $\W{B}{}{A}{}$, i.e. at the line $i+N$ of $\cL$. This corresponds to a stable eigenvalue $-\frac{1}{\tau_A}$  for $\cL$, controlling the exponential instability observed in Fig. \ref{Fig:ActivationMode} bottom. Thus, too low BCells voltages trigger a re-stabilisation of the dynamical system.
\item There are BCells  such that condition \eqref{eq:TriggerGain} holds, then gain control is activated and the system \eqref{eq:Diff_Syst_Vect_Gen} becomes non linear. Here we get out of the linear analysis and we have not been able to solve the problem mathematically. There is however a simple qualitative argument. If the cell $i$ enters in the gain control region  then the corresponding line  $i+N$ in the matrix $\W{B}{}{A}{}$ of $\cL$ is replaced by $\W{B}{i}{A}{} \,  \cG_B\pare{\A{B}{i}}$ which rapidly decays to $0$ (see e.g. Fig. \ref{Fig:AnticipationBConly} e). From the same argument as in (i) this generates a stable eigenvalue $\sim -\frac{1}{\tau_A}$ controlling as well the exponential instability.
\end{enumerate}


Eq. \eqref{eq:XalphaDriven} features therefore the direct effect of the stimulus as well as the indirect effect, via the amacrine network, corresponding to a weighted sum of propagating fronts, generated by the stimulus, and influencing a given cell through the connectivity pathways. These fronts interfere, either constructively, inducing a wave of activity enhancing the effect of the stimulus and, thereby, anticipation , or destructively somewhat lowering the stimulus effect. The fine tuning between "constructive" and "destructive" interferences depends on the connectivity matrix via the spectrum of $\cL$ and its projection vectors $\vcP_\beta$. For example, complex eigenvalues introduce time oscillations which are likely to generate destructive interferences, \textit{unless} some specific resonances conditions exist between the imaginary parts of the eigenvalues $\lambda_\beta$. Such resonances are known to exist e.g. in neural network models exhibiting a Ruelle-Takens transition to chaos \cite{ruelle-takens:71}, and they are closely related to the spectrum of the connectivity matrix \cite{cessac:19}. Although we are not in this situation here, our linear analysis clearly shows the influence of the spectrum of $\cL$, itself constrained by $\cW$, on the network response to stimuli and anticipation.


 \paragraph{Spectrum of $\cL$.}  This argumentation invites us to consider different situations where one can figure out how connectivity impacts the spectrum of $\cL$ and thereby anticipation. We therefore provide some general results about the spectrum of $\cL$ and 
potential linear instabilities before considering specific examples.  These results are proved in the appendix \ref{Sec:Spectrum}. As stated in section \ref{Sec:Connectivity_graph}, to go further in the analysis we now assume that a BCell connects only one ACell, with a weight $w^+$ uniform for all BCells, so that $\W{B}{}{A}{} = w^+ \, I_{N,N}$, $ w^+>0$. We also assume that ACells connect to BCells with a connectivity matrix $\cW$, not necessarily, symmetric, with a uniform weight $- w^-$, $ w^->0$, so that $\W{A}{}{B}{} = -w^- \, \cW$. 
 
  We note $\kappa_n, n=1 \dots N$, the eigenvalues of $\cW$ ordered as $\abs{\kappa_1} \leq \abs{\kappa_2} \leq \dots \leq \abs{\kappa_n}$ and $\vpsi_n$ is the corresponding eigenvector. We normalize $\vpsi_n$ so that $\vpsi_n^\dag.\vpsi_n=1$ where $\dag$ is the adjoint. (Note that, as $\cW$ is not symmetric in general, eigenvectors are complex). From the eigenvalues and eigenvectors of $\cW$ one can compute the eigenvalues and eigenvectors of $\cL$ (see Appendix \ref{Sec:Spectrum}), and infer stability conditions for the linear system. The main conclusions are the following:
 
 \begin{enumerate}
     \item The stability of the linear system is controlled by the reduced, a-dimensional parameter:
     \begin{equation}\label{eq:mu}
    \mu= w^- \, w^+ \, \tau^2 \geq 0,
    \end{equation}
where: 
\begin{equation}\label{eq:tau}
\frac{1}{\tau}=\frac{1}{\tau_A} - \frac{1}{\tau_B},
\end{equation}
with a degenerate case when $\tau_A=\tau_B$, considered in the appendix.

\item If $\cW$ is symmetric, its eigenvalues $\kappa_n$ are real, but the eigenvalues of $\cL$ can be real or complex. To each $\kappa_n$ correspond actually to eigenvalues $\lambda_n^\pm$ of $\cL$ (see eq. \eqref{eq:lambda_m}).

\begin{enumerate}
\item If $\kappa_n < 0$, the two corresponding eigenvalues of $\cL$ are real and one of the two corresponding eigenmodes of $\cL$ becomes unstable when:
\begin{equation}\label{eq:UnstabilitySymmetric}
w^- \, w^+ > - \, \frac{1}{\tau_A \, \tau_B} \frac{1}{\kappa_n}.
\end{equation}
\item If $\kappa_n > 0$ and if $\frac{1}{\tau} \neq 0$ the corresponding eigenvalues of $\cL$ are complex conjugate if:
\begin{equation}\label{eq:mucomplex}
\mu > \frac{1}{4 \, \kappa_n} \equiv \mu_{n,c}.
\end{equation}
The corresponding eigenmodes are always stable.
 \end{enumerate}
 
 \item If $\cW$ is asymmetric, eigenvalues $\kappa_n$ are complex, $\kappa_n=\kappa_{n,r} \, + \, i \, \kappa_{n,i}$. The eigenvalues of $\cL$ have the form $\lambda_\beta = \lambda_{\beta,r} \, + \, i \, \lambda_{\beta,i}$, $\beta=1 \dots 2N$ with:
\begin{equation}\label{eq:LambdaComplex}
\left\{
\begin{array}{llll}
\lambda_{\beta,r} &=& -\frac{1}{2 \, \tau_{AB}} \, \pm \, \frac{1}{2 \, \tau}  \, \frac{1}{\sqrt{2}} \, \sqrt{a_n+u_n}\,  ;\\
&\\
\lambda_{\beta,i} &=& \pm \, \frac{1}{2 \, \tau}  \, \frac{1}{\sqrt{2}} \, \sqrt{u_n-a_n},
\end{array}
\right.
\end{equation}
where $a_n=1- 4 \, \mu\, \kappa_{n,r}$ and $u_n=\sqrt{\pare{1- 4 \, \mu\, \kappa_{n,r}}^2 + 16 \, \mu^2 \, \kappa_{n,i}^2}$ $=\sqrt{1-8 \, \mu \, \kappa_{n,r}^2 + 16 \, \mu^2 \, \abs{\kappa_{n}}^2}$. Note that we recover the real case when $\kappa_{n,i}=0$ by setting $u_n=a_n$.\\

Instability occurs if  $\lambda_{\beta,r}>0$ for some $\beta$. This gives:
%
\begin{equation}\label{eq:UnstabilityASymmetric}
a_n+u_n > 2 \frac{\tau^2}{\tau_{AB}^2},
\end{equation}
a condition on $\mu$ depending on $\kappa_{n,r}$ and $\kappa_{n,i}$. 
 \end{enumerate}

\paragraph{Remarks:}
The introduction of the a dimensional parameter $\mu$ allows us to simplify the study of the joint influence of $w^-,w^+,\tau$ on dynamics because stability is controlled by $\mu$ only. In other words, a bifurcation condition of the form $\mu=\mu_c$ signifies that this bifurcation holds when the parameters $w^-,w^+,\tau$ lay on the manifold defined by $ w^- \, w^+ \, \tau^2=\mu_c$.\\

 We now show this in two examples of connectivity and afferent instabilities. 
 
\ssSec{Nearest neighbours connectivity}{Nearest_neighbours}

\paragraph{Eigenmodes of the linear regime.} We  consider the case where the matrix $\cW$, connecting ACells to BCells, is a matrix of nearest neighbours symmetric connections. In this case, $\cW$ can be written in terms of the discrete Laplacian $\Delta$ on a $d$ dimensional regular lattice, $d=1,\, 2$ with lattice spacing $\delta_A=\delta_B$, set here equal to $1$ without loss of generality:
\begin{equation}\label{eq:C_Lap}
\cW=2 d \, I +  \Delta.
\end{equation}
Because of this relation we will often use the terminology Laplacian connectivity for the nearest-neighbours connectivity.
We also assume that dynamics holds on a square lattice with 
null boundary conditions. That is, ACell and BCells are located on $d$-dimensional grid with indices $i_x,i_y=0 \dots L+1$ where, the voltage and activity of cells with indices $i_x=0$, $i_x=L+1$, $i_y=0$ or $i_y=L+1$, vanish. 

%
%
%

The eigenvalues and eigenvectors are explicitly known in this case. They are parametrized by a quantum number $n=n_x \in \Set{1 \dots L=N}$ in one dimension, and by two quantum numbers $\pare{n_x,n_y} \in \Set{1 \dots L=N}^2$ in two dimensions. They define a wave vector $\vk_n=\pare{\frac{n_x \pi}{L+1},\frac{n_y \pi}{L+1}}$ corresponding to wave lengths $\pare{\frac{L+1}{ n_x},\frac{L+1}{ n_y}}$. 
Hence, the first eigenmode $(n_x=1,n_y=1)$ corresponds to the largest space scale (scale of the whole retina) with the smallest eigenvalue (in absolute value) $s_{(1,1)}=2\pare{\cos\pare{\frac{\pi}{L+1}}+\cos\pare{\frac{\pi}{L+1}}-2}$. To each of these eigenmodes is related a characteristic time $\tau_n=\frac{1}{\lambda_n}$.
%
The slowest mode is the mode $\pare{1,1}$. 
In contrast, the fastest mode is the mode $(n_x=L,n_y=L)$, corresponding to the smallest space scale, the scale of the lattice spacing $\delta=1$.


Eigenvalues $\kappa_n$ can be positive or negative. Consider for example the $1$ dimensional case, where $\kappa_n=2 \,\cos\pare{\frac{n \pi}{L+1}}$. We choose $L$ even to avoid having a zero eigenvalue $\kappa_{\frac{L}{2}}$.
Eigenvalues $\kappa_n$,  $n=1 \dots \frac{L}{2}$  are positive, thus the corresponding eigenvalues $\lambda_n^\pm$ of $\cL$  are complex, and stable. The modes with the largest space scale $\frac{L}{n}$ are therefore stable for the linear dynamical system, with oscillations. Eigenvalues $\kappa_n$,  $n=\frac{L}{2}+1 \dots L$  are negative, thus the corresponding eigenvalues $\lambda_n^\pm$ of $\cL$  are real. From \eqref{eq:UnstabilitySymmetric} the mode $n$ becomes unstable when :
\begin{equation}\label{eq:InstLaplacian}
w^-\, w^+ > - \frac{1}{\tau_A \tau_B} \, \frac{1}{2 \, \cos\pare{\frac{n \pi}{L+1}}}.
\end{equation}
Therefore, the first mode to become unstable is the mode $L$ with the smallest space scale $1$ (lattice spacing). For large $L$, this happens for $w^-\, w^+ \sim \frac{1}{2}\, \frac{1}{\tau_A \, \tau_B}$. This instability induces spatial oscillations at the scale of the lattice spacing. When $w^-\, w^+$ further increases the next modes become unstable. This instability results in a wave packet following the drive (as shown in Fig. \ref{Fig:ActivationMode}). The width of this wave packet is controlled by the unstable modes and by non linear effects. We now illustrate the relations of these spectral properties with the mechanism of anticipation.

\paragraph{Numerical results.} 

\begin{figure}
\centering
\includegraphics[width=1.\textwidth]{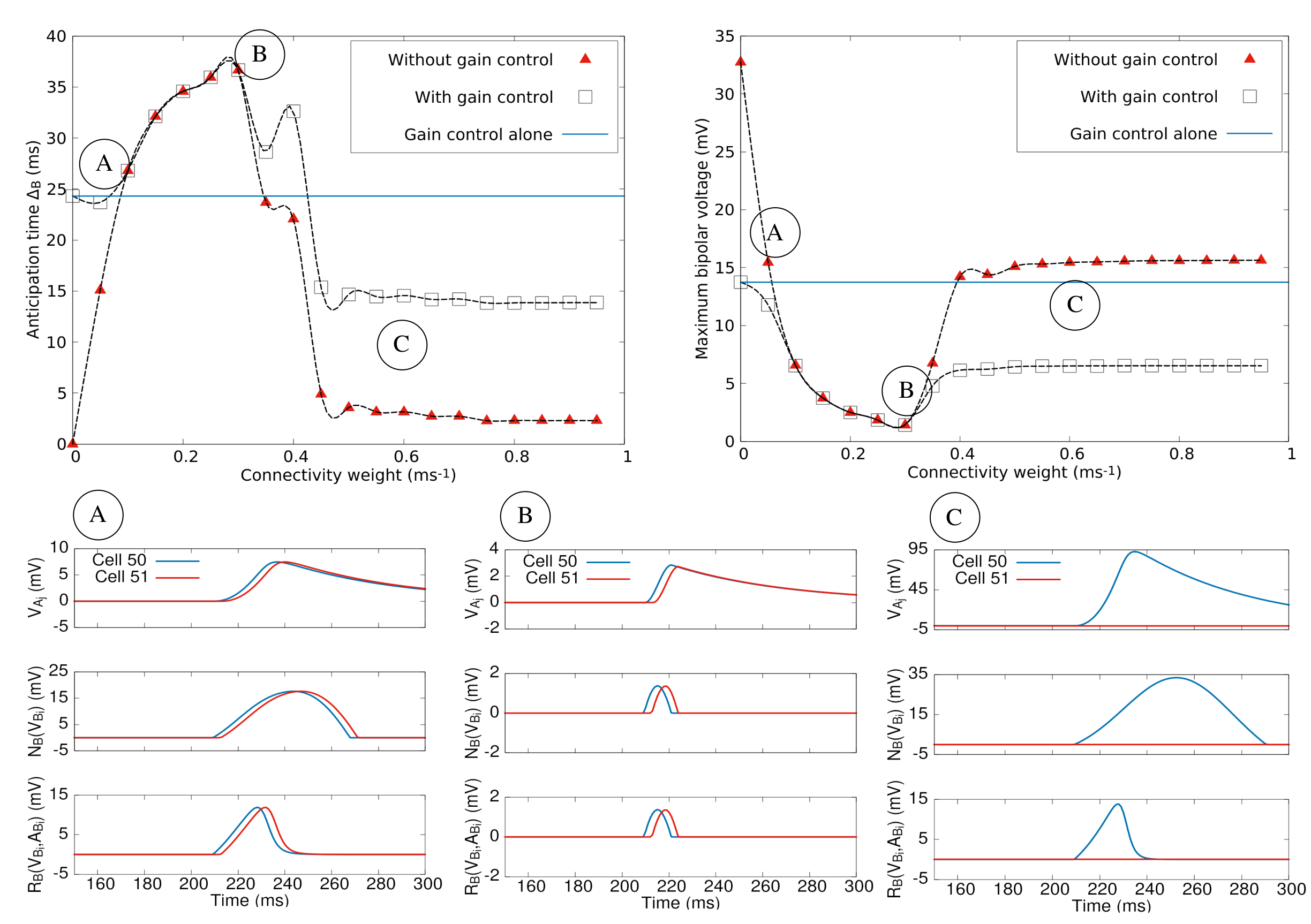}
\caption{ 
\label{Fig:AnticipationLaplacian} \textbf{Anticipation in the Laplacian (nearest-neighbours) case.} \textbf{Top.} Anticipation time and maximum bipolar response as a function of the connectivity weight $w$. The blue line corresponds to gain control alone (it does not depend on $w$). Red triangles corresponds to the effect of lateral ACells connectivity, without gain control. White squares correspond to the compound effect of ACells connectivity and gain control. The $3$ regimes A, B, C are commented in the text. 
\textbf{Bottom.} Response curves of ACells and BCells corresponding to the three regimes : (A) $w = 0.05$ $ms^{-1}$ with a small cross-inhibition, (B) $w = 0.3$ $ms^{-1}$ with an opposition  in activity between the blue (cell 50) and red cell (51), (C) $w = 0.6$ $ms^{-1}$, where the red cell (51) is completely inhibited by cell 50. 
}
\end{figure}

In all the following 1D simulations, we consider a bar  with a width $150 \, \mu m$, moving in one dimension  at constant speed $v=3 \, mm/s$. We simulate 100 BCells, 100 ACells and 100 GCells placed on a 1D horizontal grid, with a uniform spacing of $\delta_b = \delta_a = \delta_g = 30 \, \mu m$ between to consecutive cells. At time $t=0$, the first cell lies at $100 \, \mu m$ to the right of the leading edge of the moving bar.
We set $\tau_{B} = 300$ ms , $\tau_{a} = 50$ ms, $\tau_A = 100$ ms , corresponding to $\tau=150$ ms  (eq. \eqref{eq:tau}). We vary the value of weights $w^+$, $w^-$. For the sake of simplicity, we also choose $w^+ = -w^-=w$ to have only one control parameter. 
We investigate how the bipolar anticipation time $\Delta_B$ and the maximum in the response $\R{B}{}$  depend on $w$. This is summarized in Fig. ~\ref{Fig:AnticipationLaplacian} top, where we have shown the effect of gain control alone (blue horizontal line, independent of $w$), the effect of ACells lateral connectivity alone (red triangles) and the compound effect (white squares). Anticipation time is averaged over all cells. On the same figure (bottom) we see the responses of two neighbour cells lying at the center of the lattice.

As $w$ increases we observe three areas of interest: the first, (A), corresponds to a regime where ACells connectivity has a negative effect on anticipation, competing with gain control.
As $w$ is small the anticipation is controlled by the direct pathway I, II of Fig. \ref{Fig:architecture}, from BCells to GCells, with a small inhibition coming from ACells, thereby decreasing the voltage of BCells and impairing the effect of gain control. This explains why the anticipation time in the case of lateral connectivity + gain control is smaller than the anticipation time of gain control alone. The network effect (red triangles) on anticipation time increases with $w$ though. This corresponds to the "push-pull" effect already evoked above in section \ref{Sec:Amacrine_cells_layer}.  When a BCell $\Cell{B}{i}$ feels the stimulus, its activity increases favoured by the stimulus, it increases the voltage of the connected ACell, inhibiting the next Bcell $\Cell{B}{i+1}$ thereby inducing a feedback loop, the push-pull effect, enhancing the voltage of $\Cell{B}{i}$. \\ 
In zone (B) the push-pull effect becomes more efficient than gain control alone. In this region, the voltage of the BCell feeling the bar increases fast, while the voltage of its neighbours becomes more and more negative, enhancing the feedback loop. This holds \textit{until} the voltage rectification \eqref{eq:NLinRectBip} takes place. This is the time when the dynamical system gets out of $\Omega$. The push pull effect then saturates and $\V{B}{i}$ reaches a maximum, corresponding to a peak in activity. This peak is reached faster than the peak in the function $\cG_{B}(A)$. 
Thus, the peak of $\R{B}{i}(t)$ occurs at the same time as the peak of $\cN_B(\V{B}{i}(t))$, and, thus, \textit{before} the reference peak (time $t_B$ for isolated BCells defined in section \ref{Sec:Anticipation_BCell}). In other words, the ACells lateral connectivity allows the BCell to outperform the gain control mechanism for anticipation. 
As $w$ increases in zone B the push-pull effect (averaged over BCells) reaches a maximum then decreases. This is because the increase in $w$ makes the inhibitory effect of ACells stronger and stronger on silent Bcells which then remain silent longer and longer because the ACells voltage increases with $w$, and it takes longer for it to decrease and de-inhibit the neighbours.  
The silent cells are less and less sensitive to the stimulus, being strongly and durably inhibited. \\
In region C, the anticipation is again dominated by gain control. In this case, the effect on cells depends on the parity of their index. The response of BCells is either completely suppressed or identical to the response of the reference case (with gain control alone). This is why the average anticipation time with gain control is about half of the gain control without network effect.  
Cells that are inhibited do no participate to anticipation, and the others anticipate in the same way than with gain control alone. \textit{Note that this "parity" effect is due to the nearest neighbours connectivity and the symmetry of interactions.}\\

We now interpret and complete these results from the point of view of the spectrum of $\cL$ and associated dynamics. 
%
%
The fastest mode to destabilize, 
corresponds to the smallest space scale, the lattice spacing. This is a mode with alternate sign, at the scale of the lattice.  We call it the "push-pull" mode, as it is \textit{precisely}  what makes the push-pull effect. When the push-pull mode becomes unstable, the excited BCell becomes more and more excited and the next BCell more and more inhibited. However, the time it takes, $\tau_L$, has to be compared to the time where the bar stays in the RF, $\tau_{bar}$ (and more generally the time it takes to RF kernel to respond to the bar). In the case of the simulation $\sigma_{center} = 90 \, \mu m$ (see Appendix, table \eqref{Tab:params}) and $v=3 \, \mu m / ms$ giving a characteristic time $\tau_{bar}=270 \, ms$, whereas, as we observed $\tau_L < 100$ ms.
The push-pull mode is therefore quite faster than $\tau_{bar}$ so the push-pull effect takes place fast and lead to a fast exponential increase of the front depicted in Fig. \ref{Fig:ActivationMode} right. This explains the rapid increase of network anticipation effect observed in regions A, B of Fig. \ref{Fig:AnticipationLaplacian}.

%
%

\ssSec{Random connectivity}{Random_connectivity}

In this section, we study the behaviour of the model using the more realistic, probabilistic type of connectivity presented in section \ref{Sec:Connectivity_graph} and more thoroughly studied in Appendix \ref{Sec:Random_Connectivity}. Within this framework, a given ACell  $\Cell{A}{i}$ receives the upstream activity from the BCell lying at the same position, $\Cell{B}{i}$, with a constant weight $w$. The same ACell inhibits BCells with which it is coupled through the random adjacency matrix $\cW$, generated by the probabilistic model of connectivity, and the weight matrix $\W{B}{}{A}{}= -w \cW$. We recall that the connectivity depends on a scale parameter $\xi$ for the branch length), and the mean and variance $\bar{n},\sigma$ for the distribution of the number of branches. These parameters can be found in the table \eqref{Tab:params} in appendix.

\paragraph{Eigenmodes of the linear regime.}
Similarly to section \ref{Sec:Nearest_neighbours} we now analyse the spectrum of $\cL$ when $\cW$ is a random connectivity matrix. Although a couple of results can be established (using the Perron-Frobenius theorem) we have not been able to find general mathematical results on the spectrum or eigenvectors of this family of random matrices. We thus performed numerical simulations. 

\begin{figure}
\begin{center}
\includegraphics[height=6cm,width=7cm]{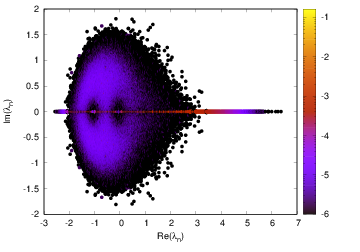}
\hspace{0.5cm}
\includegraphics[height=6cm,width=7cm]{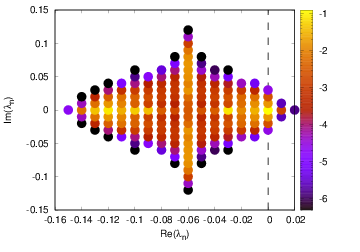}
\vspace{1cm}
\includegraphics[height=6cm,width=7cm]{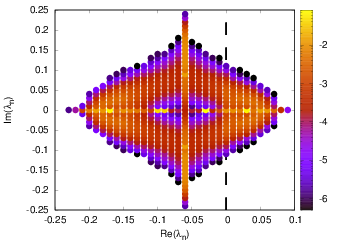}
\hspace{0.5cm}
\includegraphics[height=6cm,width=7cm]{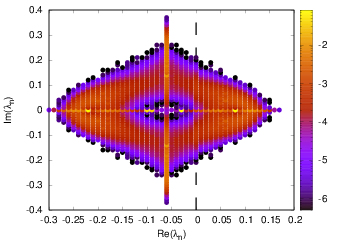}
\caption{\textbf{Spectral density of eigenvalues for $\xi=2, \bar{n}=4,\sigma_n=1$. Top Left.} 
For the matrix $\cW$ (density estimated over $10000$ samples). Density is represented in color plots, in log scale. The color bar refers to powers of $10$. 
\textbf{Top Right.} Spectral density of $\cL$ for $w=0.05$. \textbf{Bottom Left.} Spectral density of $\cL$ for $w=0.1$. \textbf{Bottom Right.} Spectral density of $\cL$ for $w=0.15$. Unstable eigenvalues are on the right to the vertical, dashed, line $x=0$ \label{Fig:RandomSpectrum}}
\end{center}
\end{figure}

The spectrum of $\cL$ is deduced from the spectrum of $\cW$ as exposed above. The spectrum of $\cW$ depends on $\bar{n},\sigma$ and $\xi$. 
In Fig. \ref{Fig:RandomSpectrum} we have plotted, on the left, an example of such spectrum. This is the spectral density (distributions of eigenvalues in the complex plane) obtained from the diagonalization of $10000$ matrices $100 \times 100$ (so the statistics is made over $10^6$ eigenvalues). We note that the largest eigenvalues is always real  positive, a straightforward consequence of Perron-Frobenius theorem \cite{gantmacher:98,seneta:06}. More generally, we observe an over-density of real eigenvalues. The same holds for random Gaussian matrices with independent entries $\cN(0,\frac{1}{N})$ \cite{edelman:97} whose asymptotic density converges to the circular law \cite{girko:84}. The shape of the spectral density in our model differs from the circular law though and it depends on the parameters $\bar{n},\sigma$ and $\xi$.

\begin{figure}
\begin{center}
\includegraphics[height=6.5cm,width=7cm]{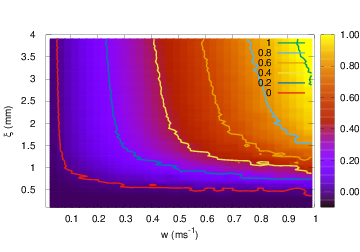}
\hspace{0.1cm}
\includegraphics[height=6.5cm,width=7cm]{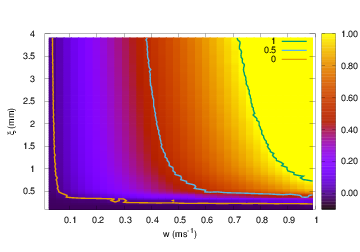}
\caption{\textbf{Heat map} for the largest real part eigenvalue in the plane $w,\xi$, for different values of $\bar{n}$. 
\textbf{Left.} $\bar{n}=1$.
\textbf{Right.} $\bar{n}=4$.
Colour lines are level lines. The level line $0$ is the frontier of instability of the linear dynamical system.
\label{Fig:HeatMap}}
\end{center}
\end{figure}

On the same figure we show the corresponding spectral density of $\cL$ obtained from eq. \eqref{eq:LambdaComplex} for  $w=0.05,0.1,015$. We have taken here $\tau_A=30, \tau_B=10$ ms to see better the transitions with $w$ (level lines in Fig. \ref{Fig:HeatMap}). 
There is an evident symmetry with respect to $\frac{1}{\tau_{AB}} = -0.066$ expected from the mathematical analysis.
We see that the largest eigenvalue is real (although it is not necessarily related to the largest eigenvalue of $\cW$). We also see that, as $w$ increases, a large number of (complex) eigenvalues become unstable. There is actually a frontier of instability that we have plotted in the plane $w,\xi$ for different values of $\bar{n}$. This is shown in Fig. \ref{Fig:HeatMap} (dashed line). The level line $0$ is the frontier of instability of the linear dynamical system. This frontier has the (empirical) form $(\xi-\xi_0).(w-w0)=c$
where $c$ has the dimension of a characteristic speed.

What matters here is that there are complex unstable eigenvalues with no specific resonance relations between them. They are therefore prone to generate destructive interferences in \eqref{eq:XalphaDriven}. 

\paragraph{Numerical results}

\begin{figure}
\centering
\includegraphics[width=1.\textwidth]{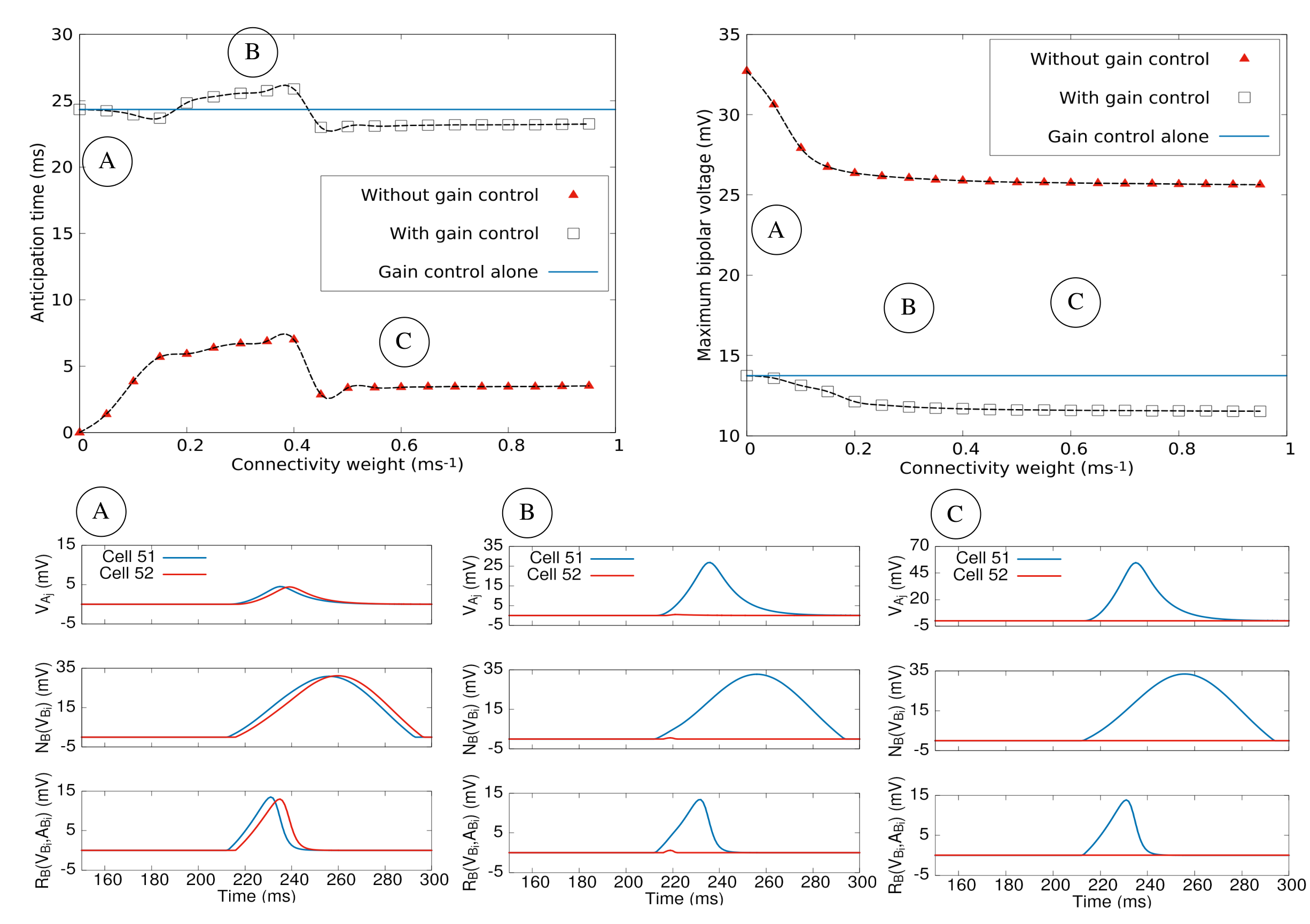}
\caption{ 
\label{Fig:AnticipationRandom} \textbf{Average anticipation in the random connectivity case}. \textbf{Top.}  Bipolar anticipation time and maximum in the response $\R{B}{}$ as a function of the connectivity weight $w$, in the case of a random connectivity graph with $\xi = 1$, $\bar{n}=2$ and $\sigma_n=1$. \textbf{Bottom.}  Response curves of ACells and BCells $51-52$ corresponding to the three regimes : (A) $w = 0.05 ms^{-1}$, (B) $w = 0.3 ms^{-1}$, (C) $w = 0.6 ms^{-1}$. 
}
\end{figure}

In fig. \ref{Fig:AnticipationRandom} we consider, similarly to Fig. \ref{Fig:AnticipationLaplacian} for Laplacian connectivity, the effect of random connectivity on anticipation, compared to pure gain control mechanism.
In contrast to the Laplacian case, we have here more parameters to handle: $\xi$, which controls the characteristic length of branches and $\bar{n},\sigma_n$ which control the number of branches distribution. We present here a few results where $\xi$ varies whereas the average number of branches $\bar{n}=2$ ($\sigma_n=1$). 
A more systematic study is done in \cite{souihel:19}. The interest of varying $\xi$ is to start from a situation which is close to the Laplacian case (characteristic distance $\xi=1$) and to increase $\xi$ to see how the size of the dendritic tree of ACells may impact anticipation.
 This is a preliminary step toward considering different physiological ACells type (e.g. narrow-, medium-, or wide-field \cite{dowling:15}). Note however that the probability of connection given the distance of cells (fixed by $\bar{n},\sigma_n$) implicitly impacts  $w$  and the anticipation effects.

The main difference with the Laplacian case is the asymmetry of connections. Here, symmetry means that if Acell $j$ connects the BCell $i$, then the Acell $i$ connects the BCell $j$ too. This does not necessarily hold for random connectivity and this has a strong impact on the push-pull effect and anticipation. So, even if the connectivity is short-range when $\xi$ is small, mainly connecting nearest neighbours, we observe already a big difference with the Laplacian case.
This is shown in Figure \ref{Fig:AnticipationRandom}, where $\xi=1$.  Similarly to the Laplacian case we observe 3 main regions depending on $w$. To have the same representation as Fig. \ref{Fig:AnticipationLaplacian} we present $\V{A}{},\N{B}{},\R{B}{}$ for two connected cells (here Acell $51$ and BCell $52$). However, in this case, connection is not symmetric: ACell $51$ inhibits BCell $52$ but  ACell $52$ does not inhibits BCell $51$. 

We observe $3$ regimes, as in the Laplacian case.
In the first region (A) ACells random connectivity has a negative effect on anticipation, as compared to gain control alone. However, since in this case the "push-pull" effect is not evoked, this decay simply comes from the fact that BCell $52$ receives an inhibition for the ACell $51$, that reduces the effect of gain control. This inhibition is though not strong enough to significantly shift the peak response, as in region (B). 

Indeed, in region (B), the inhibition of BCell $52$ is strong enough to outperform the effect of gain control. In this case, and similarly to the Laplacian case, the peak of $\R{B}{i}(t)$ occurs at the same time as the peak of $\cN_B(\V{B}{i}(t))$, and, before the reference peak. However this effect is not consistent over all cells and only occurs for BCells that receive active inhibition. This explains why the performance of the Laplacian connectivity is better, on average,  in this region. 

Finally, as $w$ grows higher, the inhibition grows stronger, completely inhibiting BCell $51$. Cells that do not receive any inhibition, as BCell $49$ in this example, keep a response that is identical to the response without ACell connectivity. The fraction of cells receiving inhibition in this case being quite small (about 15), this explains why the stationary value of anticipation is fairly close to the value with gain control alone. 

\paragraph{The role of the characteristic distance}

\begin{figure}[h!]
\begin{center}
\includegraphics[width=1.\textwidth]{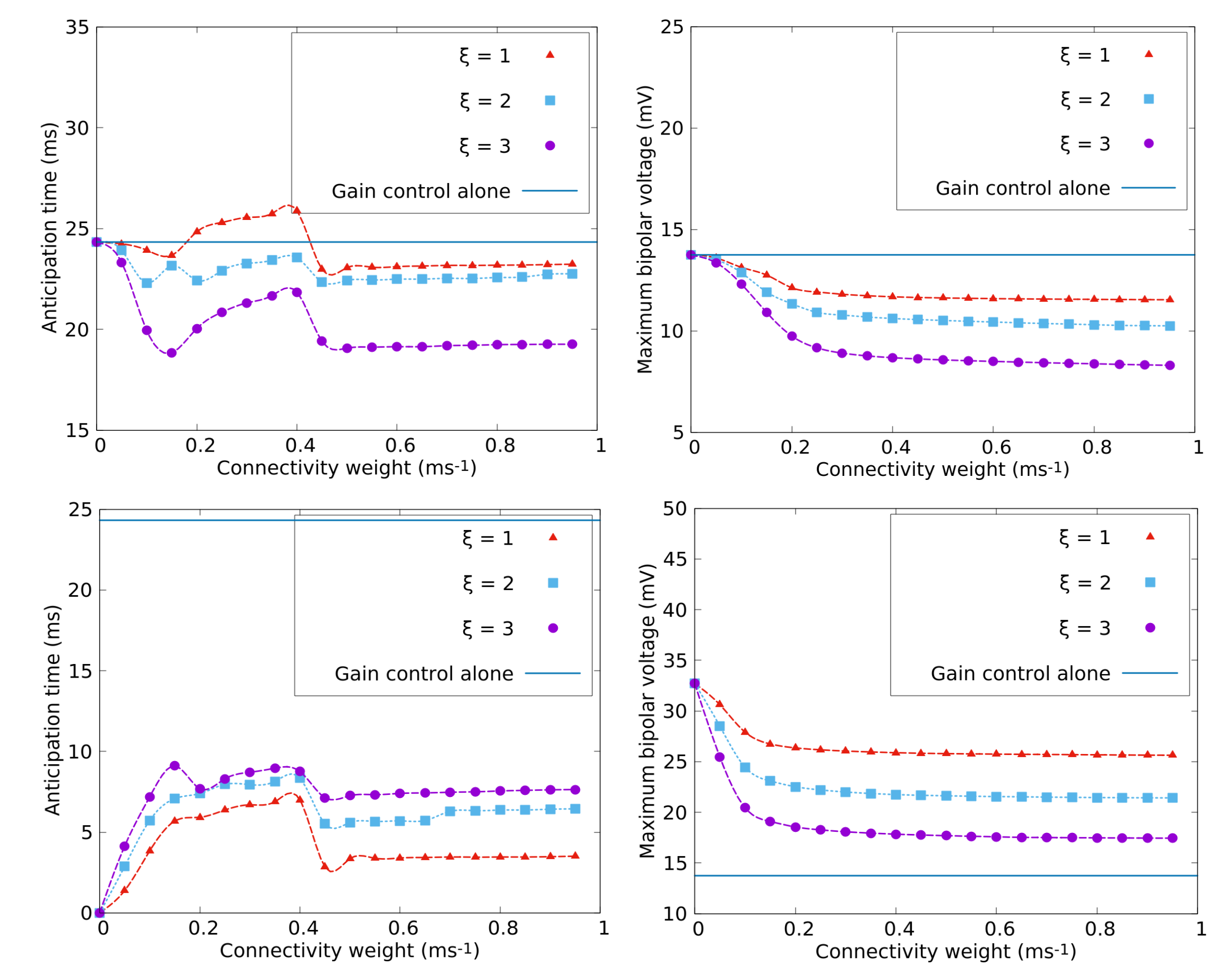}
\caption{\textbf{Role of the characteristic branch length $\xi$ on anticipation.} \textbf{Top.}  The joint effect of the random ACells connectivity and gain control on anticipation for $\xi=1,2,3$. \textbf{Left.} Average bipolar anticipation time; \textbf{Right.} Maximum value of the bipolar response $\R{B}{}$. \textbf{Bottom.}  The single effect of the random ACells connectivity on anticipation with the same representation. 
\label{Fig:Amacrine_random_xi}}
\end{center}
\end{figure}

In figure \ref{Fig:Amacrine_random_xi}, we analyse the effect of the characteristic length $\xi$ on anticipation. On the top of the figure 
we represent the joint effect of the random ACell connectivity and gain control on anticipation for three values of $\xi$. At the bottom we represent the only effect of the random ACell connectivity for the same values of $\xi$. We observe that performance in anticipation decreases with $\xi$. More precisely, we observe an anticipatory effect in this case, as shown in figure \ref{Fig:Amacrine_random_xi} bottom, but this effect is not able to compete with gain control alone. Even worse, the compound effect shown in \ref{Fig:Amacrine_random_xi} top is disastrous since increasing $w$ renders the anticipation time smaller and smaller. 

This spurious effect can be interpreted through the analysis made in section \ref{Sec:Mathematical_study}, eq. \eqref{eq:XalphaDriven}. From the spectrum of $\cL$, we see that there are unstable complex eigenvalues whose number increases with $w$. These eigenvalues are prone to generate destructive interferences, especially when their number becomes large as $w$ increases, explaining the small peak in region $B$. 
The consequence on cells activity and gain control can be dramatic as seen in the red trace of Fig. \ref{Fig:AnticipationRandom} B bottom, line $\R{B}{}$. This depends on the precise connectivity pattern when long range connections from ACells to BCells induce a desensitization of BCells, which is not counterbalanced by the push-pull effect as in the Laplacian connectivity case.

\ssSec{Conclusion}{ConclusionAC}

The two numerical examples considered in this section emphasize the role of symmetry in the synapses, and more, generally the role of complex versus real eigenvalues in the spectrum of $\cL$. Recall that, from section \ref{Sec:Mathematical_study}, if $\cW$ is symmetric complex eigenvalues are always stable, so, for the type of architecture considered here, unstable destructive interferences only occur when $\cW$ is asymmetric. This leads to several questions, potential subjects for further studies.
\begin{enumerate}
\item \textbf{How much does anticipation depend on the degree of asymmetry in the matrix $\cW$ ?} The way we generate the random connectivity in the model does not allow us to tune the degree of asymmetry (i.e. the probability that a connection $\Cell{A}{j} \to  \Cell{B}{i}$ exists simultaneously with a connection $\Cell{A}{i} \to  \Cell{B}{j}$). Therefore, one has to find a different way to generate the connectivity. From the mathematical analysis made in Appendix \ref{Sec:Random_Connectivity} a distribution depending exponentially on the distance, with a tunable probability to have a symmetric connection, could be appropriate. We don't know about any experimental results characterizing this degree of symmetry of the connections in the retina. On mathematical grounds, and from the analogy of the spectrum of $\cW$ with a circular law, one could expect the spectrum of $\cW$ to become more and more elongated on the real axis as the degree of symmetry increases, in an elliptic like law \cite{nguyen-orourke:14}. 
\item \textbf{Non linear effects.} The destructive interference effect in our model is partly due to the linear nature of the ACells dynamics. In non linear dynamics, eigenvalues of the evolution operator can display resonances conditions favouring constructive interferences. On biological grounds, it is for example known that Starburst Amacrine Cells display periodic bursting activity during development, disappearing a few days after birth \cite{zheng-lee-etal:06}. Bursting and its disappearance can be understood in the framework of bifurcation theory of a non linear dynamical system featuring these cells \cite{karvouniari-gil-etal:19}. In this setting, even if they are not bursting in the mature stage, SACs remain sensitive to specific stimulation that can temporally synchronize them, thereby enhancing the network effect, with a potential effect on anticipation. 
\end{enumerate}

\sSec{The potential role of gap junctions on anticipation}{Role_Gaps}

In this section, we study the network ability to improve anticipation in the presence of gap junctions coupling, as in eq. \eqref{eq:dVdtgapbefore}, and gain control at the level of GCells. 

We start first with mathematical results and show then simulation results.

\ssSec{Mathematical study}{Mathematical_study_Gaps}

We use a continuous space limit for  a one dimensional lattice. The extension to $2$ dimension is straightforward. Here, $x$ corresponds to the preferred direction of the direction sensitive cells. We consider a continuous spatio-temporal field $V(x,t)$, $x \in \setR$, such that $\V{G}{k} \equiv V(k \delta_G,t)$. We assume likewise that $\V{G}{k}^{(P)} \equiv V^{(P)} (k \delta_G,t)$ for some continuous function $V^{(P)}(x,t)$ corresponding to the GCells bipolar pooling input \eqref{eq:Pooling} and we take the limit $\delta_G \to 0$. In this limit eq. \eqref{eq:dVdtgapbefore} becomes:
\begin{equation}\label{eq:FieldGap}
\frac{\partial \V{G}{}}{\partial t} = f(x,t) - v_{gap} \, \frac{\partial \V{G}{}}{\partial x} + O(\delta_G^2),
\end{equation}
where $v_{gap} \equiv w_{gap} \, \delta_G$ has the dimension of a speed and $\frac{\partial V^{(P)}(x,t)}{\partial t} \equiv f(x,t)$. Finally, we note $C(x)$ the initial profile so that $V(x,t_0)=C(x)$. 

\paragraph{Solution.}

Neglecting terms of order $\delta_G^2$ the general solution  of \eqref{eq:FieldGap} is :
$$
V_G(x,t)=C(x-v_{gap}(t-t_0)) \, + \, \int_{t_0}^t 
f(x-v_{gap}(t-u),u) du.
$$
%

Eq. \eqref{eq:FieldGap} is a transport equation of ballistic type \cite{souihel:19}. For example, if we consider a stimulation of the form $V^{(P)}(x,t)=h(x-v \,t)$, where $h$ is a Gaussian pulse of the form \eqref{eq:VdriveAppGauss}, propagating with speed $v$,
%
and an initial profile $C(x)= h\pare{x-v \, t_0}$, the voltage of GCells obeys:
\begin{equation}\label{eq:SolPropGap}
V_G(x,t) 
= \, \underbrace{\frac{v}{v-v_{gap}} \,h\pare{x-vt}}_{\pi_{stim}}  - \underbrace{\frac{v_{gap}}{v-v_{gap}}\, h\pare{x-v_{gap}t-(v-v_{gap}) \, t_0}}_{\pi_{gap}}.
\end{equation}

When $v_{gap}=0$ the GCells voltage follows the stimulation i.e. $V_G(x,t)=h\pare{x-vt}$. In the presence of gap junctions there are two pulses: the first one, $\pi_{stim}$ with amplitude $\frac{v}{v-v_{gap}}$ propagating at speed $v$ and following the stimulation; the second one, $\pi_{gap}$, with amplitude $-\frac{v_{gap}}{v-v_{gap}}$, propagating at speed $v_{gap}$.

\begin{figure}
\centering
\includegraphics[width=1.\textwidth]{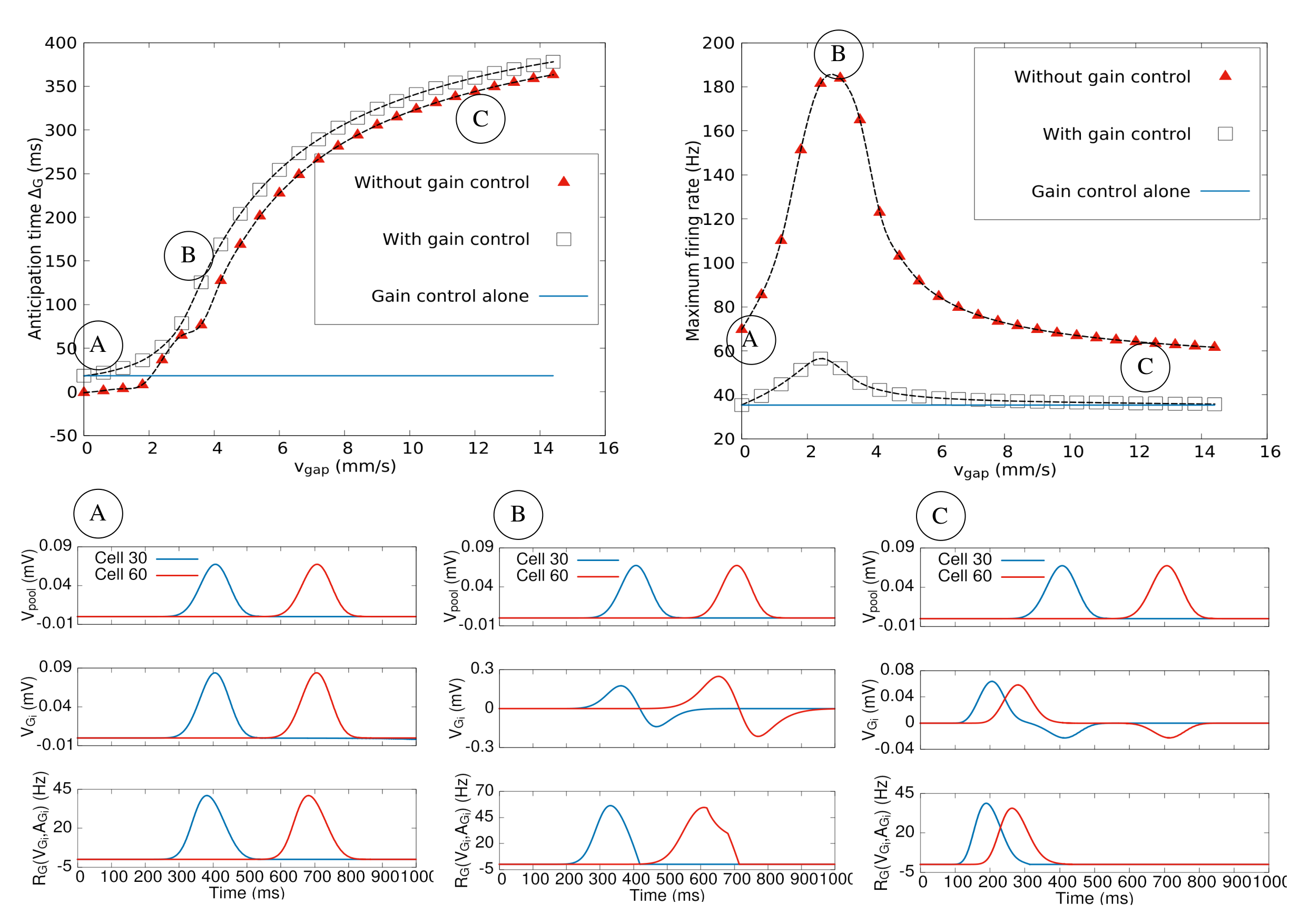}
\caption{\textbf{Anticipation  for non symmetric gap junctions. Top.} GCell anticipation time and maximum firing rate as a function of the gap junction velocity $v_{gap}$. Bottom : response curves of GCells corresponding to the three regimes : (A) $v_{gap} = 0.6$ $mm/s$, (B) $v_{gap} = 3$ $mm/s$, (C) $v_{gap} = 12$ $mm/s$. 
The curves display 3 main regimes (see text): In (A) $v_{gap} < v$ and the positive front propagates at the same speed as the pooling voltage triggered by the stimulus; In (B),  $v_{gap} = v$, the positive front and the negative fronts both propagate at the speed $v_{gap}$ and the amplitude of the positive front ($\V{G}{}(t)$) increases with $t$;  In (C), $v_{gap} > v$ and the positive front propagates faster than the stimulus so that the peak of activity arises earlier. The negative front propagates at the stimulus speed. 
\label{Fig:GapJunctionConnectivity}}
\end{figure}

We have the following cases (we take $t_0=0$ for simplicity). An illustration is given in Fig. \ref{Fig:GapJunctionConnectivity}.
\begin{enumerate}

\item If $v$ and $v_{gap}$ have the same sign:

\begin{enumerate}[(A)]
\item If $v_{gap} < v$, the front $\pi_{stim}$ is amplified by a factor $\frac{v}{v-v_{gap}}$, whereas there is a refractory front $\pi_{gap}$, proportional to $v_{gap}$, behind the excitatory pulse.
\item If $v=v_{gap}$,  $V_G(x,t) =h\pare{x-v_{gap}t}+v_{gap}(t-t_0) h'\pare{x-v_{gap}t}$ which diverges like $t$ when $t \to \infty$ and $x \to + \infty$. This divergence is a consequence of the limit $\delta_G \to 0$ in \eqref{eq:FieldGap}.
\item If $v_{gap} > v$ the amplitude of $\pi_{stim}$ follows the stimulation with a negative sign (hyper polarization) whereas $\pi_{gap}$ is \textit{ahead} of the stimulation, with a positive sign, travelling at speed $v_{gap}$.
\end{enumerate}

\item If $v$ and $v_{gap}$ have the opposite sign, we set $v=-\alpha \, v_{gap}$, with $\alpha >0$. Then, the front $\pi_{stim}$ follows the stimulus but is attenuated by a factor $\frac{\alpha}{1+\alpha}$. The front $\pi_{gap}$ propagates in the opposite direction with an attenuated amplitude $\frac{1}{1+\alpha}$.
\end{enumerate}
This shows that these gap junctions favour the response to motion in the preferred direction and attenuate the motion in the opposite direction although the attenuation is weak.
The effect is reinforced by gain control \cite{souihel:19}. The most interesting case is 1 c
 where these gap junctions can induce a wave of activation ahead of the stimulation.

\paragraph{Effect of gain control.} When the low voltage threshold  $\cN_{G}$ \eqref{eq:non_linearity_gang} and the gain control $\cG_{G}(A)$ \eqref{eq:gain_control_gang} are applied to $V_G(x,t)$ there are two effects: (i) the hyperpolarized front is cut by $\cN_{G}$; (ii) the positive pulse induces a raise in activity, which, in turn, triggers the ganglion gain control $\cG_{G}(A)$ inducing an anticipated peak in the response of the GCell, similar to what happens with BCells, with a different form for the GCell gain control though. Moreover, in contrast to pathway II of Fig. \ref{Fig:architecture} where only gain control generates anticipation, in pathway IV the wave of activity generated by gap junctions increases anticipation by two distinct effects. If $v_{gap} < v$ the cell's response propagates at the same speed as the stimulus, but its amplitude is larger than the case with no gap junction (term $\pi_{stim}$). From eq. \eqref{eq:SolPropGap} this results in an increase of $h_B$ to an effective value $h_B \frac{v}{v-v_{gap}}$ inducing an improvement in the anticipation time (with a saturation of the effect, though, as $v_{gap} \to v$).  If $v_{gap} > v$ the cell's response propagates at a larger speed than the stimulus (term $\pi_{gap}$), so that the cell responds before the time of response without gaps.  This induces as well an increase in the anticipation time.

\ssSec{Numerical illustrations}{Numerical_illustrations}

We consider a bar  with a width $200 \, \mu m$, moving in one dimension at constant speed $v=3 \, mm/s$. We simulate here 100 GCells, placed on a 1D horizontal grid, with a spacing of $ 30 \, \mu m$ between to consecutive cells. At time $t=0$, the first cell lies at $100 \, \mu m$ from the leading edge of the moving bar.

We investigate how the GCells anticipation time and GCells firing rate depend on $v_{gap}$ in Fig. ~\ref{Fig:GapJunctionConnectivity}. The top shows the effect of gain control alone (blue horizontal line, independent of $v_{gap}$ ), the effect of the asymmetric gap junction connectivity alone (red triangles) and the compound effect (white squares). Anticipation time is averaged over all GCells. On the bottom part of the figure, we show the responses of two GCells of indices 30 and 60, spaced by $ 900 \, \mu m$.

As explained in the  section \ref{Sec:Mathematical_study_Gaps}, we observe the 3 regimes A,B,C mathematically anticipated above. Note that, for these parameter values, the negative trailing front predicted in A is not visible.
%

\ssSec{Symmetric gap junctions}{Symmetric_gap_junctions} 

The asymmetry observed by Trenhlom et al. is due to the specific 
structure of the direction selective GCell dendritic tree \cite{trenholm-schwab-etal:13}. However, in general gap junctions connectivity is expected to be symmetric.
So, to be complete we consider here the effect of symmetric gap junctions on anticipation. 
It is not difficult to derive the equivalent of eq. \eqref{eq:FieldGap} in this case too. This is a diffusion equation of the form:
$$
\frac{\partial \V{G}{}}{\partial t} = f(x,t) + D_{gap} \, \Delta \V{G}{} + O(\delta_G^4),
$$
where $D_{gap}=w_{gap} \, \delta G^2$ is the diffusion coefficient and $\Delta$ is the Laplacian operator. 

The response to a Gaussian stimulus of the form \eqref{eq:VdriveAppGauss} reads:
\begin{equation}\label{eq:SolPropGapSymm}
\V{G}{}(x,y,t)=\bra{H\conv{x,y,t}f},
\end{equation}
where:
\begin{equation}\label{eq:GaussianKernelGapSym}
H(x,y,t)=\frac{e^{-\frac{x^2+y^2}{4 D_{gap} \, t }}}{4 \pi D_{gap}\, t}
\end{equation}
which is the heat equation diffusion kernel.

\begin{figure}
\centering
\includegraphics[width=1.\textwidth]{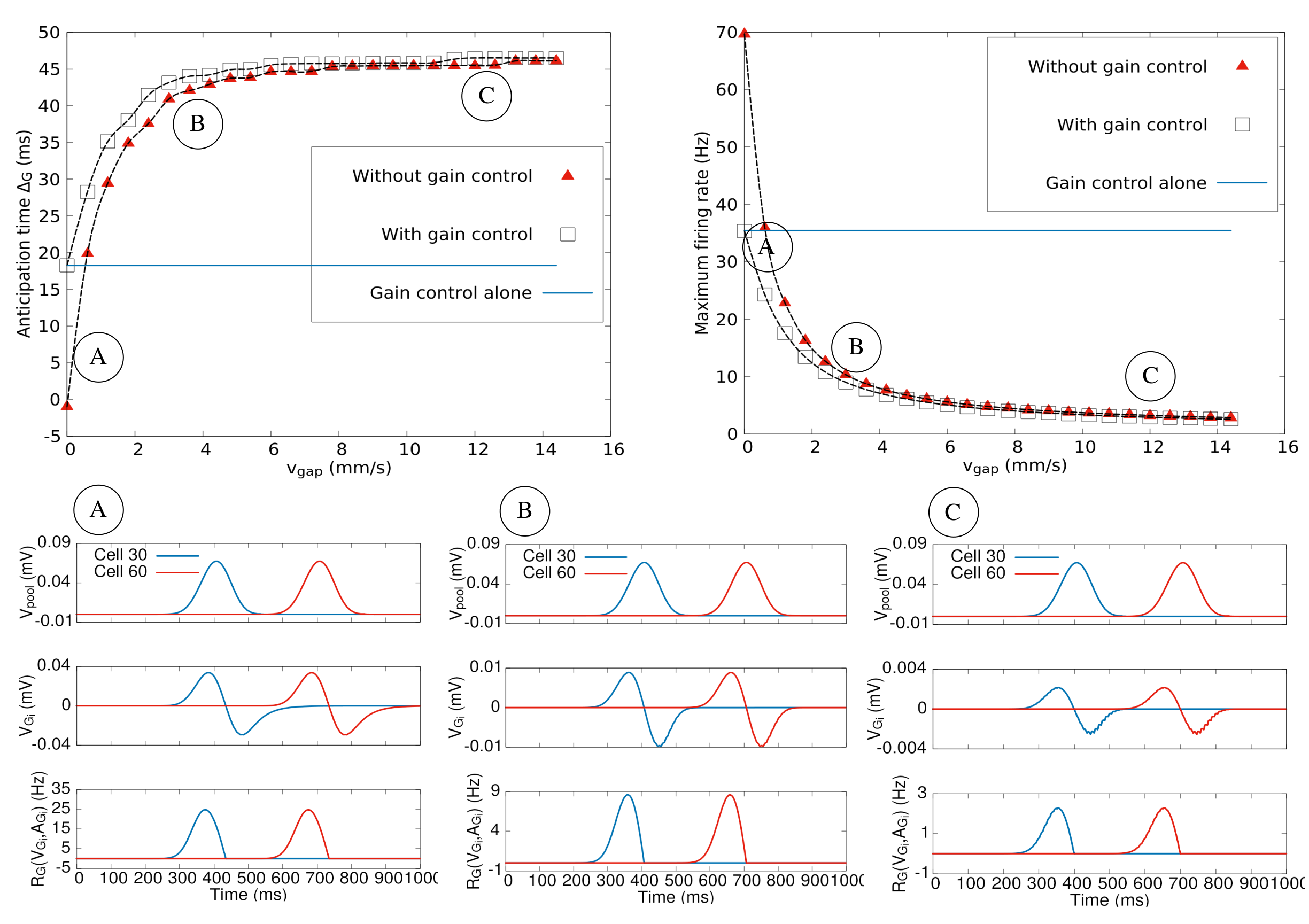}
\caption{\textbf{Anticipation for symmetric gap junctions. Top.} GCell anticipation time and maximum firing rate as a function of the gap junction velocity $v_{gap}$. \textbf{Bottom.} response curves of GCells corresponding for three values of $v_{gap}$ : (A) $v_{gap} = 0.6$ $mm/s$, (B) $v_{gap} = 3$ $mm/s$, (C) $v_{gap} = 12$ $mm/s$. For consistency, we have kept the same values as in the asymmetric case. 
Here, anticipation time grows continuously until saturation, while the maximum firing decreases like a power law as a function of $v_{gap}$. 
\label{Fig:GapJunctionConnectivity_sym}}
\end{figure}

Recall that $f(x,t) \equiv \frac{\partial V^{(P)}(x,t)}{\partial t}$. So, if $V^{(P)}(x,t)=h(x-v \,t)$, where $h$ is a Gaussian pulse of the form \eqref{eq:VdriveAppGauss} propagating with speed $v$,
$f$ is a bimodal function of the form $\frac{v}{\sigma^2} \, h(x-v \,t) \times (x-v \,t)$, the shape of which can be seen in Fig. \ref{Fig:GapJunctionConnectivity_sym} bottom, second row. The convolution with the heat kernel leads to a front propagating at the same rate as the stimulus, with a diffusive spreading  whose rate is controlled by $D_{gap}$. In particular, there is positive bump ahead of the motion, which can induce anticipation, as shown in Fig. \ref{Fig:GapJunctionConnectivity_sym} top. The effect is weak, though, essentially because the diffusive spreading makes the amplitude of the response decrease fast as a function of $D_{gap}$. 

Although this positive front, for small $D_{gap}$, increases a bit the anticipation time by accelerating the gain control triggering, rapidly the peak in the response $\R{B}{}$ is lead by the voltage peak corresponding to the positive bump, with a low voltage. 
The position of this peak is, roughly,  at a distance $\sigma=\sqrt{\sigma_{center}^2+\sigma_B^2}$ from the peak of the Gaussian pool, where $\sigma_{center}$ is the width of the center RF and $\sigma_B$ the width of the bar. This corresponds to a time $\frac{\sigma}{v}$ ahead of the peak in the drive, fixing a maximal value to the anticipation time (see the saturation of the anticipation time curve in Fig. \ref{Fig:GapJunctionConnectivity_sym} top, left). In our case, $\sigma \sim 134$ given a saturation peak at   $\frac{134 \mu m}{3 \mu m / ms}=44.84$ $ms$. A consequence of the voltage decay is the corresponding power law ($\frac{1}{\sqrt{D_{gap}}}$ for large $D_{gap}$) decay of the firing rate (Fig. \ref{Fig:GapJunctionConnectivity_sym} top, right).\\
To conclude, the situation with symmetric gap junctions is in high contrast with direction selective gap junctions where the response to stimuli was ballistic and was not decreasing with time. On this basis we consider that, for symmetric gap junctions, the anticipation effect is irrelevant, especially taking into account the smallness of the voltage response in case C.

\ssSec{Numerical results}{Numerical_Results_Gaps}

We investigate in this section how the GCell anticipation time  and GCells firing rate depend on the gap junction conductance in the case of symmetric gap junctions. In figure \ref{Fig:GapJunctionConnectivity_sym} top, we use the same representation as Fig. \ref{Fig:GapJunctionConnectivity}. For consistency with the direction sensitive case, we choose $v_{gap}=\frac{D_{gap}}{\delta G}$ as control parameter. We also take  (A) $v_{gap} = 0.6 mm/s$, (B) $v_{gap} = 3 mm/s$, (C) $v_{gap} = 12 mm/s$ in figure \ref{Fig:GapJunctionConnectivity_sym} bottom. This corresponds to a diffusion coefficient (A) $D_{gap} = 18 \times 10^{-3} mm^2/ s$, (B) $D_{gap} = 90 \times 10^{-3} mm^2/ s$, (C) $D_{gap} = 360 \times 10^{-3} mm^2/ s$

\ssSec{Conclusion}{Conclusion_gaps}

In this section we have shown how gap junctions direction sensitive cells can display anticipation due to the propagation of a wave of activity ahead of the stimulus. This effect is negligible for symmetric gap junctions. Note that symmetric gap junctions are known to favour  waves propagation, for example in the early development (stage I, see \cite{kahne-rudiger-etal:19} and reference therein for a recent numerical investigation). Here gap junctions are considered in a different context, due to the presence of a non stationary stimulus, triggering the wave.

Let us now comment our computational result. How does it fit to biological reality ? Depending on the gap-junction conductance value the propagation patterns we predict are quite different. 

What is the typical value of $v_{gap}$ in biology ? It is difficult to make an estimate from the expression $v_{gap}=\frac{g_{gap} \delta_G}{C}$. The membrane capacity $C$ and gap junctions conductance can be obtained from the literature (for connexins Cx36,  $g_{gap} \sim 10-15$ pS \cite{srinivas-rozental-etal:99}) but the distance $\delta_G$ is more difficult to evaluate. In the model, this is the average distance between GCells' soma which corresponds to $\sim 200-300$ $\mu m$. But, in the computation with gap junctions  what matters is the length of a connexin channel which is quite smaller. Taking $\delta_G$ as the distance between GCells assumes a propagation speed between somas at the speed of a connexin, which is wrong because most of the speed is constrained by the propagation of action potential along the dendritic tree. So we used a phenomenological argument (we thank O. Marre to point out it to us). The correlation of spiking activity between GCells neighbours is about $2-5 \, ms$ for cells separated by $\sim 200-300 \, \mu m$ \cite{volgyi-pan-etal:13}. This gives a speed $v_{gap}$ in the interval 
$\bra{40, 150}$ $mm/s$
which is quite fast compared to the bar speed in experiments.

So we are in case 1 b and one should observe an experimental effect ? To the best of our knowledge an effect of DSGC gap junctions
on motion anticipation has not been observed. But we don't know about experiments targeting precisely this effect. It would be interesting to block gap junctions and address Berry et al. \cite{berry-brivanlou-etal:99} or Chen et al. \cite{chen-marre-etal:13} experiments in this case. The difficulty is that  blocking gap junctions blocks many essential retinal pathways. We do not pursue this discussion further concluding that our model proposes a computational prediction that could be interesting to be experimentally investigated.

\sSec{Response to 2 dimensional stimuli}{2D_Stimuli}

In this section, we present some examples of retinal responses and anticipation to trajectories more complex than a bar moving in one dimension with a uniform speed. The aim here is not to do an exhaustive study but, instead, to assess qualitatively some anticipatory effects not considered in the previous sections. 

\ssSec{Flash lag effect}{flash_lag}

The flash lag effect is an optical illusion where a bar moving along a smooth trajectory and a flashed bar are presented to the subject, and are perceived with a spatial displacement, while they are actually aligned. A variation of this illusion consists of a bar moving in rotation, a bar flashed in angular alignment, giving rise to a perceived angular discrepancy. We have investigated this effect in our model, in the presence of the different anticipatory effects considered in the paper. 

\begin{figure}[h!]
\begin{center}
\includegraphics[width=1\textwidth]{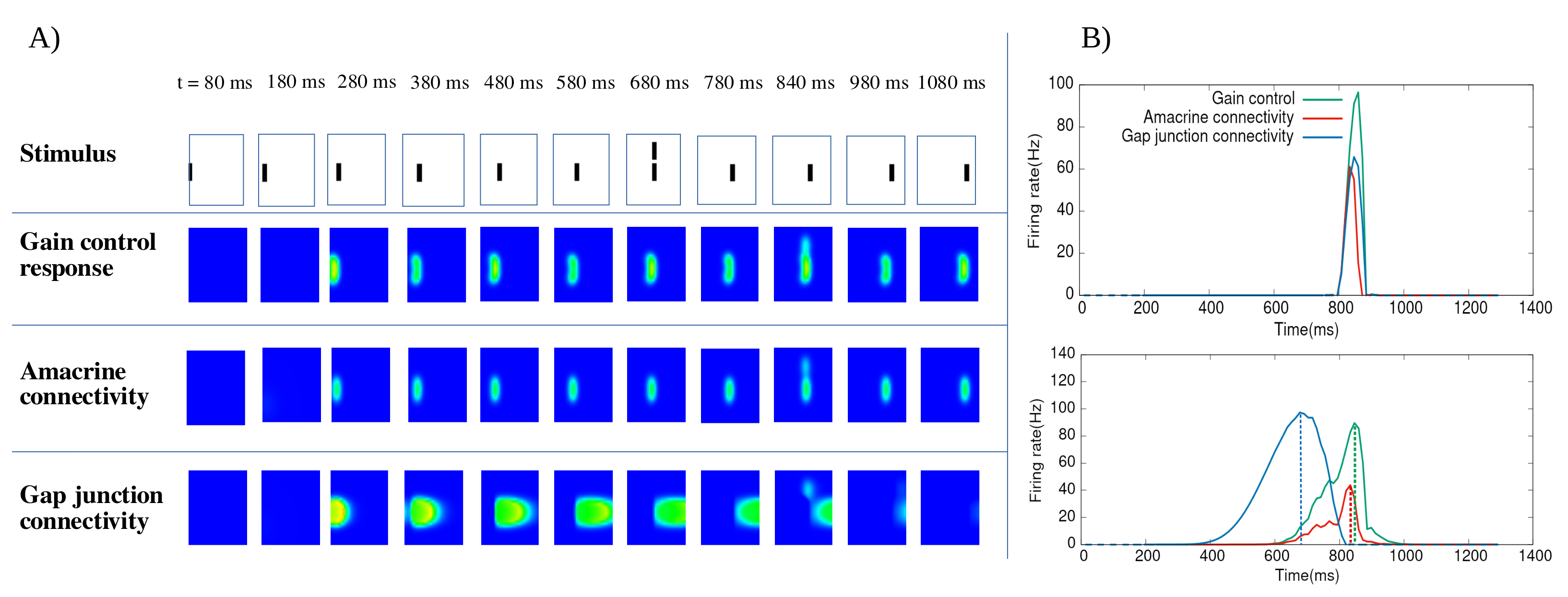}
\end{center}
\caption{\textbf{Flash lag effect with different anticipatory mechanisms.} A) Response to a flash lag stimulus : a bar moving in smooth motion, with a second bar flashed in alignment with the first bar for one time frame. The first line shows the stimulus, the second line shows the GCells response with gain control, the third line presents the effect of lateral ACells Laplacian connectivity with $w = 0.3$ $ms^{-1}$, and the last line shows the effect of asymmetric gap junctions with $v_{gap} = 9$ mm/s. B) Time course response of \textbf{(top)} a cell responding to the flashed bar, and \textbf{(bottom)} a cell responding to the moving bar. 
Dashed lines indicate the peak of each curve.
\label{Fig:flash_lag_reconstruction}}
\end{figure}

Fig. \ref{Fig:flash_lag_reconstruction} shows the response to a bar moving with a smooth motion, while a second bar is flashed in alignment with the first bar at one time frame. The first line shows the stimuli, consisting of 130 frames, of a bar moving at 2.7 mm/s, with a refreshment rate of 100 Hz. 
The second line shows the GCell response with gain control, the third line presents the effect of lateral amacrine connectivity, in the case of a Laplacian graph. Keeping the same values of parameters as in the 1D case, we set $w = 0.3$ $ms^{-1}$ corresponding to the case B in Fig. \ref{Fig:AnticipationLaplacian}.
Finally, the last line shows the effect of asymmetric gap junctions, having a preferred orientation in the direction of motion, with $v_{gap} = 9 $ mm/s. 

In the case of the gain control response, the peak of response to the moving bar is shifted by about $10$ ms in the direction of motion, as compared to the static bar. The flashed bar  elicits a lower response, given its very short appearance in comparison with the characteristic time of adaptation. We choose this time short enough to avoid gain control triggering, explaining the difference with the strong response observed by Chen et al. \cite{chen-marre-etal:13} in the presence of a still bar.

In the case of amacrine connectivity, the moving bar representation is shrunk as compared to the gain control case, given the prevalence of inhibition, while the level of activity for the flashed bar remains roughly the same. In this case, cells responding to the moving bar reach their peak activity slightly earlier (about $19$ ms for these parameters value)  than in the gain control case (Fig. \ref{Fig:flash_lag_reconstruction}, (B) top.)

Finally, asymmetric gap junction connectivity displays a wave propagating ahead of the bar, increasing the central blob, which is much larger than the size of the bar in the stimulus, while the flashed bar activity remains similar to the previous cases.

\ssSec{Parabolic trajectory}{parabolic_trajectory}

\begin{figure}[h!]
\begin{center}
\includegraphics[width=1.\textwidth]{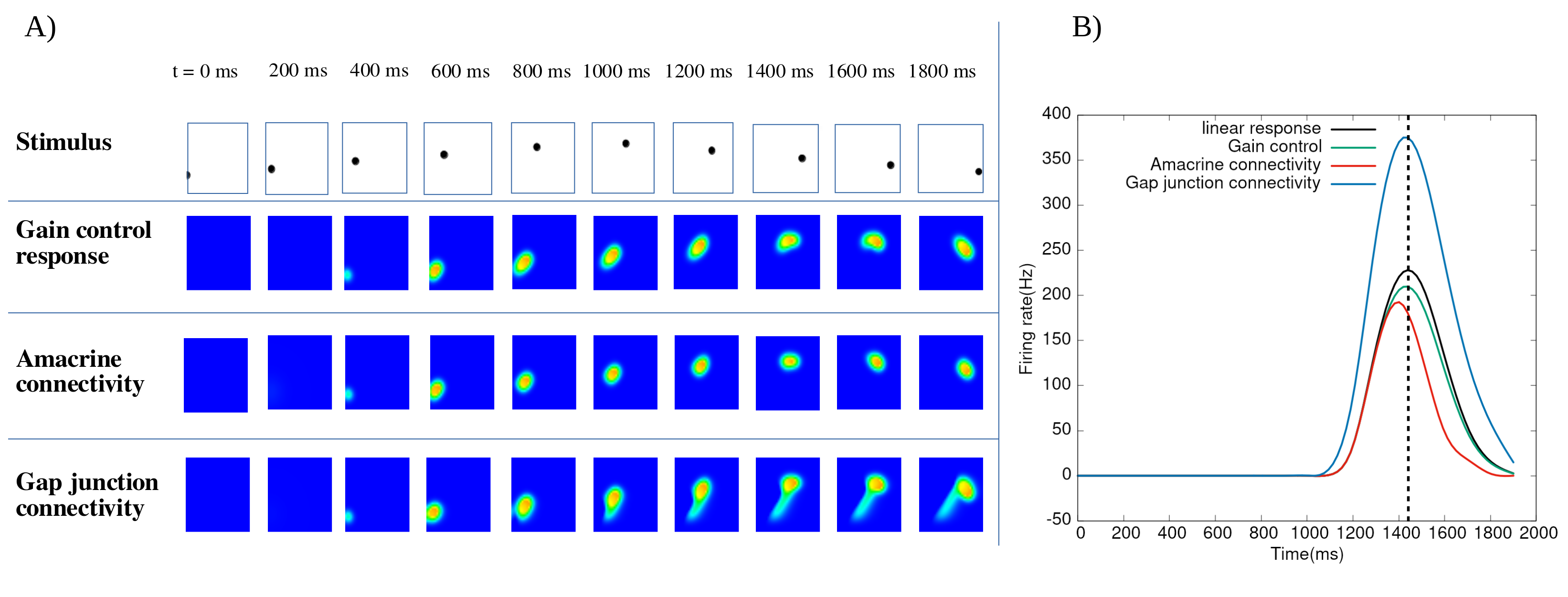}
\end{center}
\caption{\textbf{Effect of anticipatory mechanisms on a parabolic trajectory.} A) Response to a dot moving along a parabolic trajectory. The first line shows the stimulus, the second line shows the GCells response with gain control, the third line presents the effect of lateral ACells connectivity with $w = 0.3$ $ms^{-1}$, and the last line shows the effect of asymmetric gap junctions with $v_{gap} = 9$ $mm/s$ . B) Time course response of a cell responding to the dot near the trajectory turning point. Linear response corresponds to the response to the stimulus without gain control.
\label{Fig:parabolic}
}
\end{figure}

In this subsection, we assess the effect of the three anticipatory mechanisms on a parabolic trajectory. The interest is to have a trajectory with a change in direction and speed, thus an acceleration. The stimulus consists of $20$ frames, displayed at 10 Hz. The simulations parameters and connectivity weights are the same as the ones used in the previous section. 
Fig. \ref{Fig:parabolic} shows the response to a dot moving along a parabolic trajectory. 

In the case of gain control, GCells response is more elongated, which has a distortion effect on the dot representation near the turning point of the trajectory ($1400-1600$ ms). Cells responding near the trajectory turning point are still anticipating motion, as the peak response of the gain control curve is slightly shifted to the left, compared to the RF response (Fig. \ref{Fig:flash_lag_reconstruction}, B).

In the case of amacrine connectivity, the elicited response is also more localized, as compared to the gain control response, and the flow of activity follows more accurately the stimulus. This is a direct consequence of the sensitivity of the ACell connectivity model to the stimulus acceleration. In this case, the peak response is also more shifted as compared to the gain control case. 

Finally, the gap junction connectivity model performs worse in this case, giving rise to a propagating wave that doesn't follow the trajectory, since the latter is not parallel to the direction to which GCells are sensitive. Cells responding near the trajectory turning point have a higher level of activity and an increased latency, while the peak response roughly corresponds to the gain control case.


\ssSec{Angular anticipation}{angular_anticipation}

We investigate in this subsection a two dimensional example of motion where angular anticipation takes place. The stimulus consists here of 72 frames, displayed at 100 Hz, of a bar moving at a constant angular speed of $ 4.25$ rad/ms.

\begin{figure}[h!]
\begin{center}
\includegraphics[width=1.\textwidth]{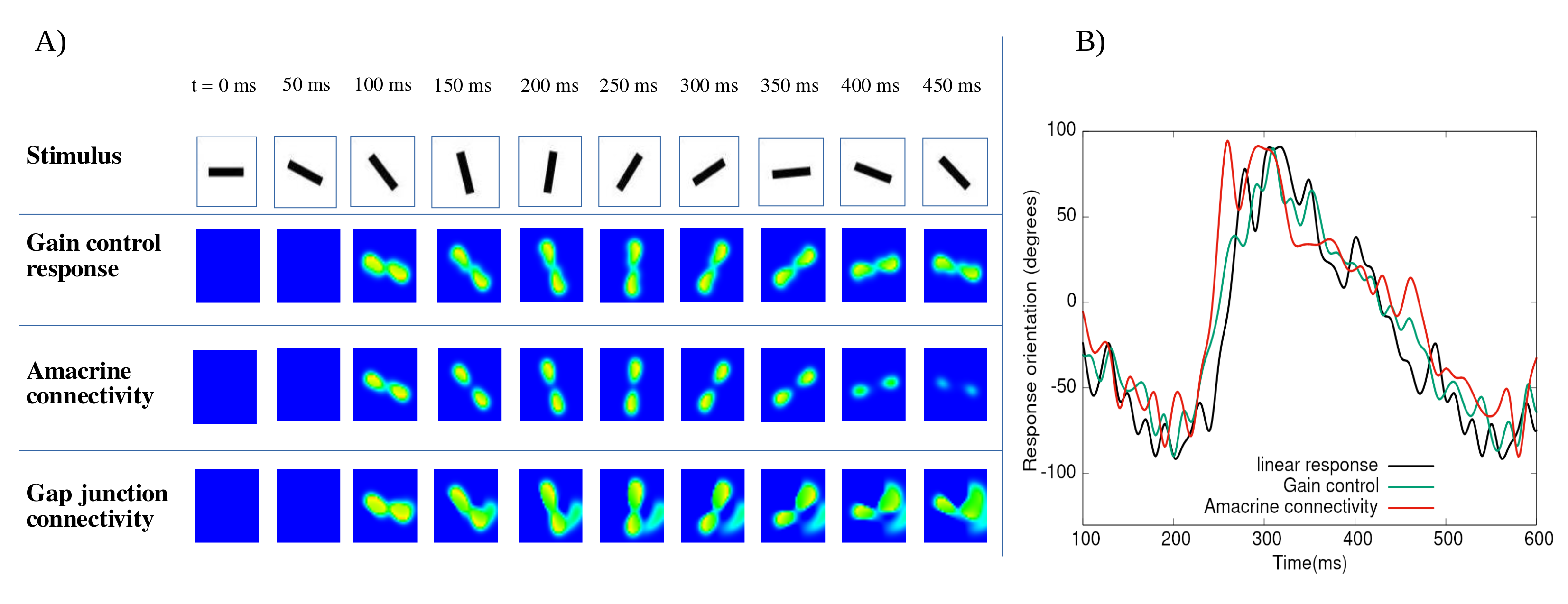}
\end{center}
\caption{\textbf{Anticipation for a rotating bar.} A) Response to a bar rotating at $ 4.25 rad/ms$. The first line shows the stimulus, the second line shows the GCells response with gain control, the third line presents the effect of lateral ACells connectivity with $w = 0.2$ $ms^{-1}$, and the last line shows the effect of asymmetric gap junctions with $v_{gap} = 9$ $mm/s$ . B) Time course response of the bar orientation in the reconstructed retinal representations.} 
\label{Fig:rotation_retina}
\end{figure}

Fig. \ref{Fig:rotation_retina} shows the retina response to a rotating bar, with the angular orientation of activity as a function of time for the different models. We used Matlab to estimate the bar orientation from the displayed activity, fitting the set of activated points by an ellipse whose principal axis determines the response orientation. 

In the three cases, one can see that the response around the center of the bar is suppressed due to gain control adaptation. While the gain control activity orientation roughly follows the linear response ( Fig. \ref{Fig:rotation_retina} (B)), the ACells response shows a slight angular shift (frames : 250-300 ms) which is also visible on the response orientation time course. The ACells angular anticipation is however only observed during the first period of the bar. Interestingly, this effect vanishes during the second rotation, due to a persistent effect of the activation function, generating a sort of a suppressive effect erasing the second occurrence of the bar (frame : 450 ms). We shall point out that the ACells connectivity weight in this simulation has been reduced to $w = 0.2$ $ms^{-1}$, since with a value of $w = 0.3$ $ms^{-1}$ used in the previous simulations, the response to the second rotation of the bar is completely suppressed. 

Finally, similarly to the parabolic trajectory, the gap junction connectivity model performs worse due to the wave propagating from left to right, distorting once more the bar shape. Consequently, the bar activity orientation in this case has been discarded in Fig. \ref{Fig:rotation_retina} B, the orientation estimate giving poor results. 


\sSec{Conclusion}{}

This section shows how lateral connectivity can play a role in motion anticipation of 2D stimuli, both in the case of the classical flash lag effect, and more complex trajectories. Indeed, for a given network setting, ACells connectivity can noticeably improve anticipation with respect to Gain Control, in all three stimuli, and has also the advantage of being sensitive to trajectory shifts (sec. \ref{Sec:parabolic_trajectory}).  

While gap junction connectivity improves anticipation when the trajectory of the bar is parallel to the preferred GCells direction, it also induces more blur around the bar, and shape distortion in the case of parabolic motion and rotation, suggesting a trade-off between anticipation and object recognition for this specific model.

\section{Discussion}

Using a simplified model, mathematically analysed with numerical simulations examples, we have been able to give strong evidences  that lateral connectivity - inhibition with ACells, gap junctions - could participate to motion  anticipation in the retina. The main argument is that a moving stimulus can - under specific conditions mathematically controlled - induce a wave of activity which propagates ahead of the stimulus thanks to lateral connectivity. This suggests that, in addition to local gain control mechanism inducing an anticipated peak of GCells activity, lateral connectivity could induce a mechanism of neural latencies reduction, similar to what is observed in the cortex \cite{benvenuti-chemla-etal:20,subramaniyan-ecker-etal:18,jancke-erlaghen-etal:04}. This is visible in particular in Fig. \ref{Fig:flash_lag_reconstruction}, where the gap junction coupling induces a wave which increases the GCell level activity before the bar reaches its RF.

Yet, this studies raise several questions and remarks.
The first one is, of course, the biological plausibility. At the core of the model, what makes the mathematical analysis tractable is the fact that we can reduce dynamics, in some region of the phase space, to a linear dynamical system. This structure is afforded by two facts: (i) Synapses are characterized by a simple convolution; (ii) Cells, especially Acells, have a simple passive dynamics, where non linear effects induced e.g. by ionic channels are neglected, as well as propagation delays.  As stated in the introduction the goal here is not to be biologically realistic, but instead, to illustrate potential general spatio-temporal response mechanisms taking into account specificities of the retina, as compared to e.g. the cortex. Essentially, most neurons are not spiking (except GCells and some type of BCells or ACells, not considered here \cite{baden-berens-etal:13}). Yet, synapses  follow the same biophysics as their cortical counterpart. As it is standard to model the whole chain of biophysical machinery triggering a post-synaptic potential upon a sharp increase of the pre-synaptic voltage by a convolution kernel \cite{destexhe-mainen-etal:94}, we adopted here the same approach. Note that it is absolutely not required, in this convolutional approach, for the pre-synaptic increase in voltage to be a spike; it can be a smooth variation of the voltage. Note also that higher order convolution kernels can be considered, integrating more details of the biological machinery. These higher order kernels are represented by higher order linear differential equations \cite{faugeras-touboul-etal:09}.  Concerning point (ii) - non linear effects are neglected - 
especially for ACells (BCells have gain control), there are not so many available models of ACells.
A linear model for predictive coding using linear ACells has been used by Hosoya et al \cite{hosoya-baccus-etal:05}. We discuss it in more detail below. The non linear models of ACells
we know has been developed to study the retina in its early stage (retinal waves) and feature either AII ACells \cite{choi-zhang-etal:14} or Starburst ACells \cite{karvouniari-gil-etal:19}. In section \ref{Sec:ConclusionAC} we have briefly commented how non-linear mechanisms could enhance resonance effects in the network and, thereby, favour the propagation of a lateral wave of activity induced by a moving stimulus. This would of course deserve more detailed study. Another potentially interesting non linear mechanism is short term plasticity, discussed below. 

The second question one may ask about the model is about the robustness of this mechanism with respect to parameters. The model contains many parameters, some of them (BCells, GCells and gain control) coming from the previous paper from Berry et al \cite{berry-brivanlou-etal:99}
and Chen et al \cite{chen-marre-etal:13}. Although they didn't perform a structural stability analysis of their model (i.e. stability of the model with respect to small variations of parameters), we believe that they are tuned away from bifurcation points so that slight changes in their (isolated cells) model parameters would not induce big changes. As we have shown the situation changes dramatically when cells are connected via lateral connectivity. Here, many types of dynamical behaviour can be expected simply by changing the connectivity patterns in the case of ACells. 
A more detailed analysis would require a closer investigation of ACells to BCells connectivity and an estimation of synaptic coupling, implying to define more specifically the type of ACell (AII, Starburst, A17, wide field, medium field, narrow field, ...) and the type of functional circuit one wants to consider. Note that ACells are difficult to access experimentally due to their location inside the retina. Even more difficult is a measurement of ACells connectivity, especially the degree of symmetry discussed in our paper. Such studies can be performed at the computational level, though, where the mathematical framework proposed here can be applied and extended. Computational results does not tell us what is reality but shed light on what it \textit{could} be.\\

We would like now to address several possible extensions of this work.

\paragraph{The retino-thalamico-cortical pathway.}
The retina is only the early stage of the visual system. Visual responses are then processed via the thalamus and the cortex. As exposed in the introduction, anticipation is also observed in V1 with a different modality than in the retina.
In this paper, our main focus was on the shift of the peak response, while when studying anticipation in the cortex, the main focus lies in the increase of the response latency, i.e. the delay between the time the bar reaches the receptive field of a cortical column, and the effective time its activity starts rising. How do these two effects combine ? How does retinal
anticipation impact cortical anticipation ? To answer these questions at a computational level one would need to propose a model of the retino-thalamico-cortical pathway which, to the best of our knowledge, has never been done. Yet, we have developed a retino-cortical (V1) model - thus, short-cutting the thalamus - based on a mean-field model of the V1 cortex, developed earlier by the groups of F. Chavane and A. Destexhe \cite{zerlaut-chemla-etal:18}\cite{chemla-reynaud-etal:19}, able to reproduce V1 anticipation as observed in VSDI imaging. The aim of this work is to understand, computationally, the effect of retinal anticipation on the cortical one, and more generally the combined effects of motion extrapolation in the retina and V1. This  is the object of a forthcoming paper  (S. Souihel, M. di Volo, S. Chemla, A. Destexhe, F. Chavane and B. Cessac., in preparation). See \cite{souihel:19} for preliminary results.

\paragraph{Retinal circuits.} BCells, ACells and GCells are organized into multiple, local, functional circuits with specific connectivity patterns and dynamics in response to stimuli. Each circuit is related to a specific task, such as light intensity or contrast adaptation, motion detection, orientation, motion direction and so on. Here, we have considered a circuit allowing the retina to detect a moving object on a moving background, where motion sensitive retinal
cells remain silent under global motion of the visual
scene, but fire when the image patch in their receptive
field moves differently from the background. From our study we have emitted the hypothesis that this circuit, spread over the retina, could improve motion anticipation thanks to what we have called the "push-pull" effect. Yet, other circuits could be studied in their role to process motion and anticipation. We especially think of the ON-OFF cone and rod-cone pathways responsible for the separation of highlights and shadows, allowing to provide information to the
GCells concerning brighter than
background stimuli (ON-center) or darker than
background stimuli (OFF-center) \cite{nelson-kolb:04}. This circuit involves both gap junctions and ACells (AII) connectivity and our model could allow to study its dynamics in the presence of a moving object.

\paragraph{Adaptation effects.} In a paper from 2005, Hosoya  et al \cite{hosoya-baccus-etal:05} have studied dynamic predictive coding in the retina and shown how  spatio-temporal receptive fields of retinal GCells change after a few seconds in a new environment, allowing the retina to adjust its processing dynamically when encountering changes in its visual environment. They have shown that an Amacrine network model with plastic synapses can account for the large variety of observed
adaptations. They feature a linear network model of ACells, similar to ours, with, in addition, anti-Hebbian plasticity. Their mathematical analysis, based on linear algebra, allows to determine the behaviour of the model in terms of eigenvalues and eigenvectors.
However, their analysis does not carry out to the gain control introduced by Berry et al, which, as we show, renders quite more complex the spectral analysis. It would therefore be interesting to explore how plasticity in the ACells synaptic network, conjugated with local gain control, contributes to anticipation.

\paragraph{Correlations.} 
The trajectory of a moving object - which is, in general, quite more complex than a moving bar with constant speed - involves long-range correlations in space and in time. Local information about this motion is encoded by retinal GCells. Decoders based on the firing rates of these cells can extract some of the motion features \cite{palmer-marre-etal:15,salisbury-palmer:16,deny-ferrari-etal:17,sederberg-maclean-etal:18,srinivasan-laughlin-etal:82,hosoya-baccus-etal:05,kastner-baccus:13}. Yet lateral connectivity  plays a central role in motion processing (see e.g.  \cite{gollisch-meister:10}). One may expect it to induce spatial and temporal correlations in spiking activity, as an echo, a trace, of the object's  trajectory. These correlations cannot be read in the variations of firing rate; they also cannot be read in synchronous pairwise correlations as the propagation of information due to lateral connectivity necessarily involves \textit{delays}. This example raises the question about what  information can be extracted from spatio-temporal correlations in a  network of connected neurons submitted to a transient stimulus. What is the effect of the stimulus on these correlations? How can one handle this information from data where one has to measure \textit{transient} correlations ?
This question has been addressed in \cite{cessac-ampuero-etal:20}. The potential impact of these spatio-temporal correlations
on decoding and anticipate a trajectory will be the object of further studies.



\paragraph{Orientation selective cells.} Our model affords the possibility to consider BCells with orientation sensitive RF. The potential role of such BCells for predictive coding has been outlined by Johnston et al \cite{johnston-seibel-etal:19}. In their model individual Gcells receive excitatory BCell inputs tuned to different orientations, generating a dynamic predictive code, while feed-forward inhibition generates a high-pass filter that only transmits the initial activation of these inputs, removing redundancy. Should such circuits play a role in motion anticipation ? We didn't elaborate on this in the present paper, leaving it to a potential forthcoming work. Another, important question, is "how to model a retinal network with cells having different orientation selectivity ?" A V1 cortical model has been proposed by Baspinar et al. \cite{baspinar-citti-etal:18} for the generation of orientation preference maps, considering both orientation and scale features. Each point (cortical column) is characterized by intrinsic variables, orientation and scale and the corresponding RF is a rotated Gabor function. The visual stimulus is lifted in a 4-dimensional space, characterized by coordinate variables, position, orientation and scale. The authors infer, from the V1 connectivity a "natural" geometry from which they can apply methods from differential geometry. This type of mathematical construction could be interesting to investigate in the case of the retina with families of orientation selective cells, although the retinal connectivity between these cells is not the same as the V1 orientation preference map structure \cite{bosking-zhang-etal:97,bosking-crowley-etal:02,tversky-miikkulainen:02,xu-bosking-etal:04}.

\paragraph{Biologically inspired vision systems} 
When the retina receives a visual stimulus, it determines which component of it is significant and needs to be further transmitted to the brain. This efficient coding heuristic has inspired many recent studies in developing biologically inspired systems, both for static image and motion representations \cite{oliveira-roque:02,wei-zuo:15,xu-park-etal:19}. Two major applications of biologically inspired vision systems are retinal prostheses \cite{morillas-romero-etal:07,parikh-itti-etal:10,tran:15} and navigational robotics \cite{delbruck-martin-etal:16, lehnert-escobar-etal:19}. Focusing on the second field of application, the ability of a mobile device to navigate in its environment is of utmost interest, especially in order to avoid dangerous situations such as collisions. To be able to move, the robot requires a mapped representation of its environment, but also the ability to interpret and process this representation. Motion processing mechanisms such as anticipation can be thus implemented to assess the efficiency of bio-inspired vision in obstacle avoidance.

\section{Declarations}

\subsection{Ethical Approval and Consent to participate}

Not applicable.

\subsection{Consent for publication}

The Authors transfer to Springer the non-exclusive publication rights and they warrant that their contribution is original and that they have full power to make this grant.

\subsection{Availability of supporting data}

The model source code is available upon request. 

\subsection{Competing interests}

The Authors state that they do not have any conflicts of interest to disclose.

\subsection{Funding}

This work was supported by the National Research Agency (ANR),
in the project "Trajectory",\\ 
https://anr.fr/Project-ANR-15-CE37-0011, funding Selma Souihel's PhD, and by the interdisciplinary Institute for Modelling in Neuroscience and Cognition (NeuroMod http://univ-cotedazur.fr/en/idex/projet-structurant/neuromod ) of the Université Côte d’Azur.

\subsection{Authors' contributions}

Both authors contributed to the final version of the manuscript. Bruno Cessac supervised the project.
 
\subsection{Acknowledgment}

We warmly acknowledge Olivier Marre and Frédéric Chavane for their insightful comments, as well as Michael Berry, Matthias Hennig, Benoit Miramond, Stephanie Palmer and Laurent Perrinet for their thorough feedback as Jury members of Selma Souihel's PhD.  

\bibliographystyle{abbrv}
\bibliography{odyssee}

\newpage

\begin{appendices}

\Sec{Parameters of the model}{parameters}

\begin{center}
\begin{tabular}{|l|c|c|r|}
    \hline
    \textbf{Function} & \textbf{Parameter} & \textbf{Value} & \textbf{Unit} \tabularnewline
    \hline
    $\bm{\KS{B}{}}$ (eq. \eqref{eq:KS}) &  &  &  \tabularnewline
    \hline
    & $\sigma_{1}$ \mbox{(center)} & 90 & $\mu m$ \tabularnewline
    \hline
    & $\sigma_{2}$  \mbox{(surround)} & 290 & $\mu m$ \tabularnewline
    \hline
    & $A_{1}$  \mbox{(center)} & 1.2 & $mV$ \tabularnewline
    \hline
    & $A_{2}$  \mbox{(surround)} & 0.2 & $mV$ \tabularnewline
    \hline
    $\bm{\cK_T(t)}$ (eq. \eqref{eq:KT}) &  &  &  \tabularnewline
    \hline
    & $\mu_{1}$ & 60 & $ms$ \tabularnewline
    \hline
    & $\mu_{2}$ & 180 & $ms$ \tabularnewline
    \hline
    & $\sigma_{1}$ & 20 & $ms$ \tabularnewline
    \hline
    & $\sigma_{2}$ & 44 & $ms$ \tabularnewline
    \hline
    & $K_1$  & 0.22 & unitless \tabularnewline
    \hline
    & $K_2$  & 0.1 & unitless \tabularnewline
    \hline
    \textbf{BCells dynamics} &  &  & \tabularnewline
    \hline
    & $\tau_a$ & 100 & $ms$ \tabularnewline
    \hline
    & $h_B$ & $6.11 e^{-3}$ & $mV^{-1}.ms^{-1}$ \tabularnewline
    \hline
    & $\theta_B$ & 5.32  & $mV$ \tabularnewline
    \hline
    & $\tau_B$ & 200 & $ms$ \tabularnewline
    \hline
    \textbf{ACells dynamics} &  &  & \tabularnewline
    \hline
    & $\tau_A$ & 200 & $ms$  \tabularnewline
    \hline
    & $w$ & [0, 1] & $ms^{-1}$  \tabularnewline
    \hline
    \textbf{ACells connectivity} &  &  &  \tabularnewline
    \hline
    & $\xi$ & $\left\{1,2,3,4\right\}$ & $mm$  \tabularnewline
    \hline
    & $\bar{n}$ & $\left\{1,2,3,4\right\}$ & unitless  \tabularnewline
    \hline
    & $\sigma_n$ & 1 & unitless  \tabularnewline
    \hline
    \textbf{GCells dynamics} &  &  & \tabularnewline
    \hline
    & $a_{p}$ & $0.5$ &  unitless \tabularnewline
    \hline
     & $\sigma_{p}$ & 90 & $\mu m$  \tabularnewline
    \hline
    & $\tau_G$ & 189.5 & $ms$ \tabularnewline
    \hline
    & $h_G$ & $3.59 e^{-4} $ & unitless \tabularnewline
    \hline
    & $\alpha_G$ & $1110 $ & $Hz/mV$ \tabularnewline
    \hline
    & $\theta_G$ & $0$ & $mV$ \tabularnewline
    \hline
    & $N_{G}^{max}$ & 212 & $Hz$  \tabularnewline
    \hline
\end{tabular}
\label{Tab:params} 
\end{center}

\Sec{Spatio-temporal filtering}{Spatio-temp_filtering}

\sSec{Receptive Fields}{Receptive_Fields}

The spatial kernel of the Bcell $i$ is modelled with a difference of Gaussians (DOG): 
\begin{equation}\label{eq:KS}
 \KS{B}{i}(x,y) = \frac{A_1}{2 \pi \sqrt{\det C_1}}\, e^{-\frac{1}{2} \, \tilde{X_i}.C_1^{-1}.X_i} \,-\,  \frac{A_2}{2 \pi \sqrt{\det C_2}}\, e^{-\frac{1}{2} \, \tilde{X_i}.C_2^{-1}.X_i},
\end{equation}
where $X_i=\vect{x-x_i\\y-y_i}$, $\, \widetilde{} \,$ denotes the transpose, $x_i$ and $y_i$ are the coordinates of the receptive field center which coincide with the coordinates of the cell, $C_1, C_2$ are positive definite matrix whose main principal axis represent the preferred orientation. For  circular DOGs(no preferred orientation) $C_1 \equiv \sigma_1^2 \, \cI, C_2 \equiv \sigma_2^2 \, \cI$ where $\cI$ is the identity matrix in 2 dimensions.
The two Gaussians of the DOG are thus concentric. They have the same principal axes. $X_i$ has the physical dimension of a length ($mm$) thus the entries of $C_a, a=1,2$ are expressed in $mm^{2}$. $A_a, a=1 \dots 2$ have the dimension of $mV$ so that the convolution \eqref{eq:Vdrive} has the dimension of a voltage. 

We model the temporal part of the RF with a difference of non concentric Gaussians whose integral on the time domain is zero. This kernel well fits the shape of the temporal projection of the bipolar RF observed in experiments \cite{souihel:19}. 

\begin{equation}\label{eq:KT}
\cK_T(t) = \pare{
\frac{K_1}{\sqrt {2\pi} \sigma_1 } \, 
e^{ - \frac{\pare{t- \mu_1 }^2 }{2\sigma_1 ^2 }} \, -  \, \frac{K_2}{\sqrt {2\pi} \sigma_2 } \, 
e^{ - \frac{\pare{t- \mu_2 }^2 }{2\sigma_2 ^2 }}} \, H(t)
\end{equation}
where $H(t)$ is the Heaviside function. The parameters $\mu_b,\sigma_b$, $b=1,2$ have the dimension of a time ($s$) whereas $K_b$ are dimensionless. The following condition must hold to ensure the continuity of $\cK_T(t)$ at zero: 
\begin{equation}\label{eq:KTzero}
\frac{K_1}{\sigma_1}e^{-\frac{\mu_1^2}{2\sigma_1^2}} = \frac{K_2}{\sigma_2}e^{-\frac{\mu_2^2}{2\sigma_2^2}}.
\end{equation}
Thus, $\K{B}{i}(x,y,0)=0$. In addition, we require that the integral of a constant stimulus converges to zero, so that the cell is only reactive to changes. This reads:
\begin{equation}\label{eq:normalizedintegral}
K_1 \, \Pi\pare{\frac{\mu_1}{\sigma_1}}=K_2 \, \Pi\pare{\frac{\mu_2}{\sigma_2}},
\end{equation}
where:
\begin{equation}\label{eq:Pi}
\Pi(x)=\frac{1}{\sqrt{2 \pi}} \int_{- \infty}^x e^{-\frac{y^2}{2}} \, dy,
\end{equation}
is the cumulative distribution function of the standard Gaussian probability. 


\sSec{Numerical convolution}{Numerical_Convolution}

Here we describe the method use to numerically integrate the  convolution \eqref{eq:Vdrive} of the receptive field $\KS{B}{i}$ \eqref{eq:KS} in the spatial domain with a stimulus $\cS$. 
For the sake of clarity we restrict the computation to one Gaussian in the DOG. The extension to a difference of Gaussian is straightforward. In the following, we consider a spatially discretized stimulus.
When dealing with a 2D stimulus, we have to integrate over two axis. In the case where the eigenvectors of the 2D of Gaussians are the axis of integration, the spatial filter is separable in the stimulus coordinate system. Considering the stimulus as a grid of pixels, we can integrate using the following discretization : let $L_x$ be the size of the stimulus along the x axis in pixels, $L_y$ its size along the y axis, and $\delta$  the pixel length.
We set $S_{ij}(t) \equiv S(i\delta, j \delta, u)$, with $i=0, \dots, \frac{L_x}{\delta}$ and $j=0, \dots, \frac{L_y}{\delta}$. The spatial convolution becomes then :

\begin{align*}
\bra{\K{B}{i} \conv{x,y} \cS}(t) &= \frac{1}{2\pi \sigma_x \sigma_y} \, \int\int_{\mathbb{R}^2}S(x,y,t) e^{- \frac{(x- x_0)^2}{2\sigma_x^2} - \frac{(y- y_0)^2}{2\sigma_y^2}}dx dy\\
&=\sum_{i,j}S_{ij}(t) \bra{erf(\frac{i+\delta-x_0}{\sqrt{2}\sigma_x})-erf(\frac{i-x_0}{\sqrt{2}\sigma_x})}
\bra{erf(\frac{j+\delta-y_0}{\sqrt{2}\sigma_y})-erf(\frac{j-y_0}{\sqrt{2}\sigma_y})}
\end{align*}

In the case where the eigenvectors of the 2D of Gaussians are not the axes of integration, the spatial filter is not separable in the stimulus coordinates system. 
There exists methods that perform the computation by making a linear combination of basis filters \cite{freeman-adelson:91}, others that use Fourier based deconvolution techniques \cite{unser:94}, and others using recursive filtering techniques \cite{deriche:87}. However, these methods are of high computational complexity. We choose instead to use a computer vision method from Geusenroek et al. \cite{geusebroek-smeulders-etal:03}.

\begin{figure}[h]
\begin{center}
\includegraphics[width=0.5\textwidth]{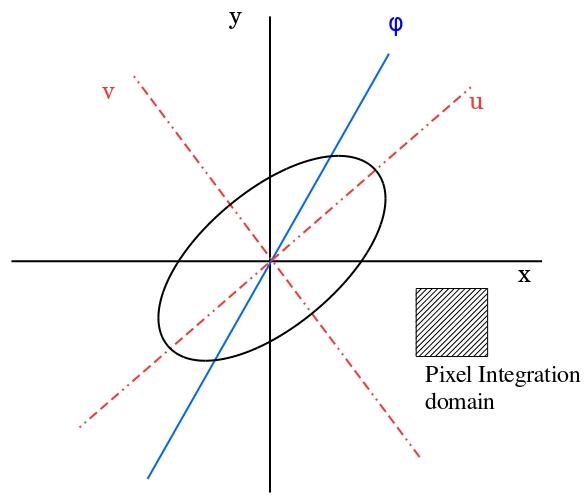}
\caption{\textbf{Filter transformation description} \cite{geusebroek-smeulders-etal:03}. The original system of axes is represented by x and y, and the ellipse system of axes by u and v. $\phi$ represents the angle of the second axis of the non-orthogonal basis. 
The integration domain of a pixel is limited by four lines of equations : $x = i\delta$, $x = (i+1)\delta$, $y = j\delta$ and $y = (j+1)\delta$. Rewriting these fours equations in the new system of axes through a coordinate change enables us to write the equation \eqref{eq:Anisotropic_convolution}
\label{Fig:CoordinateChange}. }
\end{center}
\end{figure}

It is based on a projection in a non-orthogonal basis, where the first axis is $x$ and the second is parametrized by a angle $\phi$ (see Fig. \ref{Fig:CoordinateChange}).
The new standard deviations read : 
$$
\sigma_{x'}=\frac{\sigma_{x}\sigma_{y}}{\sqrt{\sigma_{x}^2\cos{\theta}^2+\sigma_{y}^2\sin{\theta}^2}}
$$
$$
\sigma_{\phi}=\frac{\sqrt{\sigma_{y}^2\cos{\theta}^2+\sigma_{x}^2\sin{\theta}^2}}{\sin{\phi}}
$$
with 
$$
\tan(\phi)=\frac{\sigma_{y}^2\cos{\theta}^2+\sigma_{x}^2\sin{\theta}^2}{(\sigma_{x}^2-\sigma_{y}^2)\cos{\theta}\sin{\theta}}
$$
and $\sigma_{x} \neq \sigma_{y}$ (in the orientation sensitive case).

We adapt the implementation to the spatially discretized stimulus, using an integration scheme similar to the one introduced in the separable case. The spatial convolution -reads now: 
\begin{equation}\label{eq:Anisotropic_convolution}
\begin{array}{lll}
&\bra{\K{B}{i} \conv{x,y} \cS}(x_0,y_0,t) =\\
& \sigma_{x'}\sqrt[]{\frac{\pi}{2}}\sum_{(i;j)\in[0,s_x]\times[0,s_y]}\int_{y\frac{\delta}{\sin(\phi)}}^{(y+1)\frac{\delta}{\sin(\phi)}}C_{ij}e^{ \frac{(y'- y_0)^2}{2\sigma_{\phi}^2}} 
[erf(\frac{(-\cos(\phi)y'+x+1)\delta-x_0}{\sqrt{2}\sigma_{x'}})-erf(\frac{(-\cos(\phi)y'+x)\delta-x_0}{\sqrt{2}\sigma_{x'}})] dy'
\end{array}
\end{equation}
The integral is then computed numerically. The advantage of this formulation is to replace a two dimension integration by 
a one dimensional.

\Sec{Random connectivity}{Random_Connectivity}

Here we define the random connectivity matrix from ACell to BCells considered in section \ref{Sec:Connectivity_graph}. 
 Each cell (ACell and BCell) has a random number of branches (dendritic tree), each of which has a random length and a random angle with respect to the horizontal axis. The length of branches $L$ follow an exponential distribution:
\begin{equation}\label{eq:fL}
f_L(l)=\frac{1}{\xi}e^{-\frac{l}{\xi}}, \quad l \geq 0.
\end{equation}
  with  spatial scale $\xi$.  
The number of branches $n$ is also a random variable, Gaussian with mean $\bar{n}$ and variance $\sigma_n$. 
The angle distribution is taken to be isotropic in the plane, i.e. uniform on $[0,2 \pi[$. 
When a branch of an ACell A intersects a branch of a BCell B there is a chemical synapse from A to B.

Here, we assume that both cells types have the same probability distributions for branches, thus neglecting the actual shape of ACell and BCell dendritic trees. On biological grounds, this assumption is relevant if we consider the shape of BCell dendritic tree \textit{in the Inner Plexiform Layer (IPL)} (see e.g \url{https://webvision.med.utah.edu/book/part-iii-retinal-circuits/roles-of-ACell-cells/} Fig. 5, 16, 17). While out of the IPL BCells have the form of a dipole, in the IPL their dendrites have a form well approximated by our $2$-dimensional model. A potential refinement would consist of considering different set of parameters in the probability laws respectively defining BCell and ACell dendritic tree. 

\begin{figure}
\centering
\includegraphics[height=4cm,width=6cm]{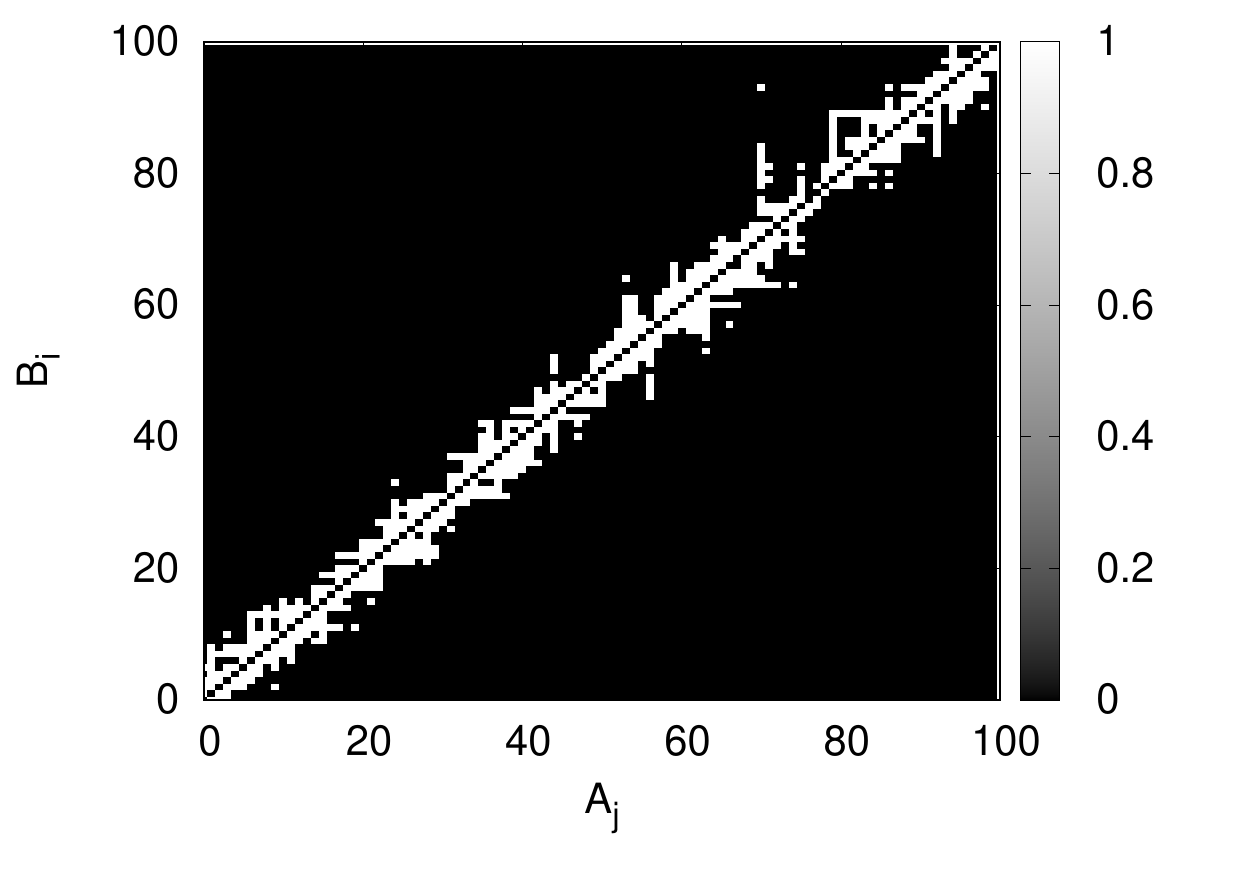}
\hspace{1cm}
\includegraphics[width=6cm,height=4cm]{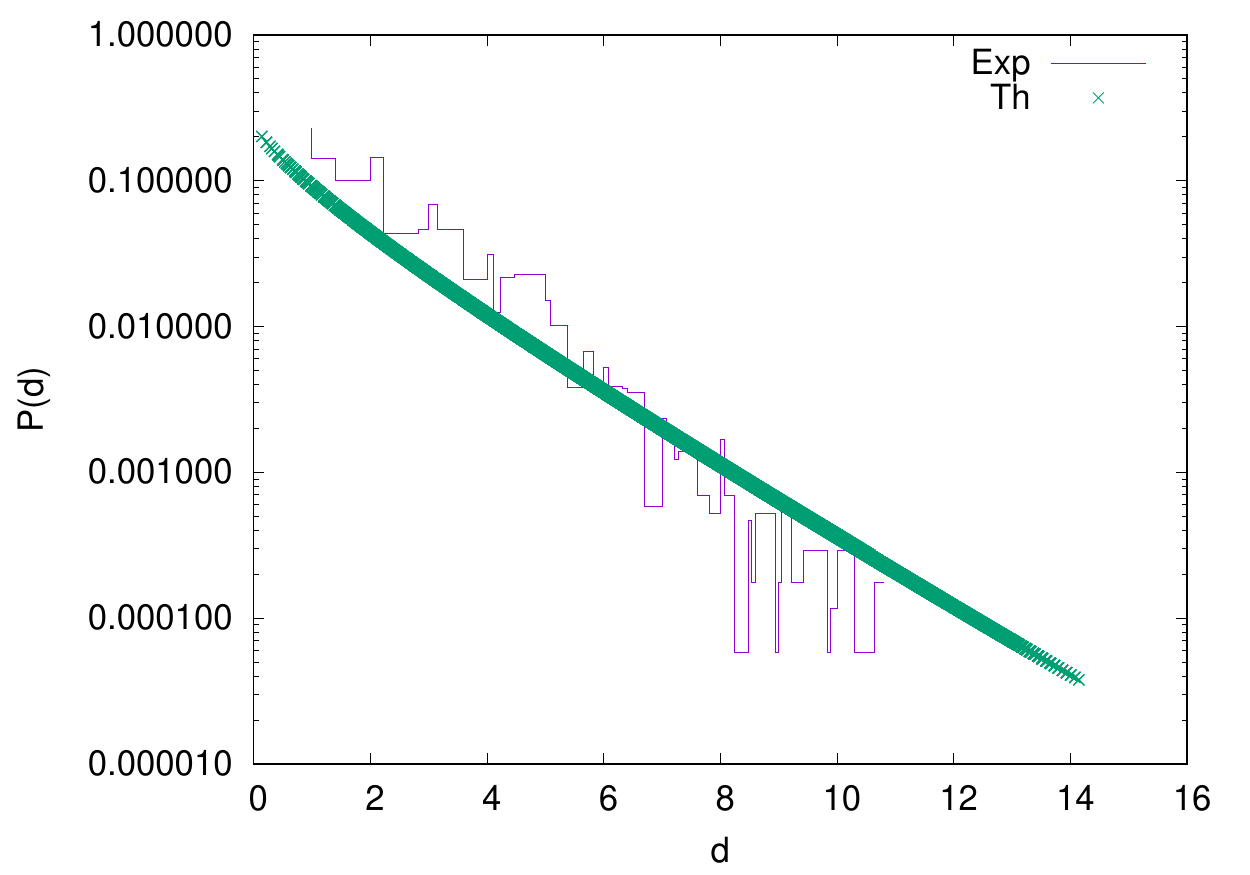}
\caption{\textbf{Random connectivity. Left.} Example of a random connectivity matrix from ACells $\Cell{A}{j}$ to BCells cells $\Cell{B}{i}$. White points correspond to connection from $\Cell{A}{j}$ to $\Cell{B}{i}$. \textbf{Right.} Probability $P(d)$ that two branches intersect as a function of the distance between two cells. 'Exp' corresponds to numerical estimation and 'Th' corresponds to the theoretical prediction. Here, $\xi=2$.}
\label{Fig:PlotProbaDistance}
\end{figure}

We show, in Fig. \ref{Fig:PlotProbaDistance} an example of connectivity matrix produced this way, as well as the probability that two branches intersect as a function of the distance of the two cells.

We now compute this probability. 
We use the standard notation in probability theory where the random variable is written in capitals and its realization in small letter. Thus,  $F_X(x) =\Prob{X < x}$ is the cumulative distribution function of the random variable $X$ and $f_X(x)=\frac{d F_X}{dx}$ its density.

We consider the oriented connection between the cell $A$ (ACell), of coordinates $(x_A,y_A)$ to a cell $B$ (BCell), of coordinates $(x_B,y_B)$, so that the distance between the two cells 
is $d_{AB}=\sqrt{\pare{x_B-x_A}^2 + \pare{y_B-y_A}^2}$. 

The vector $\vec{AB}$ makes an oriented angle $\eta=\ang{AB}{Ax}$ with the positive horizontal axis, where:
\begin{equation}\label{eq:eta}
\eta = 
\left\{
\begin{array}{lll}
     &  \arctan \frac{y_B-y_A}{x_B-x_A}, \quad &\mbox{if } x_B > x_A\\
     &  \pi+\arctan \frac{y_B-y_A}{x_B-x_A}, \quad &\mbox{if } x_B < x_A.
\end{array}
\right.
\end{equation}
Here, we neglect the effects of boundaries - taking, e.g., an infinite lattice, or periodic boundary conditions - so that the probability to connect $A$ to $B$ is invariant by rotation.  Thus, we compute this probability in the first quadrant $x_B > x_A$, $y_B > y_A$. In this case $\eta=\arctan \frac{y_B-y_A}{x_B-x_A}$.

\begin{figure}
\begin{center}
\includegraphics[width=0.5\textwidth]{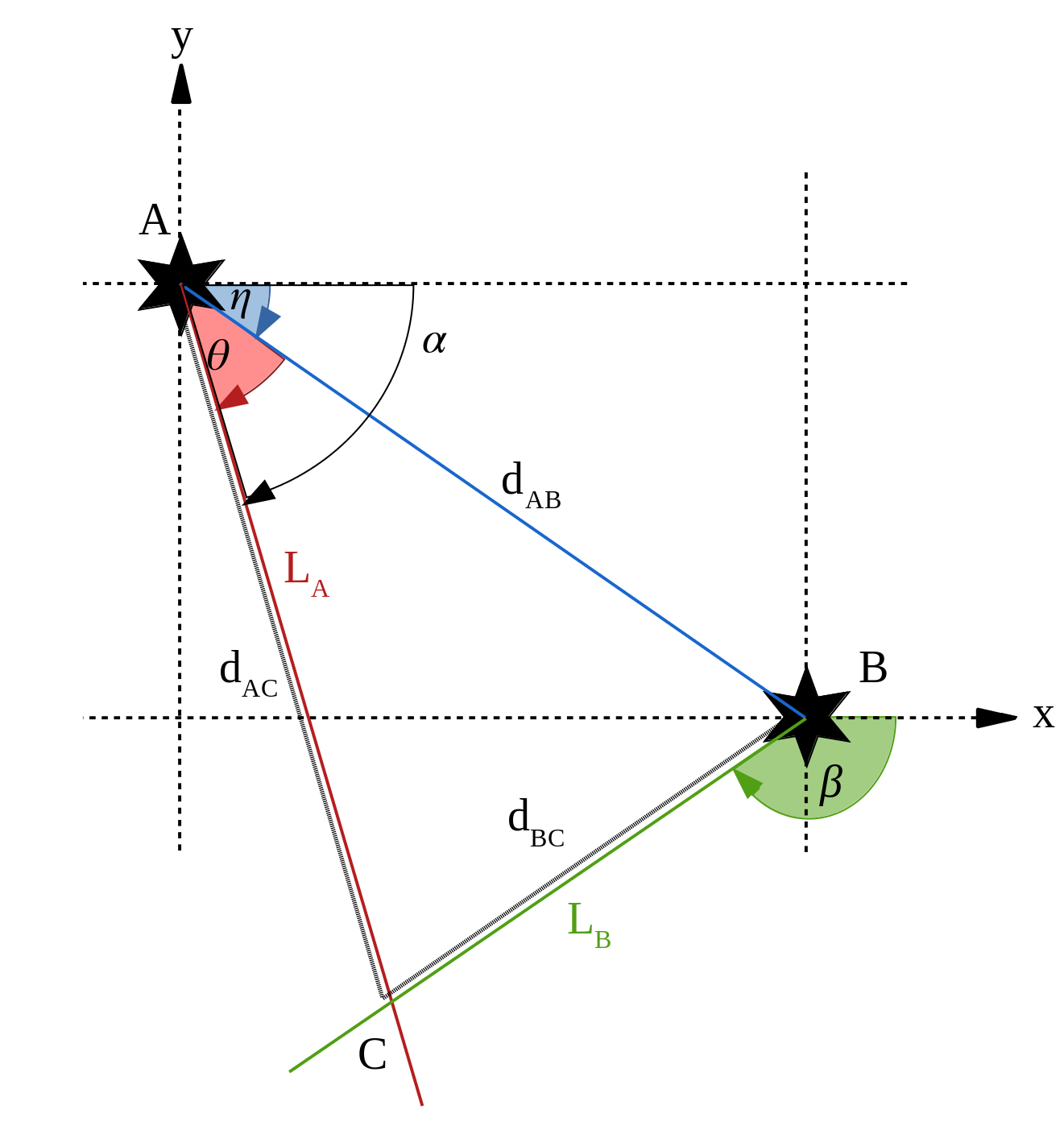}
\caption{\textbf{Geometry  of connection between $2$ neurons.} $\alpha$ ($\beta$) is the angle of the neuron $A$'s branch with length $L_A$ (neuron $B$'s branch with length $L_B$) with respect to the horizontal axis. $\theta$ is the angle between the segment connecting $AB$ and the branch $A$. C represents the virtual point that lies at the intersection of the branches of length $L_A$ and $L_B$. $d_{AB}$ (resp. $d_{AC}$, $d_{BC}$) denotes the distance between A and B (resp. A, C and B,C). Note that  $d_{AC} \leq L_A$, $d_{BC} \leq L_B$.
\label{Fig:Branches}}
\end{center}
\end{figure}

Each cell has a random number of branches (dendritic tree), each of which has a random length and a random angle with respect to the horizontal axis (Fig. \ref{Fig:Branches}). 
The length of branches $L$ follow the exponential distribution \eqref{eq:fL}:
$
f_L(l)=\frac{1}{\xi}e^{-\frac{l}{\xi}}, \quad l \geq 0,
$
with repartition function:
\begin{equation}\label{eq:FL}
F_L(l)=1-e^{-\frac{l}{\xi}}. 
\end{equation}
The spatial scale $\xi$ favours short range connections. The number of branches N distribution follows an normal distribution with mean $\bar{n}$ and variance $\sigma_n$. The angle distribution is taken to be isotropic in the plane, i.e. uniform on $[0,2 \pi[$.\\

We compute  the probability that a branch of ACell $A$, of length $L_A$, intersects, at point $C$, a branch of BCell $B$, of length $L_B$. We note $\alpha$, the oriented angle $\ang{Ax}{AC}$; $\beta$, the oriented angle $\ang{Bx}{BC}$; $\theta$, the oriented angle $\ang{AB}{AC}$. In the first quadrant, $\alpha=\theta+\eta$. Note that the condition to be in the first quadrant constraints $\eta$
but not $\alpha$.

From the sin rule we have:
$
\frac{\sin\pare{\beta-\alpha+\theta}}{d_{AC}}=\frac{\sin \theta}{d_{BC}}=\frac{\sin\pare{\beta-\alpha}}{d_{AB}}$. This holds however if $A,B,C$ is a triangle, that is, if the two branches are long enough to intersect at $C$ which reads: $0 \leq d_{AC} = \frac{\sin\pare{\beta-\alpha+\theta}}{\sin\pare{\beta-\alpha}} \, d_{AB} \leq L_A$ and $0 \leq d_{BC} =\frac{\sin\theta}{\sin\pare{\beta-\alpha}} \, d_{AB}   \leq L_B$. These are necessary and sufficient conditions for the branches to intersect. 

Note that the positivity of these quantities imposes conditions linking the angles $\alpha,\beta,\eta$ with $\theta=\alpha-\eta$. 
\begin{enumerate}
\item If $\sin\pare{\beta-\alpha} > 0$
 $\Leftrightarrow 0 < \beta - \alpha < \pi$ 
$\Leftrightarrow \alpha < \beta  < \pi+\alpha$
we must have $\sin \theta > 0$  $\Leftrightarrow 0 < \theta=\alpha - \eta < \pi$ so that 
$\eta < \alpha < \pi+\eta$,
and, $\sin\pare{\beta-\alpha+\theta}>0$ $\Leftrightarrow 0 < \beta-\alpha+\theta=\beta - \eta < \pi$ so that 
$\eta < \beta < \pi+\eta$ (because $\eta \geq 0$). 
All these constraints are satisfied if $ \eta < \alpha < \beta < \min\pare{\pi+\alpha, \pi+\eta}=\pi + \eta$.
\item If $\sin\pare{\beta-\alpha} < 0$
 $\Leftrightarrow - \pi < \beta - \alpha < 0$ 
$\Leftrightarrow -\pi+\alpha < \beta  < \alpha$
we must have $\sin \theta < 0$  $\Leftrightarrow -\pi < \theta=\alpha - \eta < 0$ so that 
$-\pi + \eta < \alpha < \eta$,
and, $\sin\pare{\beta-\alpha+\theta}<0$ $\Leftrightarrow -\pi < \beta-\alpha+\theta=\beta - \eta <0$ so that 
$-\pi+\eta < \beta <  \eta$. 
All these constraints are satisfied if $ \max\pare{-\pi+\alpha,-\pi+\eta} = -\pi+\eta < \beta < \alpha < \eta $. 
\end{enumerate}

Modulo these conditions,
the conditional probability $\rho_{c|\pare{\alpha,\beta}}$ to have intersection given the angles $\alpha,\beta$ is:
$$
\rho_{c|\pare{\alpha,\beta}}=\Probc{Connection}{\alpha,\beta}
$$
$$= \Probc{0 \leq \frac{\sin\pare{\beta-\alpha+\theta}}{\sin\pare{\beta-\alpha}} \, d_{AB} \leq L_A,0 \leq \frac{\sin\theta}{\sin\pare{\beta-\alpha}} \, d_{AB}   \leq L_B}{\alpha,\beta}.
$$
Using the cumulative distribution function \eqref{eq:FL} of the exponential distribution, and the independence of $L_A,L_B$ this gives:
$$
\rho_{c|\pare{\alpha,\beta}}=\pare{1-F_L\pare{\frac{\sin\pare{\beta-\alpha+\theta}}{\sin\pare{\beta-\alpha}} \, d_{AB}}}\,\pare{1-F_L\pare{\frac{\sin\theta}{\sin\pare{\beta-\alpha}} \, d_{AB}}}
=e^{-\frac{d_{AB}}{\xi} \, \frac{\sin\pare{\frac{\alpha+\beta}{2} - \eta}}{ \sin\pare{\frac{\beta-\alpha}{2}}}}
$$

The probability to connect the two branches in the first quadrant is then:
\begin{equation}\label{eq:rhoc}
\rho_c(d_{AB},\eta) = \frac{1}{4 \, \pi^2} 
\pare{
\int_{\alpha=\eta}^{\pi+\eta} \,   \int_{\beta=\alpha}^{\pi+\eta} \,
e^{-\frac{d_{AB}}{\xi} \, \frac{\sin\pare{\frac{\alpha+\beta}{2} - \eta}}{ \sin\pare{\frac{\beta-\alpha}{2}}}}\, d\alpha \, d\beta
+
\int_{\alpha=-\pi+\eta}^{\eta} \,   \int_{\beta=-\pi+\eta}^{\alpha} \, 
e^{-\frac{d_{AB}}{\xi} \, \frac{\sin\pare{\frac{\alpha+\beta}{2} - \eta}}{ \sin\pare{\frac{\beta-\alpha}{2}}}}\, d\alpha \, d\beta,
}
\end{equation}
%
%
which depends on the distance between the two cells and their angle $\eta$, depending parametrically on the characteristic length $\xi$. Note that the condition of positivity of the sine ratio ensures an exponential decay of the probability as $d_{AB}$ increases.\\

\textbf{Remark.} The positivity of arguments in the exponential implies that:
$$4 \, \pi^2 \, \rho_c(d_{AB},\eta) \leq \int_{\alpha=\eta}^{\pi+\eta} \,   \int_{\beta=\alpha}^{\pi+\eta} \, d\alpha \, d\beta
+
\int_{\alpha=-\pi+\eta}^{\eta} \,   \int_{\beta=-\pi+\eta}^{\alpha} \, d\alpha \, d\beta
=\pi^2,
$$
so that $\rho_c(d_{AB},\eta) \leq \frac{1}{4}$.\\

\Sec{Linear analysis}{Linear_Analysis}

\sSec{General solution of the linear dynamical system}{General_Solutions}

Here we consider the dynamical system \eqref{eq:Diff_Syst_Vect_Lin},
$\frac{d \vcX}{dt} = \cL.\vcX  + \vcF(t)$,
whose general solution is:
$$
\vcX(t)=\int_{t_0}^t e^{\cL(t-s)}.\vcF(s) \, ds,
$$

The behaviour of this integral depends on the spectrum of $\cL$. 
The difficulty is that $\cL$ is \textit{not} diagonalisable (because of the activity term $h_B \, I_{N,N}$).
We write it in the form:
$$
\cL=
\underbrace{\pare{\begin{array}{cccccc}
&\overbrace{\pare{\begin{array}{ccccc}
&-\frac{I_{N,N}}{\tau_B} & &\W{A}{}{B}{} \\
&\W{B}{}{A}{} & & -\frac{I_{N,N}}{\tau_A} 
\end{array}
}}^{\cM}
&& 0_{N,N}  && 0_{N,N}\\
& 0_{N,N}  && 0_{N,N} &&-\frac{I_{N,N}}{\tau_a} 
\end{array}
}
}_{\cD}
+
\underbrace{\pare{\begin{array}{cccccc}
&0_{N,N}& &0_{N,N}&& 0_{N,N}\\
&0_{N,N} & & 0_{N,N} && 0_{N,N}\\
& h_B \, I_{N,N}  && 0_{N,N} &&0_{N,N} 
\end{array}
}
}_{\cJ}
$$

We assume that the matrix $\cM$ is diagonalisable. Even in this case, $\cL$ is not diagonalisable because of the Jordan matrix $\cJ$. 
We note $\vphi_\beta$, the normalized eigenvectors of $\cM$ and $\lambda_\beta$ the corresponding eigenvalues, with $\beta=1 \dots 2N$. 
The eigenvalues of $\cD$ are then the $2N$ eigenvalues of $\cM$
plus $N$ eigenvalues $-\frac{1}{\tau_a}$. We note them $\lambda_\beta$ too, with $\lambda_\beta=-\frac{1}{\tau_a}$, $\beta=2N+1 \dots 3N$. The eigenvectors of $\cD$ have the form: 
$$
\vcP_{\beta} = 
\left\{ 
\begin{array}{cccc}
&\vect{\vphi_\beta \\ \vz_N}, &\quad \beta=1 \dots 2N;\\
&\ve_\beta, &\quad \beta=2N+1 \dots 3N,
\end{array}
\right.
$$
where $\vz_N$ is the $N$ dimensional vector with entries $0$ and $\ve_\beta$ is the canonical basis vector in direction $\beta$.
The matrix $\cP$ made by the columns $\vcP_{\beta}$ is the matrix which diagonalizes $\cD$. We note $\Lambda=\cP^{-1}\cD \cP$ the diagonal form where $\Lambda=Diag\Set{\lambda_\beta, \beta=1 \dots 3N}$. $\cP$ and $\cP^{-1}$ have the block form:
\begin{equation}\label{eq:Passage}
\cP=
\pare{\begin{array}{cccccc}
& \Phi && 0_{2N,N}\\
& 0_{N,2N} && I_{N,N}
\end{array}
}; \quad \cP^{-1}=
\pare{\begin{array}{cccccc}
& \Phi^{-1} && 0_{2N,N}\\
& 0_{N,2N} && I_{N,N}
\end{array}
},
\end{equation}
 where $\Phi$ is the matrix whose columns are the eigenvectors $\vphi_\beta$ of $\cM$. This form implies that $\cP_{\alpha\beta}=\cP^{-1}_{\alpha\beta}=\delta_{\alpha\beta}$, for $\beta=2N+1 \dots 3N$. \\
 
 We now compute $e^{\cL \, t}$ using the series expansion $e^{\cL \, t} 
 = \sum_{n=0}^{+\infty} \frac{t^n}{n!} \, \pare{\cD+\cJ}^n$.
 Using the relations:
 %
$$ 
 \cJ^2=0_{3N,3N}; \quad \cD^n.\cJ= \pare{-\frac{1}{\tau_a}}^n \, \cJ,
 $$
 %
 one proves that:
 %
 $$
 \pare{\cD+\cJ}^n = \cD^n + \cJ. \sum_{k=0}^{n-1} \pare{-\frac{1}{\tau_a}}^k \, \cD^{n-1-k}. 
 $$
%

Therefore:
%
$$
e^{\cL \, t} = e^{\cD \, t} + \cJ. \sum_{n=1}^{+\infty} \frac{t^n}{n!} \, \sum_{k=0}^{n-1} \pare{-\frac{1}{\tau_a}}^k \, \cD^{n-1-k}. 
$$

We use the matrices $\cP$, $\cP^{-1}$ to write it in the form:
$$
e^{\cL \, t} = \cP.e^{\Lambda \, t}.\cP^{-1} + \cJ. \sum_{n=1}^{+\infty} \frac{t^n}{n!} \, \sum_{k=0}^{n-1} \pare{-\frac{1}{\tau_a}}^k \, \cP.\Lambda^{n-1-k}.\cP^{-1}.
$$

From relation \eqref{eq:GenSolSDLin} we obtain, for the entries of $\vcX(t)$:
\begin{equation}\label{eq:Xalpha}
\cX_\alpha(t) = \sum_{\beta,\gamma=1}^{3N} \cP_{\alpha\beta} \cP^{-1}_{\beta \gamma} \int_{t_0}^t e^{\lambda_\beta(t-s)} \, \cF_\gamma(s) \, ds + \sum_{\delta,\beta,\gamma=1}^{3N} 
\cJ_{\alpha,\delta} \, \cP_{\delta\beta} \cP^{-1}_{\beta \gamma}
\, \int_{t_0}^t \sum_{n=1}^{+\infty} \frac{(t-s)^n}{n!} \, \sum_{k=0}^{n-1} \pare{-\frac{1}{\tau_a}}^k \lambda_\beta^{n-1-k} \, \cF_\gamma(s) \, ds.
\end{equation}

We consider the first term of this equation. We use $\cF_\gamma=\F{B}{i}$, $\gamma=i=1 \dots N$ (BCells). We recall that, from  \eqref{eq:FBip}, $\F{B}{i}(t)= \frac{V_{i_{drive}}}{\tau_B} \, + \, \frac{d V_{i_{drive}}}{d t}$, so that:
\begin{equation}\label{eq:intFgamma}
\int_{t_0}^t e^{\lambda_\beta(t-s)} \,  \cF_\gamma(s) \, ds=
V_{\gamma_{drive}}(t)+ \pare{\frac{1}{\tau_B} + \lambda_\beta } \, \int_{t_0}^t e^{\lambda_\beta(t-s)} \, V_{\gamma_{drive}}(s)\, ds.
\end{equation}
%
Moreover, 
$$
\sum_{\beta=1}^{3N}\sum_{\gamma=1}^{3N} \cP_{\alpha\beta} \cP^{-1}_{\beta \gamma}V_{\gamma_{drive}}(t)
=\sum_{\gamma=1}^{3N} V_{\gamma_{drive}}(t) \pare{\sum_{\beta=1}^{3N} \cP_{\alpha\beta} \cP^{-1}_{\beta \gamma}} = \sum_{\gamma=1}^{3N} V_{\gamma_{drive}}(t)  \delta_{\alpha\gamma} = V_{\alpha_{drive}}(t),
$$
We extend the definition of the drive term \eqref{eq:Vdrive}  to $3N$-dimensions such that $V_{\alpha_{drive}}(t)=0$ if $\alpha > N$.
Thus:
$$
\sum_{\beta,\gamma=1}^{3N} \cP_{\alpha\beta} \cP^{-1}_{\beta \gamma} \int_{t_0}^t e^{\lambda_\beta(t-s)} \, \cF_\gamma(s) \, ds  = V_{\alpha_{drive}}(t) \, + \,   
\sum_{\beta=1}^{3N} \pare{\frac{1}{\tau_B} \, + \, \lambda_\beta } \, \sum_{\gamma=1}^{N} \cP_{\alpha\beta} \cP^{-1}_{\beta \gamma} \int_{-\infty}^t e^{\lambda_\beta(t-s)} \, V_{\gamma_{drive}}(s)\, ds.
$$
%

 We decompose the sum over $\beta$ in $3$ sums:
$\beta=1 \dots N$ corresponding to BCells; $\beta=N+1 \dots 2N$ corresponding to ACells;
$\beta=2N+1 \dots 3N$ corresponding to activities of BCells.
We define (eq. \eqref{eq:EBB} in the text):
$$
\E{B}{B,\alpha}(t)=\sum_{\beta=1}^{N} \pare{\frac{1}{\tau_B} \, + \, \lambda_\beta } \, \sum_{\gamma=1}^{N} \cP_{\alpha\beta} \cP^{-1}_{\beta \gamma} \int_{t_0}^t e^{\lambda_\beta(t-s)} \, V_{\gamma_{drive}}(s)\, ds, \quad \alpha=1 \dots N,
$$
corresponding to the indirect effect, via the ACells connectivity, of the drive on BCells voltages. The term
$$
\E{B}{A,\alpha}(t)=\sum_{\beta=N+1}^{2N} \pare{\frac{1}{\tau_B} \, + \, \lambda_\beta } \, \sum_{\gamma=1}^{N} \cP_{\alpha\beta} \cP^{-1}_{\beta \gamma} \int_{t_0}^t e^{\lambda_\beta(t-s)} \, V_{\gamma_{drive}}(s)\, ds, \quad \alpha=N+1 \dots 2N,
$$ 
(eq. \eqref{eq:EBA} in the text) corresponds to the effect of BCell drive on ACell voltages. The third term :
$$
\sum_{\beta=2N+1}^{3N} \pare{\frac{1}{\tau_B} \, + \, \lambda_\beta } \, \sum_{\gamma=1}^{N} \cP_{\alpha\beta} \cP^{-1}_{\beta \gamma} \int_{t_0}^t e^{\lambda_\beta(t-s)} \, V_{\gamma_{drive}}(s)\, ds = 0, \quad \alpha=2N+1 \dots 3N.
$$
because $\cP^{-1}_{\beta \gamma}=\delta_{\beta \gamma}$.\\

To compute the second term in eq. \eqref{eq:Xalpha}, we first first remark that $\cJ_{\alpha,\delta}=0$, if $\alpha=1 \dots 2N$, and $\cJ_{\alpha,\delta}=h_B \, \delta_{\alpha-2N,\delta}$, if $\alpha=2N+1 \dots 3N$, so that this term is non zero only if $\alpha=2N+1 \dots 3N$ (BCells activities). Also $\cF_\gamma \neq 0$ for $\gamma=1 \dots N$, while $
 \cP^{-1}_{\beta \gamma} = \delta_{\beta \gamma}$ for $\beta=2N+1 \dots 3N$ . Therefore, for $\alpha=2N+1 \dots 3N$ the second term in $\cX_\alpha(t)$ is:
$$
h_B \, \sum_{\beta=1}^{2N} \sum_{\gamma=1}^{N} 
\cP_{\alpha-2N \beta} \cP^{-1}_{\beta \gamma}
\, \int_{t_0}^t \sum_{n=1}^{+\infty} \frac{(t-s)^n}{n!} \, \sum_{k=0}^{n-1} \pare{-\frac{1}{\tau_a}}^k \lambda_\beta^{n-1-k} \, \cF_\gamma(s) \, ds
$$

We now simplify the series:
$$
\sum_{n=1}^{+\infty} \frac{(t-s)^n}{n!} \, \sum_{k=0}^{n-1} \pare{-\frac{1}{\tau_a}}^k \lambda_\beta^{n-1-k} = \sum_{n=1}^{+\infty}  \frac{(t-s)^n}{n!} \, \lambda_\beta^{n-1} \, \sum_{k=0}^{n-1} \pare{-\frac{1}{\tau_a \, \lambda_\beta}}^k 
$$
$$
=\sum_{n=1}^{+\infty}  \frac{(t-s)^n}{n!} \, \lambda_\beta^{n-1} \, \pare{\frac{1-\pare{-\frac{1}{\tau_a \, \lambda_\beta}}^n}{1+\frac{1}{\tau_a \, \lambda_\beta}} 
} 
$$
$$
=\frac{1}{\lambda_\beta} \, \frac{1}{1+\frac{1}{\tau_a \, \lambda_\beta}} \, \bra{ \sum_{n=1}^{+\infty}  \frac{ \pare{ \lambda_\beta \,(t-s)}^n}{n!}  \, -   \sum_{n=1}^{+\infty}  \frac{(-\frac{t-s}{\tau_a})^n}{n!}}
$$
$$
= \frac{1}{\lambda_\beta+\frac{1}{\tau_a \, }} \,  \bra{e^{\lambda_\beta \,(t-s)} \, - \, e^{-\frac{t-s}{\tau_a}}}
$$

The time integral is computed the same way as eq. \eqref{eq:intFgamma}:
$$
\int_{t_0}^t \sum_{n=1}^{+\infty} \frac{(t-s)^n}{n!} \, \sum_{k=0}^{n-1} \pare{-\frac{1}{\tau_a}}^k \lambda_\beta^{n-1-k} \, \cF_\gamma(s) \, ds= 
\frac{1}{\lambda_\beta+\frac{1}{\tau_a \, }} \,  \bra{\int_{t_0}^t \, e^{\lambda_\beta \,(t-s)}  \, \cF_\gamma(s) \, ds - \int_{t_0}^t \, e^{-\frac{t-s}{\tau_a}} \, \cF_\gamma(s) \, ds}
$$
$$
=\frac{1}{\lambda_\beta+\frac{1}{\tau_a \, }} \,  \bra{ \pare{\frac{1}{\tau_B} + \lambda_\beta } \, \int_{t_0}^t e^{\lambda_\beta(t-s)} \, V_{\gamma_{drive}}(s)\, ds - \pare{\frac{1}{\tau_B} - \frac{1}{\tau_a} } \, \int_{t_0}^t  e^{-\frac{t-s}{\tau_a}} \,  V_{\gamma_{drive}}(s)\, ds}.
$$
%
Similarly to eq.  \eqref{eq:EBB}, \eqref{eq:EBA} in the text we introduce:
$$
\E{B}{a,\alpha}(t)= h_B \, \sum_{\beta=1}^{2N} \sum_{\gamma=1}^{N} 
\cP_{\alpha-2N \beta} \cP^{-1}_{\beta \gamma}
\, \frac{1}{\lambda_\beta+\frac{1}{\tau_a \, }} \,  \bra{
\begin{array}{lll}
 &\pare{\frac{1}{\tau_B} + \lambda_\beta } \, \int_{t_0}^t e^{\lambda_\beta(t-s)} \, V_{\gamma_{drive}}(s)\, ds\\
  -& \pare{\frac{1}{\tau_B} - \frac{1}{\tau_a} } \, \int_{t_0}^t  e^{-\frac{t-s}{\tau_a}} \,  V_{\gamma_{drive}}(s)\, ds.
\end{array}
}, \quad \alpha=2N+1 \dots 3N,
$$
corresponding to the action of BCells and ACells on the activity of BCells, via the network effect. 
Let us consider in more detail the second term. From \eqref{eq:Passage}, $\sum_{\beta=1}^{2N} \cP_{\alpha-2N \beta} \cP^{-1}_{\beta \gamma} = \delta_{\alpha-2N \gamma}$ thus:
$$
\sum_{\beta=1}^{2N} \sum_{\gamma=1}^{N} 
\cP_{\alpha-2N \beta} \cP^{-1}_{\beta \gamma} \, \int_{t_0}^t  e^{-\frac{t-s}{\tau_a}} \,  V_{\gamma_{drive}}(s)\, ds
=  \sum_{\gamma=1}^{N}  \pare{\, \int_{t_0}^t  e^{-\frac{t-s}{\tau_a}} \,  V_{\gamma_{drive}}(s)\, ds \underbrace{\sum_{\beta=1}^{2N} \ 
\cP_{\alpha-2N \beta} \cP^{-1}_{\beta \gamma}}_{\delta_{\alpha-2N \gamma}}}
$$
$$
=  \int_{t_0}^t  e^{-\frac{t-s}{\tau_a}} \,  V_{\alpha-2N_{drive}}(s)\, ds  \equiv A^0_{\alpha-2N}(t),
$$
(eq. \eqref{eq:A0} in the text).

This finally leads to (\eqref{eq:EBa} in the text):
$$
\E{B}{a,\alpha}(t)= h_B \, \pare{\sum_{\beta=1}^{2N} \sum_{\gamma=1}^{N} 
\cP_{\alpha-2N \beta} \cP^{-1}_{\beta \gamma}
\, \frac{\lambda_\beta+\frac{1}{\tau_B}}{\lambda_\beta+\frac{1}{\tau_a}}  \, \int_{t_0}^t e^{\lambda_\beta(t-s)} \, V_{\gamma_{drive}}(s)\, ds
 + \frac{-\frac{1}{\tau_B} + \frac{1}{\tau_a}}{\lambda_\beta+\frac{1}{\tau_a}}\, A^0_{\alpha-2N}(t)}, \quad \alpha=2N+1 \dots 3N,
$$

\sSec{Spectrum of $\cL$ and stability of the dynamical system \eqref{eq:Diff_Syst_Vect_Lin}}{Spectrum}

Here, we assume that a BCell connects only one ACell, with a weight $w^+$ uniform for all BCells, so that $\W{B}{}{A}{} = w^+ \, I_{N,N}$, $ w^+>0$. We also assume that ACell connect to BCell with a connectivity matrix $\cW$, not necessarily, symmetric, with a uniform weight $- w^-$, $ w^->0$, so that $\W{A}{}{B}{} = -w^- \, \cW$. We have shown in the previous section that the $2N$ first eigenvalues and eigenvectors of $\cL$ are given by the $2N$ eigenvalues and eigenvectors of $\cM$
which reads now:
\begin{equation}\label{eq:M}
\cM=
\pare{\begin{array}{cccccc}
&-\frac{I_{N,N}}{\tau_B} & &-w^- \, \cW\\
& w^+ \,I_{N,N} & & -\frac{I_{N,N}}{\tau_A}
\end{array}
}.
\end{equation}
We now show that this specific structure allows compute the spectrum of $\cM$ in terms of the spectrum of $\cW$.

\ssSec{Eigenvalues and eigenvectors of $\cM$}{EigenvaluesM}

We note $\kappa_n, n=1 \dots N$, the eigenvalues of $\cW$ ordered as $\abs{\kappa_1} \leq \abs{\kappa_2} \leq \dots \leq \abs{\kappa_n}$ and $\vpsi_n$ is the corresponding eigenvector. We normalize $\vpsi_n$ so that $\vpsi_n^\dag.\vpsi_n=1$ where $\dag$ is the adjoint. (Note that, as $\cW$ is not symmetric in general, eigenvectors are complex).\\
We shall neglect the case where, simultaneously, $\frac{1}{\tau}=0$ and $\kappa_n=0$ for some $n$. 

\textbf{Proposition.} For each $n$, there is a pair of eigenvalues $\lambda_n^\pm$  and eigenvectors $\vphi_n^\pm=c_n^\pm \, \vect{\vpsi_n \\ \rho_n^\pm \vpsi_n}$ of $\cM$ with $c_n^\pm=\frac{1}{\sqrt{1+\pare{\rho_n^\pm}^2}}$ (normalisation factor), and:
\begin{equation}\label{eq:beta}
\rho_n^\pm= 
\left\{
\begin{array}{lll}
\frac{1}{2 \, \tau \, w^- \, \kappa_n }\pare{1 \, \pm \, \sqrt{1- 4 \, \mu\, \kappa_n }},& \quad \kappa_n \neq 0, \frac{1}{\tau} \neq 0;\\
w^+ \, \tau, & \quad \kappa_n =0,\frac{1}{\tau} \neq 0;\\
\pm \sqrt{- \frac{w^+}{w^-} \, \frac{1}{\kappa_n}},& \quad \frac{1}{\tau}=0.
\end{array}
\right.
\end{equation}
where:
$$
\frac{1}{\tau}=\pare{\frac{1}{\tau_A} - \frac{1}{\tau_B}}.
$$
and:
$$
\frac{1}{\tau_{AB}}=\pare{\frac{1}{\tau_A} + \frac{1}{\tau_B}}.
$$

Eigenvalues are given by:
\begin{equation}\label{eq:lambda_m}
\lambda_n^\pm = 
\left\{
\begin{array}{llll}
-\frac{1}{2 \, \tau_{AB}} \mp \frac{1}{2 \, \tau} \, \sqrt{1- 4 \, \mu\, \kappa_n },& \quad \frac{1}{\tau} \neq 0;\\
&& \\
-\frac{1}{\tau_A} \, \mp \, \sqrt{-w^- \, w^+ \kappa_n}, 
& \quad \frac{1}{\tau} = 0.
\end{array}
\right.
\end{equation}
with:
$$\mu= w^- \, w^+ \, \tau^2 \geq 0,
$$
%

As a consequence, in addition to the $N$ last eigenvalues $-\frac{1}{\tau_A}$,  $\cL$ admits $2N$ eigenvalues given by \eqref{eq:lambda_m}, while the $2N$ first columns of the matrix $\cP$
(eigenvectors of $\cL$) are: 
\begin{equation}\label{eq:phibeta}
\vcP_\beta=\frac{1}{\sqrt{1+\pare{\rho_n^-}^2}} \, \vect{\vpsi_n \\ \rho_n^- \vpsi_n \\ \vz_N}; \quad 
\vcP_{\beta+N}=\frac{1}{\sqrt{1+\pare{\rho_n^+}^2}} \, \vect{\vpsi_n \\ \rho_n^+ \vpsi_n \\ \vz_N},
 \quad \beta=n=1 \dots N.
\end{equation}
For the $N$ last eigenvectors $\vcP_\beta=\ve_\beta, \beta=2N+1 \dots 3N$.

\paragraph{Remark.} The structure of these eigenvectors is quite instructive. Indeed, the factors $\rho_n^{\pm}$ control the projection of the eigenvectors $\vcP_\beta$ on the space of ACells, thereby tuning \textit{the influence of ACells via lateral connectivity}. 

\paragraph{Proof.}
We use here the generic notation $\lambda_\beta,\vphi_\beta$, $\beta=1 \dots 2N$ for the eigenvalues and associated eigenvectors of $\cM$. If we assume that $\vphi_\beta$ is of the form $\vphi_\beta= \vect{\vpsi_n \\ \rho \vpsi_n}$ for some $n$, then, we have:
$$
\cM.\vphi_\beta = \pare{\begin{array}{cccccc}
&-\frac{I}{\tau_B} & &-w^- \, \cW\\
& w^+ \,I & & -\frac{I}{\tau_A}
\end{array}
}.\vect{\vpsi_n \\ \rho \vpsi_n}
= \vect{\pare{-\frac{1}{\tau_B}  -w^- \, \rho \, \kappa_n}.\vpsi_n\\ \pare{-\frac{\rho}{\tau_A} + w^+ }.\vpsi_n} = \lambda_\beta \, \vect{\vpsi_n \\ \rho \vpsi_n},
$$
which gives:
$$
\left\{
\begin{array}{lll}
\pare{-\frac{1}{\tau_B}  -w^- \, \rho \, \kappa_n} =\lambda_\beta \\ 
\pare{-\frac{\rho}{\tau_A} + w^+ } = \lambda_\beta \rho,
\end{array}
\right.
$$
%
%
%
leading to:
$$
w^- \, \kappa_n \, \rho^2 -  \frac{1}{\tau} \, \rho + w^+ =0,
$$
where:
$$
\frac{1}{\tau}=\frac{1}{\tau_A} - \frac{1}{\tau_B}.
$$
This gives, if $\kappa_n \neq 0$ and $\frac{1}{\tau} \neq 0$: 
%
%
$$
\rho_n^\pm=\frac{1}{2 \, \tau \, w^- \, \kappa_n }\pare{1 \, \pm \, \sqrt{1- 4 \, \mu\, \kappa_n }},
$$
where:
$$
\mu= w^- \, w^+ \, \tau^2 \geq 0.
$$
Thus, for each $n$, there are two eigenvalues:
$$
\lambda_n^\pm = -\frac{1}{2 \, \tau_{AB}} \mp \frac{1}{2 \, \tau} \, \sqrt{1- 4 \, \mu\, \kappa_n },
$$
with:
$$
\frac{1}{\tau_{AB}}=\frac{1}{\tau_A} + \frac{1}{\tau_B}.
$$
Note that $\frac{1}{\tau_{AB}} \geq \frac{1}{\tau}$.

If $\kappa_n=0$, $\frac{1}{\tau} \neq 0$, $\rho_n^\pm=w^+ \, \tau$.
Then:
$$
\lambda_\beta =-\frac{1}{\tau_B}  -w^- \, \rho \, \kappa_n
$$
Finally, if $\kappa_n \neq 0$, $\frac{1}{\tau}=0$, ($\tau_A=\tau_B$), $\rho_n^\pm=-\frac{w^+}{w^-}\,\frac{1}{\kappa_n}$ and $\lambda_n^\pm=-\frac{1}{\tau_B} \, \pm \, \sqrt{-w^- \, w^+ \kappa_n}  $.
If $\kappa_n=0$, $\frac{1}{\tau}=0$ there is no solution for $\rho$ .\\

\textbf{End of proof.}

\paragraph{Remark.} When $\mu=0$, $\cM$ is diagonal: the $N$ first eigenvalues are $-\frac{1}{\tau_B}$, the $N$ next eigenvalues are $-\frac{1}{\tau_A}$. We have, in this case: $\lambda_n^+ = -\frac{1}{\tau_B}  $ and $\lambda_n^- = -\frac{1}{\tau_A}$. Therefore, in order to be coherent with this diagonal form of $\cL$ when $\mu=0$ we order eigenvalues and eigenvectors of $\cM$ such that the $N$ first eigenvalues are $\lambda_\beta=\lambda_n^+, \beta=1 \dots N$, and the $N$ next are  $\lambda_\beta=\lambda_n^-, \beta=N+1 \dots 2N.$

\ssSec{Stability of eigenmodes}{Stability}

\paragraph{Stability of eigenmodes when $\cW$ is symmetric.} 

If $\cW$ is symmetric, its eigenvalues $\kappa_n$ are real,
but $\lambda_\beta$, $\beta=1 \dots 2N$ can be real or complex, depending on $\kappa_n$, as $\mu$ is positive. \\
We have four cases:
\begin{itemize}

\item $\kappa_n < 0$. Then, from \eqref{eq:lambda_m}, $\lambda_\beta$s are real and there are two cases. If $\frac{1}{\tau}>0$ the eigenvalues $\lambda_\beta$, $\beta=1 \dots N$ can have a positive real part (unstable) while $\lambda_\beta$, $\beta=N+1 \dots 2N$  has always  a negative real part (stable);  for $\frac{1}{\tau} <0$ the situation is inverted. In both case, the eigenvalue $\lambda_\beta$ has a positive real part if:
$$
\mu > - \frac{1}{\kappa_n} \, \frac{\tau_A \tau_B}{\pare{\tau_B -\tau_A}^2}  \equiv \mu_{n,u},
$$
which reads as well, using the definition of $\mu$:
\begin{equation}\label{eq:AppUnstabilitySymmetric}
w^- \, w^+ > - \, \frac{1}{\tau_A \, \tau_B} \frac{1}{\kappa_n}
\end{equation}
Thus, $\tau_A, \tau_B$ play a symmetric role.
If $\frac{1}{\tau}=0$ ($\tau_A=\tau_B$), all eigenvalues 
are real. Eigenvalues $\lambda_n^-$ are all stable. The eigenvalue $\lambda_n^+$ becomes unstable if:
$w^- \, w^+ > - \, \frac{1}{\tau_A^2} \frac{1}{\kappa_n}$, corresponding to \eqref{eq:AppUnstabilitySymmetric}.

\paragraph{Proof}
 There are two cases.
\begin{itemize}
\item $\frac{1}{\tau} > 0 \Leftrightarrow \tau_A < \tau_B$.
$$\lambda_\beta=-\frac{1}{2 \, \tau_{AB}} \pm \frac{1}{2 \, \tau} \, \sqrt{1- 4 \, \mu\, \kappa_n } > 0$$
$$
\Leftrightarrow \pm \sqrt{1- 4 \, \mu\, \kappa_n } >  \frac{\tau}{\tau_{AB}}
$$
Only $+$ is possible (because $\tau$ and $\tau_{AB}$ are positive). This gives:
$$
1- 4 \, \mu\, \kappa_n >  \frac{\tau^2}{\tau_{AB}^2}
$$
$$
1- \frac{\tau^2}{\tau_{AB}^2} = - 4 \, \frac{\tau_A \tau_B}{\pare{\tau_B -\tau_A}^2}> 4 \, \mu\, \kappa_n  
$$
which is possible because $\kappa_n < 0$. Thus, $\lambda_\beta$ is unstable if:
$$
\mu > - \frac{1}{\kappa_n} \, \frac{\tau_A \tau_B}{\pare{\tau_B -\tau_A}^2}  \equiv \mu_{n,u}.
$$
\item $\frac{1}{\tau} < 0 \Leftrightarrow \tau_A > \tau_B$.
$$-\frac{1}{2 \, \tau_{AB}} \pm \frac{1}{2 \, \tau} \, \sqrt{1- 4 \, \mu\, \kappa_n } > 0$$
$$
\Leftrightarrow \pm \sqrt{1- 4 \, \mu\, \kappa_n } <  \frac{\tau}{\tau_{AB}}
$$
Only $-$ is possible (because $\tau<0$). This gives:
$$
1- 4 \, \mu\, \kappa_n >  \frac{\tau^2}{\tau_{AB}^2},
$$
the same condition as in the previous item.
\item If $\frac{1}{\tau}=0$,  $\lambda_\beta=-\frac{1}{\tau_A} \, \mp \, \sqrt{-w^- \, w^+ \kappa_n}$ so that eigenvalues are real. The eigenvalue with the minus sign ($\lambda_n^-$ )  are all stable. The eigenvalue $\lambda_n^+$ becomes unstable if:
$$
w^- \, w^+ > - \frac{1}{\tau_A^2} \frac{1}{\kappa_n}.
$$
\end{itemize}
\textbf{End of proof.}

\item $\kappa_n > 0$. Then $\lambda_\beta$, $\beta=1 \dots 2N$ are real or complex. If $\frac{1}{\tau} \neq 0$ they are complex if:
$$
\mu > \frac{1}{4 \, \kappa_n} \equiv \mu_{n,c}.
$$
In this case the real part is $-\frac{1}{2 \, \tau_{AB}}$, the imaginary part is $\pm \frac{1}{2 \, \tau} \, \sqrt{1- 4 \, \mu\, \kappa_n }$, and all eigenvalues are stable. If $\mu \leq \mu_{n,c}$ eigenvalues $\lambda_\beta$ are real and all modes are stable as well. Indeed:

\begin{itemize}
\item If $\frac{1}{\tau}=0$, all eigenvalues are equal to $-\frac{1}{2 \, \tau_{AB}}$, hence are stable.
\item If $\frac{1}{\tau} > 0 \Leftrightarrow \tau_A < \tau_B$.
$$-\frac{1}{2 \, \tau_{AB}} \pm \frac{1}{2 \, \tau} \, \sqrt{1- 4 \, \mu\, \kappa_n } > 0$$
$$
\Leftrightarrow \pm \sqrt{1- 4 \, \mu\, \kappa_n } >  \frac{\tau}{\tau_{AB}}
$$
which is not possible because $\frac{\tau}{\tau_{AB}} > 1$ whereas $\sqrt{1- 4 \, \mu\, \kappa_n } < 1$.
\item If $\frac{1}{\tau} < 0 \Leftrightarrow \tau_A > \tau_B$.
$$-\frac{1}{2 \, \tau_{AB}} \pm \frac{1}{2 \, \tau} \, \sqrt{1- 4 \, \mu\, \kappa_n } > 0$$
$$
\Leftrightarrow \pm \sqrt{1- 4 \, \mu\, \kappa_n } <  \frac{\tau}{\tau_{AB}}
$$
Only $-$ is possible because $\tau < 0$. 
$$
1- 4 \, \mu\, \kappa_n > \frac{\tau^2}{\tau_{AB}^2}
$$
which is not possible because $\frac{\tau}{\tau_{AB}} > 1$ whereas $\sqrt{1- 4 \, \mu\, \kappa_n } < 1$.
%
\end{itemize}

\end{itemize}

\paragraph{Stability of eigenmodes when $\cW$ is asymmetric.}

If $\cW$ is asymmetric, eigenvalues $\kappa_n$ are complex, $\kappa_n=\kappa_{n,r} \, + \, i \, \kappa_{n,i}$. We write $\lambda_\beta = \lambda_{\beta,r} \, + \, i \, \lambda_{\beta,i}$, $\beta=1 \dots 2N$ with:
$$
\left\{
\begin{array}{llll}
\lambda_{\beta,r} &=& -\frac{1}{2 \, \tau_{AB}} \, \pm \, \frac{1}{2 \, \tau}  \, \frac{1}{\sqrt{2}} \, \sqrt{a_n+u_n}\,  ;\\
\lambda_{\beta,i} &=& \pm \, \frac{1}{2 \, \tau}  \, \frac{1}{\sqrt{2}} \, \sqrt{u_n-a_n},
\end{array}
\right.
$$
where $a_n=1- 4 \, \mu\, \kappa_{n,r}$ and $u_n=\sqrt{\pare{1- 4 \, \mu\, \kappa_{n,r}}^2 + 16 \, \mu^2 \, \kappa_{n,i}^2}$ $=\sqrt{1-8 \, \mu \, \kappa_{n,r}^2 + 16 \, \mu^2 \, \abs{\kappa_{n}}^2}$. Note that we recover the real case when $\kappa_{n,i}=0$ by setting $u_n=a_n$.\\

Instability occurs if 
%
$$a_n+u_n > 2 \frac{\tau^2}{\tau_{AB}^2},$$
a condition depending on $\kappa_{n,r}$ and $\kappa_{n,i}$. 

%

\end{appendices}

\end{document}